# *Performance-based Screening of Porous Materials for Carbon Capture*

Amir H. Farmahini[1]*, Shreenath Krishnamurthy[2], Daniel Friedrich[3], Stefano Brandani[4], Lev Sarkisov[2,4]*

[1] Department of Chemical Engineering and Analytical Science, School of Engineering, The University of Manchester, Manchester M13 9PL, United Kingdom

[2] Process Technology Department, SINTEF Industry, Oslo 0373, Norway

[3] School of Engineering, Institute for Energy Systems, The University of Edinburgh, EH9 3FB, United Kingdom

[4] School of Engineering, Institute of Materials and Processes, The University of Edinburgh, Sanderson Building, EH9 3FB, United Kingdom

*Corresponding authors: Amir H. Farmahini a.farmahini@manchester.ac.uk; Lev Sarkisov Lev.Sarkisov@manchester.ac.uk

**Abstract**

Computational screening methods have been accelerating discovery of new materials and deployment of technologies based on them in many areas from batteries and alloys to photovoltaics and separation processes. In this review, we focus on post-combustion carbon capture using adsorption in porous materials. Prompted by unprecedented developments in material science, researchers in material engineering, molecular simulations, and process modelling have been interested in finding the best materials for carbon capture using energy efficient pressure-swing adsorption processes. Recent efforts have been directed towards development of new multiscale and performance-based screening workflows where we are able to go from the atomistic structure of an adsorbent to its equilibrium and transport properties for gas adsorption, and eventually to its separation performance in the actual process. The objective of this article is to review the current status of these emerging approaches, explain their significance for materials screening, while at the same time highlighting the existing pitfalls and challenges that limit their application in practice and industry. It is also the intention of this review to encourage cross-disciplinary collaborations for the development of more advanced screening methodologies. For this specific reason, we undertake an additional task of compiling and introducing all the elements that are needed for the development and operation of the performance-based screening workflows, including information about available materials databases, state-of-the-art molecular simulation and process modelling tools and methods, and the full list of data and parameters required for each stage.

## Table of Contents

## 1. Introduction

Recent discoveries in material science and advances in computational chemistry are having a profound impact on the way we approach design and optimization of chemical processes, devices, and technologies.

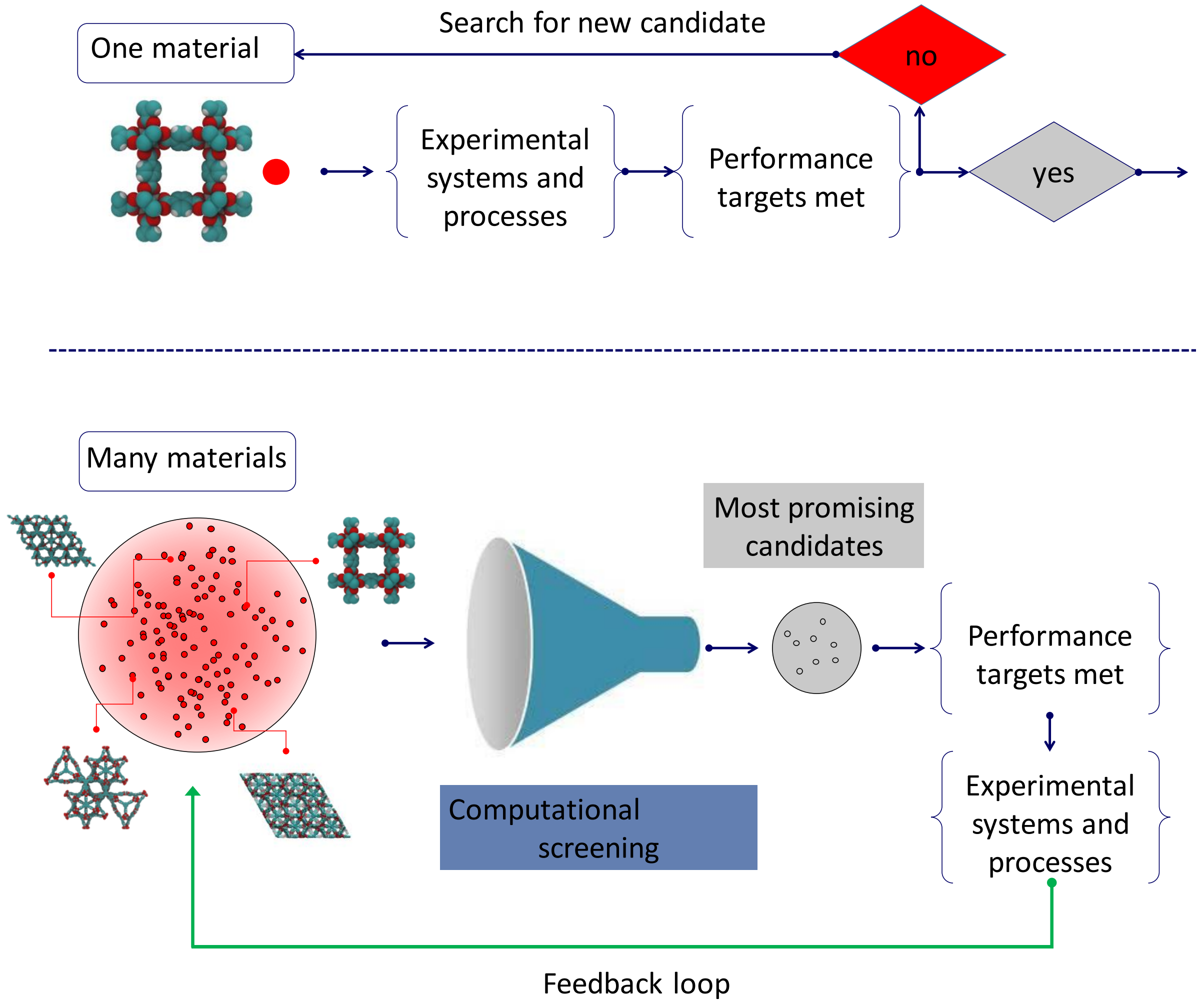

**Figure 1.** Traditional (top) and emerging (bottom) approaches to material selection for an application. Within the emerging approaches, a significant role is played by computational screening of a large database of materials, with the experimental effort focused only on the most promising candidates.

Traditionally, the workflow for the design of a process or a device would focus on a small number of materials available for experimentation and testing, as shown in the top panel of Figure 1. If performance of the material was not satisfactory, the experience gained in the process and the intuition of the investigator would guide the search for another material to be tried or suggest some modification of the existing material.

Unprecedented developments in material science in the last 20-30 years have challenged this approach. Indeed, over this period, several new classes of materials have been discovered with each class encompassing hundreds or even thousands of members. Testing all these materials in relevant experiments, according to the traditional workflow, is prohibitive in terms of cost and effort. Alternatively, performance of the materials can be first tested using a computer model with a view to focus the experimental phase only on the most promising candidates. Moreover, using computational methods allows chemists and materials scientists to explore the performance of hypothetical, not yet synthesized materials. This is important for both the new classes of materials and for the well-known classes, where the phase space is significant (*i.e.* alloys). Within the new workflow, the process starts from the assembly of a large database of materials (real, hypothetical or both), shown in the bottom of Figure 1 as a cloud of points. Their performance is then assessed using computational modelling. The most promising candidates are passed on to the experimental phase for the validation and testing. In the feedback loop, the information obtained at the experimental stage is used to search for specific properties and functionalities within the database of materials to further enhance performance of the process.

This is a new strategy for *in silico* discovery of new materials and high-throughput screening of materials for various applications. A review article by Curtarolo *et al.* [1] identifies the following areas where this strategy is likely to make the most significant impact: alloys, solar materials, photocatalytic water splitting, materials for carbon capture and sequestration, nuclear detection and scintillators, topological insulators, piezoelectric and thermoelectric materials, materials for catalysis, energy storage and batteries. These developments also come with new challenges, *e.g.*, how to organize and share large material databases; how to navigate the clouds of materials properties to identify the most promising candidates, how to relate material properties to their actual performance at the process level. Some of these challenges have been recognized through forming large scale collaborative projects, such as the Material Genome Initiative [2] and the Materials Cloud project [3].

Carbon capture reviewed in the article by Curtarolo *et al.* is an example of a chemical separation process [1]. Significant reduction of carbon emissions from power plants has been on the top of the agenda in the scientific and technology policies of the major economies in the world. Most decarbonisation scenarios show that carbon capture is needed to reach net zero emissions [4]. The main challenge in the implementation of carbon capture technologies for the existing plants is a significant additional energy (and, ultimately, financial) cost associated with the process. Adsorption and membrane separations have been considered as energy efficient alternatives to the traditional amine-solution based processes. Similar factors have been driving developments in other chemical separation processes: as has been recently discussed by Sholl and Lively [5], overall these processes consume 15% of the worldwide energy and, naturally, there is a significant incentive to reduce this impact by developing more efficient alternatives.

At the heart of an adsorption and membrane process is the material used as an adsorbent or a membrane. The efficiency of the process hinges on the characteristics of this material and the interplay between the material characteristics and process configuration. Recently, several new family of porous materials such as Metal-Organic Frameworks (MOFs) [6-8], Zeolitic Imidazolate Frameworks (ZIFs) [9], Covalent Organic Frameworks (COFs) [10], Porous Organic Cages (POCs) [11], Porous Aromatic Frameworks (PAFs) [12], and polymers such as Porous Polymer Networks (PPNs) [13] and Polymers with Intrinsic Microporosity (PIMs) [14, 15] have been discovered. A common motif, associated with these families, is a large number of (synthesized and hypothetical) members available within each family, as well as tunability and exquisite control of structural characteristics of the materials such as surface area, pore size distribution (PSD) and surface chemistry. This has prompted an extensive research effort to explore these new landscapes of structures to identify new porous materials with superior characteristics for adsorption applications, such as carbon capture.

The initial efforts in this field were led by the molecular simulation community, with various computational tools being used to obtain structural (*e.g.* surface area, porosity) and functional characteristics (*e.g.* equilibrium adsorption data) of the materials. These properties or metrics were then used to explore possible correlations between them and the function of the material in the actual application. An important question emerged from these early computational screening studies concerns the process descriptors or performance metrics: what descriptors and metrics should one actually adopt for ranking and selection of materials for a specific application? A useful metric must somehow reflect the essence of the process under consideration. For example, for methane storage the realistic metric is the working capacity, in other words the specific amount of methane released by the material when pressure is reduced from the storage pressure to the lowest pressure in the device, as oppose to the absolute capacity, corresponding to the lowest pressure being zero.

If for some applications, such as gas storage, a single metric may suffice the selection process, for other more complex, dynamic processes this is not possible. This was eloquently demonstrated by Rajagopalan *et al.* [16] by comparing a broad range of traditional and new separation performance metrics developed over the years and the actual performance of the material in the process simulation using post-combustion $CO_2$ capture as a case study.

In fact, a significant amount of literature and studies have been accumulated over the years on design and optimization of pressure, vacuum, temperature, concentration, electric, and microwave swing adsorption processes, from simplified equilibrium models to more advanced numerical approaches [17-28]. Typically, these studies focus on a particular process configuration and conditions, while the cycle configuration is optimized to meet specific process objectives. In the case of the post-combustion carbon capture application, the objectives (or constrains of the process) are 90% recovery of the $CO_2$ from the feed with 95% purity, as recommended by the DOE based on the emission control targets and storage requirements [29]. The efficiency of the process and hence performance of the material for the process can then be assessed from the perspective of two metrics: productivity, in other words the amount of $CO_2$ captured per unit of time by a unit of volume of the adsorbent, and energy penalty, which is the energy required to capture a mole of $CO_2$ in the process. These two metrics are in competition with each other and a complex trade-off between them cannot be captured using simplified equilibrium-based figures of merits.

The co-current developments in computational screenings based on molecular simulations and in advanced process simulations invariably led to the following proposition: what if the screening of porous materials for dynamic adsorption processes can be implemented using realistic process simulations while the microscale properties of materials are provided by molecular simulations? This multiscale screening protocol is schematically depicted in **Figure *2***. According to this diagram, molecular simulations can be used to obtain equilibrium data (*e.g.* adsorption isotherms), dynamic properties (*e.g.* micropore diffusivity) or other materials characteristics (*e.g*. thermal properties), if needed. This information is then fed into a process simulator and the performance of the materials is assessed using the metrics previously developed for dynamic adsorption process analysis.

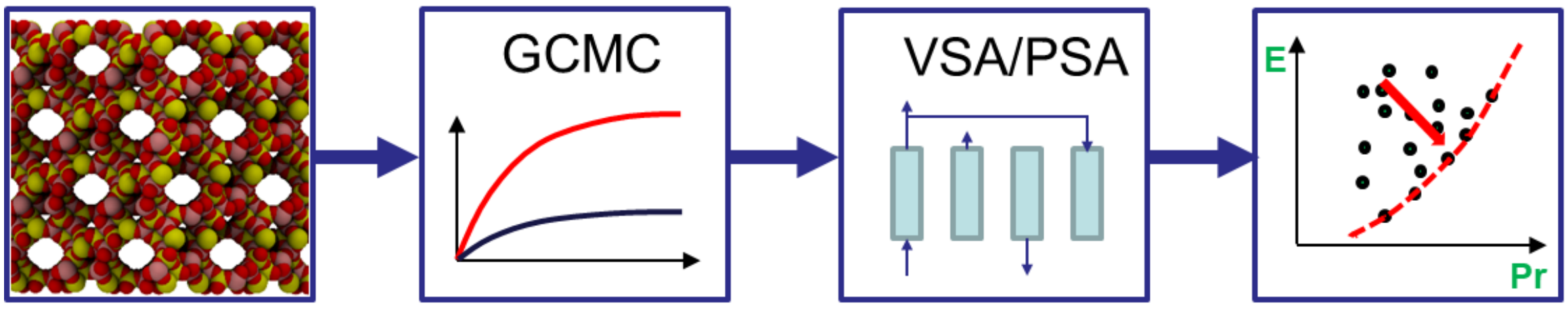

**Figure 2.** Multiscale workflow concepts in Vacuum Swing Adsorption (VSA) and Pressure Swing Adsorption (PSA) engineering. The starting point of the workflow is the structure of the porous material (either experimental or hypothetical, on the left). Molecular simulations are used to obtain equilibrium adsorption and kinetics data. Process simulations are performed for various cycle configurations. Finally, on the right, performance of the material is assessed in terms of energy (E) – productivity (Pr) trade-offs, with the red arrow in the graph indicating progression of this assessment towards the Pareto front (dashed red line).

The first examples of such a multiscale approach were published in two pioneering studies by Hasan *et al.* [30] for *in silico* screening of zeolite materials in the context of carbon capture, and by Banu *et al.* [31] for computational screening of MOFs for hydrogen purification. The early endeavours into the field of performance-based materials screening also exposed a number of challenges. These challenges are associated with consistent and reliable transfer of data and information between the different levels of the simulation (*e.g.* from molecular simulations to process simulations), sensitivity of the process simulation predictions to the properties that cannot be obtained from molecular simulations, lack of experimental validation of the process simulation predictions, the accuracy of the produced material rankings, and propagation of errors, just to name a few.

Early studies indicate that multiscale approaches where one is able to seamlessly progress from a material structure to its performance in the actual process or device will become of immense importance in the near future. With the advent of machine learning and Quantum-Mechanical methods we are witnessing the dawn of material-driven process design, which will have a profound impact on a number of technologies and applications. Hence, this review is prompted by recognition of the importance of this emerging field for materials screening and discovery, and the challenges that have been already encountered in the early studies. Here, not only we aim to provide a critical review of the topic by discussing previous contributions and developments of the field, but also to offer a practical guide and a single source of information for both "users" and "developers" of the performance-based materials screening workflows. The users can include chemists and materials scientists working on the development and characterization of new adsorbents. They can simply use screening workflows to evaluate performance of their newly synthesized (or yet to be synthesized) materials in a target application. The developers, on the other hand, include computational chemists and molecular modellers (who develop molecular models and force fields for molecular simulations); experts in the field of process modelling and optimization (who develop new methods for simulation of the actual processes), and data scientists taking on the development of advanced machine-learning frameworks to better explore materials-performance space.

Although some of the background information provided as part of this review is also available in the classical textbooks and field-specific review articles, it is not always straightforward for various practitioners coming from different backgrounds to quickly extract, compile and synthesize the information needed for the advancement of this highly interdisciplinary field. Moreover, it is important to put different elements of the materials screening workflow (being simulation methods or tools) into the same perspective, and highlight their relations with respect to one another in order to demonstrate the difficulties that arise when they are integrated into a single workflow. Hence, we have undertaken the task to compile and synthesize all the elements and ingredients needed for the development of the aforementioned screening workflows for wide range of readers of this review.

We note that although this review deliberately focuses on the post-combustion carbon capture using Pressure Swing Adsorption (PSA) and Vacuum Swing Adsorption (VSA) processes, the multiscale workflow developed for this purpose and the challenges associated with advancement of this approach will be similar for a wide range of other separations processes such as hydrogen separation, oxygen purification, air separation and so on.

Throughout this review, we aim to highlight the fact that development of accurate and efficient multiscale workflows for realistic screening of porous materials can only be successful, if scientists working on different elements of these workflows are aware of the requirements of other parts. We also hope that the current review can encourage more cross-disciplinary collaborations in this emerging field and lead to the development of multiscale screening tools to be used in a variety of settings, from chemistry labs to chemical engineering pilot plants. With this in mind, the specific objectives of this article are as follows:

*i. Critically review recent contributions and major developments in the field of performance-based materials screening for post-combustion carbon capture using PSA/VSA processes.*

*ii. Provide a practical guide and a single source of information on the principles of molecular and process simulations, full list of data and parameters required at each stage, sources of data, and sources of uncertainties.*

*iii. Review the key challenges in the implementation of the multiscale screening strategies and how they can be tackled.*

*iv. Outline the existing gaps, and propose directions for future developments and trends in this emerging field.*

The review is divided into eight main sections. After this introduction, Sections 2, 3, 4 and 5 will cover the application in question (post-combustion carbon capture), explain different elements of pressure/vacuum swing adsorption processes, discuss hierarchy of metrics that can be used for selection and screening of porous materials for this application, and provide a historical perspective on how computational screening methods evolved over the last 10 years towards current multiscale workflows. We also critically review the methods proposed and used so far in application to materials screening. Section 6 mirrors in its structure the multiscale workflow depicted in **Figure *2***. Here, we will cover practical aspects associated with material databases and the tools available for structural characterization of materials that are currently collected by these databases. Next, we will move to introduce the fundamentals of molecular simulations and process modelling. We will explain how these elements should be used together and as part of a multiscale workflow for materials screening. For each method, we will also introduce available simulation tools and software packages that can be used for performing these types of simulations. Our emphasis will be on explaining what data are required at each stage and what information is obtained at each level, but we will also discuss the gaps in the methods that need to be addressed. In section 7 we review current progress and state-of-the-art in the process-level studies of VSA/PSA systems for carbon capture, including advanced process configurations for this task. In Section 8, we reflect on the overall picture emerging from the multiscale, performance-based screening of porous materials for carbon capture, explore challenges in this field and provide our suggestions for addressing them, which we hope will stimulate further cross-disciplinary approaches and collaborations. Finally, in Section 9, we finish the review with a brief discussion on future opportunities and possible directions of research in multiscale, performance-based screening of materials for carbon capture and other adsorptive separations.

## 2. Post-combustion Carbon Capture

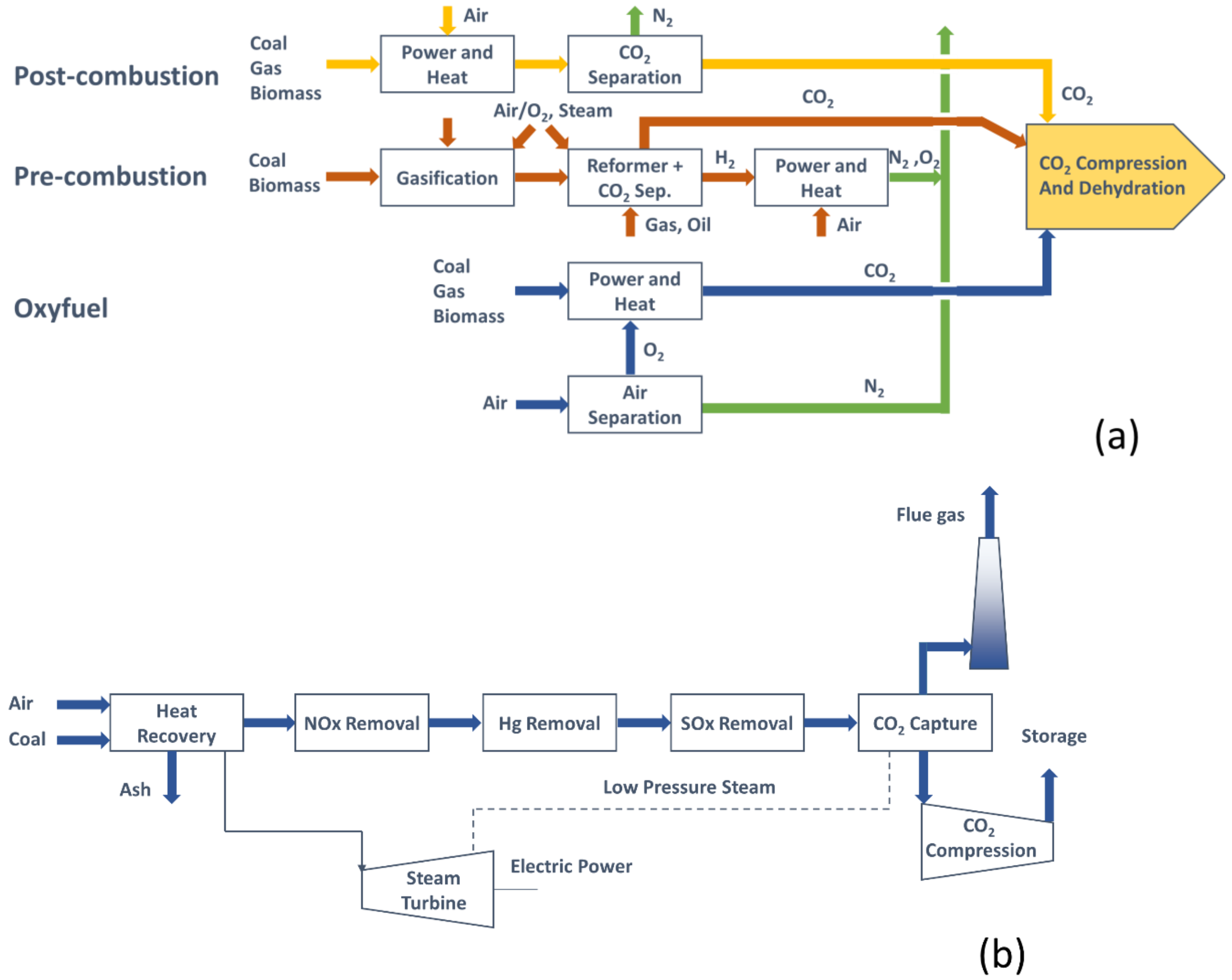

**Figure 3.** Different routes to carbon capture from power plants (a), schematic illustration of post-combustion CCS plant (b). Adapted from Metz, *et al.* [4].

Carbon capture and sequestration (CCS) [32-35] remains one of the key priorities in addressing the global climate change. This is the area where additional energy penalty associated with preventing carbon dioxide emission from power plants is the most significant barrier to the implementation of CCS technology, and any advance in this domain will likely have a profound impact on our ability to control atmospheric carbon dioxide levels. For this reason, CCS has been one of the most explored applications in the context of computational screening of new materials, such as zeolites, MOFs, ZIFs and others [36, 37]. This is also the area where the multiscale screening approaches have made the most significant progress. Hence, CCS and in particular post-combustion capture is the logical focus of this review.

Given the intended target audience of this review (as outlined in the introduction), it is useful to introduce the basic concepts of post-combustion carbon capture, while referring the interested reader to the more specialized and extensive sources on the topic [38-45].

The 2005 IPCC [4] committee identified three possible technologies for carbon capture from power plants, the most significant stationary $CO_2$ emitter globally: pre-combustion carbon capture, oxy-fuel process and post-combustion carbon capture (**Figure *3*** a). In the pre-combustion capture fuel reacts with oxygen (or air) and steam. This produces so-called syngas (synthesis gas) composed predominantly of carbon monoxide and hydrogen. In the water-shift reactor, this mixture reacts with steam to produce carbon dioxide and more hydrogen. Carbon dioxide is then separated from the mixture and the remaining purified hydrogen is used as a clean fuel in various processes. The idea of the oxyfuel process is to use pure oxygen for combustion. This oxygen is produced in the air separation step, which naturally comes with energy cost. However, as the process produces pure carbon dioxide during the combustion step, it does not require any carbon dioxide separation step, saving the costs down the line. Finally, in the post-combustion process carbon dioxide separation is applied to the flue gas from a standard power plant.

Post-combustion capture is the only technology that can be retrofitted onto the existing power plants and therefore is a promising approach in short and medium terms. In fact, detailed analysis of the US National Energy Technology Laboratory's (NETL) CCS database shows that there are currently more than 30 active post-combustion carbon capture plants around the world [46]. This is illustrated in Figure 4. In addition, post-combustion capture can be applied to hard-to-decarbonise emissions such as those from industrial processes and to power plants converted to bioenergy (BECCS) which would enable negative emissions.

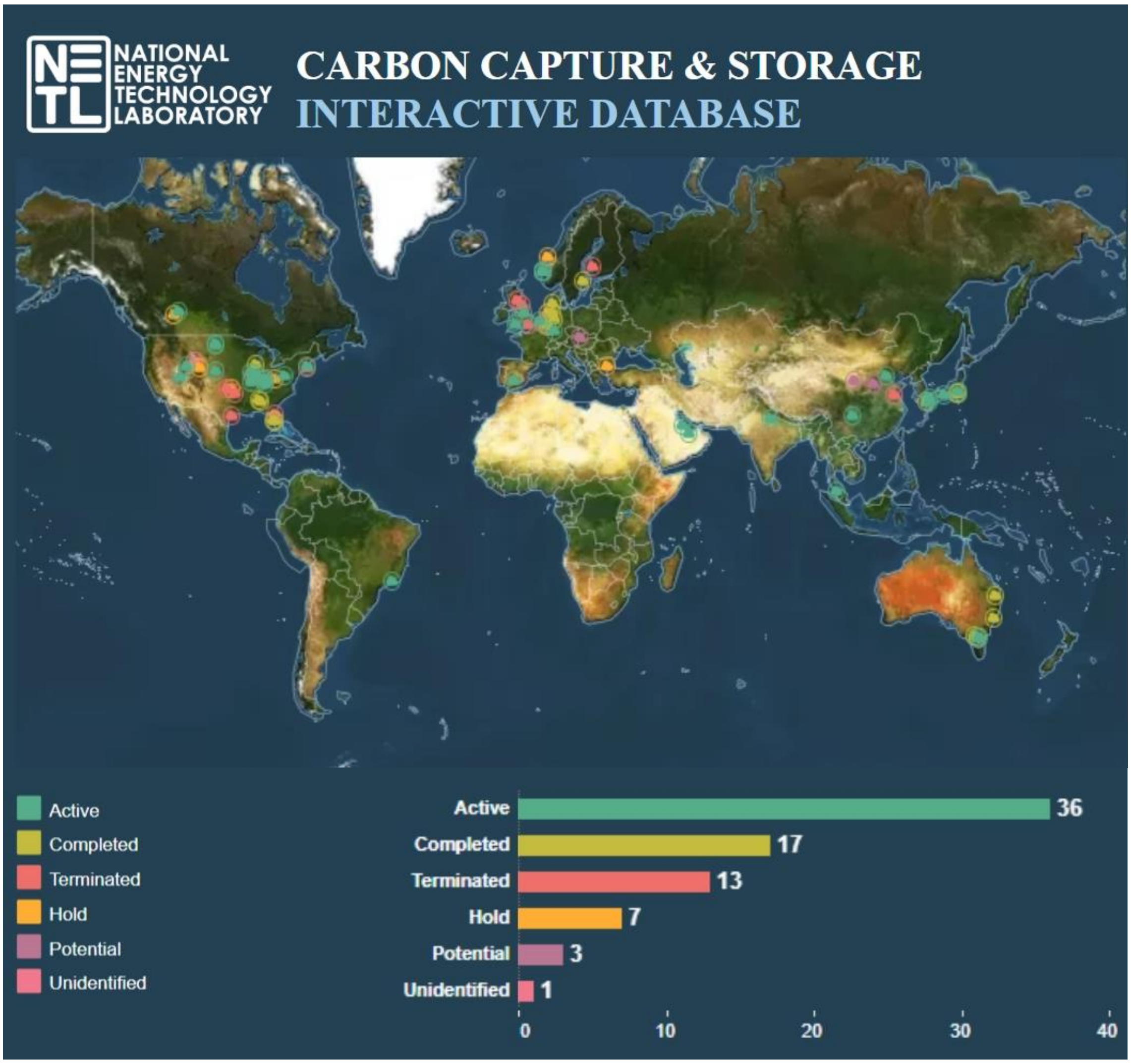

**Figure 4.** Active post-combustion carbon capture plants around the world as shown by green circles. Reprinted from the NETL CCS database [46].

The composition of the flue gas is typically 15-16 vol% $CO_2$, 5-7 vol% $H_2O$, 3-4 vol% $O_2$, and 70-75 vol% $N_2$ for coal-fired power plants. In addition, the flue gases may contain trace amounts (tens and hundreds of ppm) of carbon monoxide, $SO_x$ and $NO_x$. This stream is at 1 bar and 50-75°C [47]. We note, however, that most of the design efforts focus on a simplified separation operation involving only a binary mixture of $CO_2$ and $N_2$ at 1 bar and temperatures below 40°C.

A viable carbon capture technology must remove 90% of carbon dioxide from this flue gas and produce it with 95% purity [29]. The purity constraint is mostly dictated by the requirement to compress the product $CO_2$ gas to 150 bar for further transportation or geological storage. Higher proportion of nitrogen in this stream would incur higher compression costs. These targets set the basis for the comparison of technologies proposed for this task.

Traditional approaches for carbon capture from power plant streams employ solvent-based (*e.g.* amine) absorption processes. It is estimated that the best absorption technologies incur a parasitic energy penalty of about 1.3 MJ per kg $CO_2$ captured [48]. This is associated with a significant energy demand of the solvent regeneration step. Any new technology proposed for carbon capture must demonstrate that it is economically more viable (*i.e.* has lower energy penalty) than the reference, state-of-the-art amine absorption processes.

## 3. Pressure and Vacuum Swing Adsorption for Post-Combustion Carbon Capture

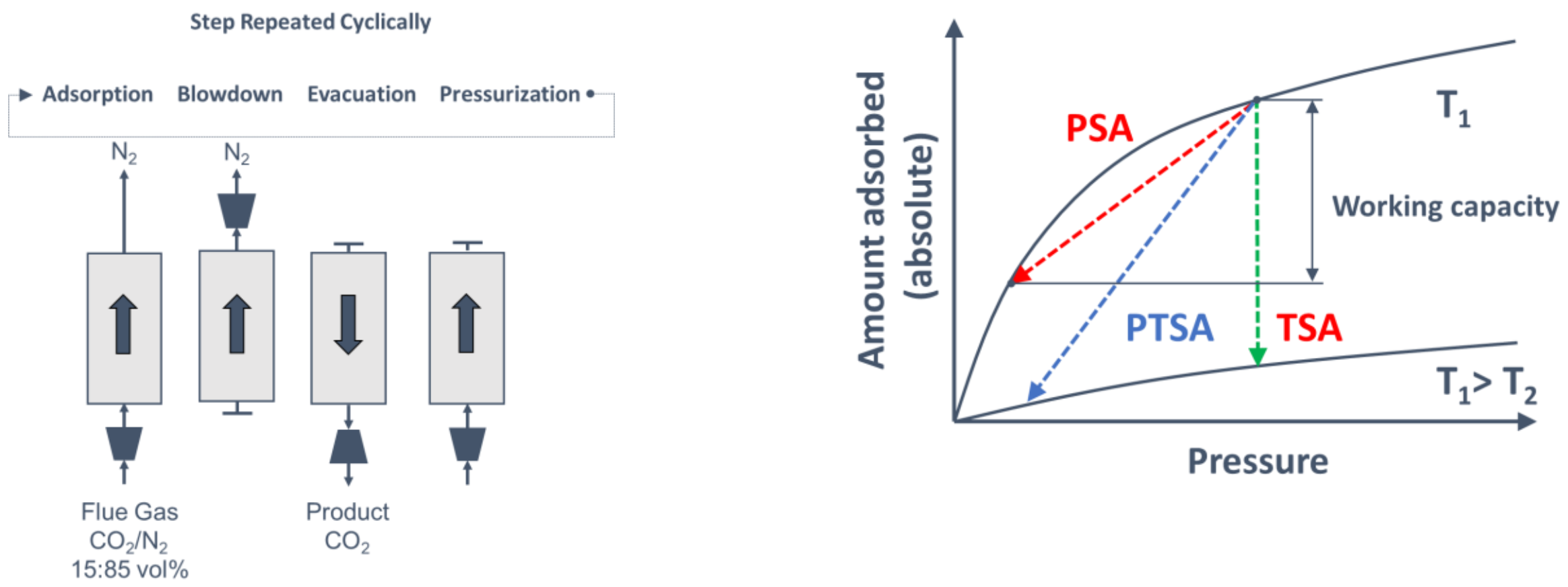

**Figure 5.** A schematic 4-step VSA cycle for separation of $CO_2$ and $N_2$ (a), difference of PSA/PTSA/TSA processes illustrated using equilibrium adsorption isotherms of $CO_2$ and $N_2$ (b).

The main objective of this section is to introduce the key concepts and terminology associated with the pressure/vacuum and temperature swing adsorption processes that are required later in the article. The essential principle behind adsorption separation is that the components of the gas or liquid mixture somehow interact differently with the porous material and this difference can be exploited to separate them. Depending on the nature of this difference, we can distinguish three classes of adsorption-based separation processes: i) kinetic separations, in which diffusion of molecules of the gas mixture in and out of the material happens at significantly different rates; ii) molecular sieving, where one of the components of the mixture is simply too bulky to fit in the pores of the structure while molecules of the other component are able to permeate through the porous structure; or iii) equilibrium separations, where one of the components interacts more strongly with the porous structure via intermolecular interactions. The PSA/VSA processes under consideration in this article belong to this class of processes that constitute the largest group of the industrial adsorption-based separation processes.

To illustrate the principles of a PSA processes, let us consider the diagram in **Figure *5*** (a), which shows different phases of a typical PSA cycle. The main element of this diagram is the adsorption column (schematically shown as just a rectangular box) filled with the porous material, or *adsorbent*. In the first step (adsorption) the feed is introduced in the column. Stronger interacting components (called *heavy components*) are preferentially adsorbed by the porous material in the column, changing the composition of the gas phase. As a result, the product gas stream leaving the column on the other side (so called, *raffinate*) is rich in the *light components* (weakly adsorbing components of the mixture). At some point in time, the adsorbent becomes saturated and will not be able to adsorb anymore of the heavy components. At this point, the adsorption step should be stopped, and the column should go through the regeneration or desorption phase. This phase may consists of a preliminary pressure reduction step (the blowdown step) followed by further reduction of pressure (the evacuation or extraction step), moving the process to the conditions associated with the low loadings on the isotherm and causing desorption of the heavy component (**Figure *5*** (b)). The column is then re-pressurized and goes through the adsorption step again.

The difference in the equilibrium amount adsorbed between the adsorption and desorption cycle is called the *working capacity*. If the PSA system is cycling between ambient pressure and vacuum, then it is called a *Vacuum Swing Adsorption* (VSA) process. The main additional energy cost of PSA/VSA processes is associated with pulling the vacuum (VSA) and compression (PSA). Hence, the work of vacuum pumps and compressors become a key ingredient in the assessment of economic viability of the PSA/VSA processes.

As can be seen from the simplistic description above, the PSA/VSA process is a cyclic process, where the basic unit of the process, the adsorption column, goes through the repeating phases of adsorption and desorption. In the example above, we used pressure swing on the adsorption isotherm to regenerate the column as depicted in **Figure *5*** (b). Alternatively, we could have used higher temperature for regeneration. Indeed, as adsorption from the gas phase is an exothermic process, a higher temperature will shift the equilibrium to lower loadings, leading to desorption. This process is called *Temperature Swing Adsorption* (TSA). A combination of Pressure and Temperature Swing is also possible (PTSA) and the trajectory of conditions associated with this process is also shown in **Figure *5*** (b). Here, it is useful to note that ***Figure 5*** (b) represents an ideal case for PSA/VSA and TSA processes. In reality, PSA/VSA processes are not completely isothermal, and the TSA processes are not fully isobaric. This must be considered when idealized models are used for materials screening based on these processes.

For the PSA/VSA adsorption process to operate continuously, the actual plant consists of several columns going through various stages of the cycle. The number of units and how they are arranged is called *process configuration*. The types of steps involved, the timing of the steps within a single cycle, their duration and other parameters constitute a *cycle configuration*. Developing process and cycle configurations, in order to lower energy penalty and increase productivity constitute the main objective of the PSA/VSA design process.

In case of the post-combustion separation process of a binary mixture, carbon dioxide is the heavy component and nitrogen is the light component. Unlike purification adsorption processes, such as hydrogen production from steam methane reformer off-gas where the main product is the light component, in carbon capture we are interested in the heavy component with specific constraints on its quality and this makes design of the process more complex. Zeolite 13X is the most explored material for this application, both in process modelling and in pilot plant studies. This material is hydrophilic and will adsorb water present in the flue gas, leading to higher cost of the process. The brief introduction provided in this section serves only to establish the most essential elements of the PSA/VSA processes, while for the more extensive reviews of this technology for carbon capture the reader is referred to more specialized and extensive sources [43, 49-52].

## 4. Hierarchy of Performance Metrics for Materials Screening

In Section 2 we described the problem in hand: to capture $CO_2$ from flue gas of a power plant with 90% recovery and 95% purity. Imagine now that we want to identify the best adsorbent material for this from a cloud of many thousands of possible porous materials. To do so, we need a suitable performance indicator (*i.e.* metric) which can correctly quantify separation performance of porous materials, and also is able to sufficiently discriminate between similar materials with different performance. A large number of performance indicators has been proposed for this purpose. In this section, we review the most important of these indicators as reported in the literature, focusing predominantly on their nature, classification, and availability. The information provided here will form the basis of the discussion in the next section where we will illustrate how application of these metrics in the field of computational material screenings evolved over the years leading to wider adoption of the process-level metrics for materials ranking.

Colloquially speaking, one would want to select the best material for a particular application simply by looking at its structure. The specific structural characteristics of a material may include its porosity, density, surface area, pore size distribution (PSD) and so on, see for example Refs [53, 54]. These properties can be either obtained from the experiments, as a part of the standard characterization procedure for every newly synthesized material, or from the computational characterization methods that will be discussed later in this article. We call this group of metrics *intrinsic structural material metrics* (ISMMs). These structural metrics do not tell us anything about how material interacts with its environment. Functional behaviour of materials is described by adsorption equilibrium data (*e.g.* adsorption isotherms, Henry's constants of adsorption, adsorption capacity), transport characteristics (*e.g.* diffusivity), thermal properties (*e.g.* heat capacity, thermal conductivity), see for example Refs [55-58]. These properties constitute another group of metrics that can be termed *intrinsic functional material metrics* (IFMMs).

In separation applications, adsorption is a competitive process between two or more adsorbing species. Naturally, to characterize this competition we need a metric which can compare the behaviour of the material with respect to the competing species. For example, selectivity is the ratio of loadings for two gases which at low pressure can be expressed simply as the ratio of the two Henry's constants. Selectivity is the simplest metric from the group of *hybrid metrics* (HMMs), which combine various adsorbent metrics mentioned above, to more accurately discriminate between adsorbents with different separation performances. Examples of these metrics include: adsorption figure of merits (AFM) [59], sorbent selection parameter (SSP) [60], separation factor (SF) [61], adsorbent performance indicator (API) [62], and adsorbent performance score (APS) [63]. Mathematical definitions of these metrics are provided in **Table 1**.

One important step in the development of more realistic metrics for material screening was the realization that selectivity and working capacity are not necessarily representative of the economic drivers of gas separation processes [16]. To address this limitation, new screening metrics were developed to exploit the correlations between adsorption characteristics of porous materials and the plant-wide economic appraisal of the separation process. The first prominent example of such evaluation metrics is the separation performance metric (SPP) by Braun *et al.* [64], which was developed to represent the most important economic drivers for separation of $CO_2$ from natural gas mixtures. It assumes equilibrium adsorption and desorption in the PSA/TSA/PTSA processes in order to calculate the value of an objective function, which accounts for the amount of captured target gas (*e.g.* $CH_4$), amount of adsorbent material used, and the total energy required for the separation process [64]. The assumption of a process performing fully under equilibrium represents an ideal case scenario, however this condition is not always achieved in the dynamic separation processes such as PSA/VSA. The other limitation of SPP metric is that instead of using conventional cost indicators (*e.g.* capital and operating costs), SPP assumes that all process costs scale with the amount of adsorbent ($M_{ads}$) used in the separation unit [64]. As has been discussed in the same publication, there are cases where a large portion of the capital costs does not depend on the amount of material used in the process, and if these contributions of the capital cost become significantly larger, the amount of material used in the separation unit will become irrelevant [64]. Comparison of SPP, SSP and API metrics with detailed process modelling indicates that for $CO_2/CH_4$ separation, SPP surpasses the other two evaluation metrics in terms of accuracy [64].

Another important example of new evaluation metrics is the parasitic energy (PE) which was first used by Lin *et al*. [65] and Huck *et al*. [66] for evaluation of different classes or porous materials for post-combustion carbon capture. In their analysis, the additional energy required for adsorption carbon capture process consists of: (1) energy to heat the adsorbent material, (2) energy to supply the heat of desorption which is equal to the heat of adsorption, and (3) energy needed to compress $CO_2$ to 150 bar which is a standard requirement for transport and storage [65]. Based on this, the authors formulated a simplified expression for the parasitic energy of a CCS process as a combination of the thermal energy requirement and the compressor work [65]. In the definition of parasitic energy provided by Lin *et al*. [65] equilibrium adsorption and desorption is assumed. As mentioned before, this may not be always the case in dynamic PSA/VSA systems. The parasitic energy curve is however shown to be a useful metric for assessing performance of large groups of porous materials, examples of which are illustrated in **Figure *6*** for all-silica zeolites and hypothetical ZIFs [65].

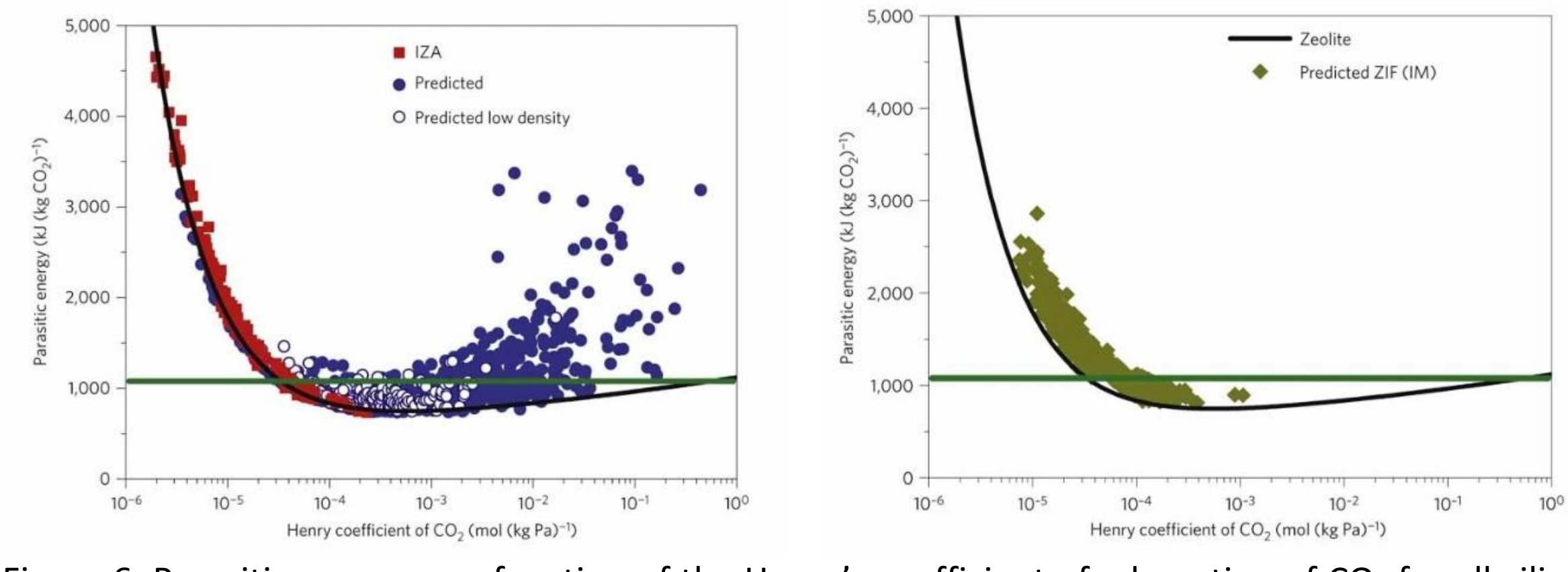

Figure 6. Parasitic energy as a function of the Henry's coefficient of adsorption of $CO_2$ for all-silica zeolites (left) and hypothetical ZIFs (right). The green lines is the parasitic energy of the current

monoethanolamine (MEA) absorption technology. Reprinted with permission from Lin *et al.*, Nature Materials, 2012. 11(7): p. 633-41. Copyright (2012), Springer Nature [65].

Inadequacy of screening metrics which are solely linked to the adsorbent properties and not their performance at the process level has been recently demonstrated by Rajagopalan *et al.* [16] using a case study for post-combustion $CO_2$ capture. Without intending to repeat the entire argument here, one may consider an example selectivity of a candidate material for $CO_2/N_2$ separation using PSA process. On its own, a high value of selectivity is unlikely to be enough to select the material for $CO_2$ separation. For instance, if the material has very low capacity the operation is likely to be very costly, despite high selectivity of the material. This study clearly demonstrates that for complex, dynamic adsorption processes such as PSA/VSA processes for carbon capture, the realistic performance of a specific material must be assessed in the actual process, by performing process simulation and optimization under realistic conditions. For this purpose, a new class of evaluation metrics is required. The metrics used to assess performance of porous materials at the process level are therefore called *process-level metrics* (PLMs) in this review. In this case, a trade-off curve between overall energy penalty of the process and its productivity is used as an evaluation metric for materials screening [16, 67, 68]. Energy penalty and productivity not only are more realistic measures of process performance, they are also more directly related to the economic drivers of the separation process. Therefore, the next natural step in developing realistic evaluation metrics for materials screening is to link the existing process modelling platforms to techno-economic analyses of the process because the ultimate goal of any separation unit is to achieve the design objective at the lowest cost [69-71]. Khurana and Farooq have extended this concept to include a comprehensive costing framework for the entire carbon capture plant [69, 70]. Their integrated optimization framework looks at the separation cost in terms of \$/tonne of $CO_2$ captured or \$/tonne of $CO_2$ avoided, where the latter is defined as the difference between emissions of two power plants, one without a capture unit and the other with a capture unit but both producing the same net amount of electricity [69]. Fully integrated techno-economic analysis of carbon capture plants or any other industrial separation facility can be a daunting task for the purpose of screening of large groups of adsorbent materials that are currently available. As a result of this limitation, more recent studies have attempted to develop *general evaluation metrics* (GEM) that are strongly correlated with the results of the detailed techno-economic analyses [72]. Usually, this is achieved by combining all previously known evaluation metrics into a more general one (*i.e.* GEM) and then reducing complexity of the GEM by removing the elements whose contribution into the correlation coefficient is insignificant [72, 73]. Importance of each feature in the GEM developed by Leperi *et al.* [72] for evaluation of materials performance for post-combustion carbon capture is illustrated in Figure *7*.

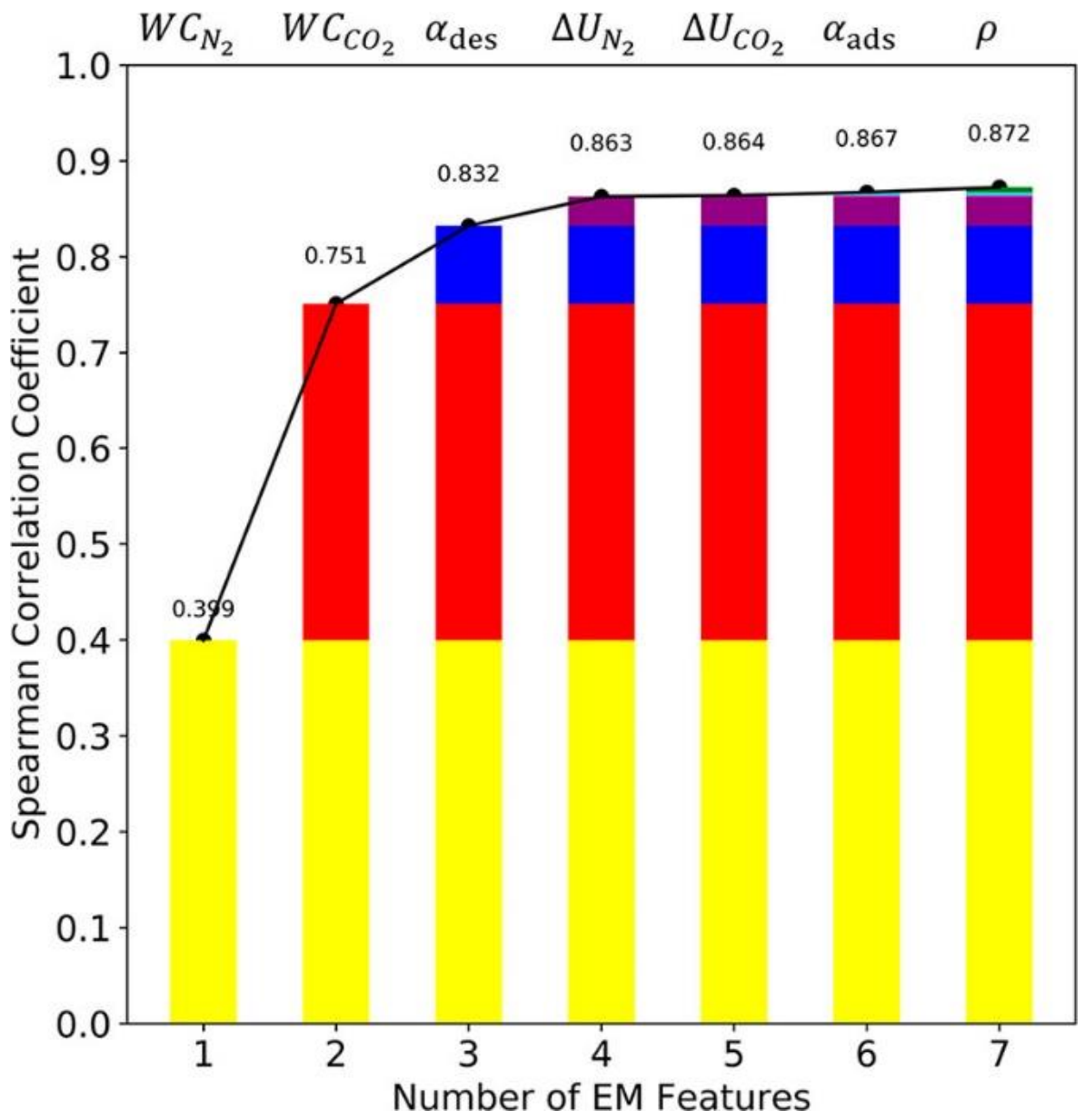

Figure 7. Importance of each feature in the GEM developed by Leperi *et al.* [72] for the Spearman correlation coefficient (SCC). The higher the value of SCC, the more reliable the metric is for predicting the cost of $CO_2$ capture. From right to left, the features are adsorbent density, selectivity at adsorption conditions, internal energy of adsorption for $CO_2$, internal energy of adsorption for $N_2$, selectivity at desorption conditions, working capacity of $CO_2$, and working capacity of $N_2$. SCC for each column is calculated with the feature listed on top plus the features listed in the previous columns. For example, the three GEM features used to calculate SCC in the third column are $N_2$ working capacity, $CO_2$ working capacity, and selectivity at desorption condition. Reprinted with permission from Leperi *et al.*, ACS Sustainable Chemistry & Engineering, 2019. 7(13): p. 11529-11539. Copyright (2019), American Chemical Society. [72].

Leperi *et al.* [72] have shown that this approach is quite promising for the development of universal screening metrics that simultaneously take into account most important characteristics of the process associated with adsorbent material, process optimization and overall economic cost of the plant. Development of new GEMs can particularly benefit from recent advances in machine-learning techniques, if adequately large datasets of techno-economic forecasts were available for training the GEM function.

From the provided review of the hierarchy of metrics, one could make an impression that if the most accurate assessment of the material performance is provided by the detailed process and plant models, then this should be the standard level of description in all material screening protocols. This, however, does not take into account, the computational cost of these metrics. Once the equilibrium adsorption data are available, the hybrid methods provide effectively an instant assessment of the material candidate. Process simulation of a single cycle configuration for a PSA/VSA process may be done in a few minutes on a conventional CPU, whereas cycle optimization for the best performance may take many hours to complete. This computational price tag applied to thousands and tens of thousands of materials would still make routine use of screening of all materials at the process level unaffordable. Hence, this is still an ongoing area of research to develop a multistage screening process, where efficiency of process optimization are improved using novel numerical techniques, or alternatively some preliminary screening is done using hybrid metrics/simplified process models, while accurate process modelling and optimization is only carried out for a selected group of promising materials. The role of emerging numerical techniques for process optimization and screening of large group of materials is discussed in the following sections.

**Table 1.** Performance indicators (performance evaluation metrics)

| Index | Metric Class | Screening Metric | Definition | Reference |
|---|---|---|---|---|
| 1 | ISMM | Pore volume | - - - | - - - |
| 2 | ISMM | Porosity | - - - | - - - |
| 3 | ISMM | Surface area | - - - | - - - |
| 4 | ISMM | Pore limiting diameter | - - - | - - - |
| 5 | ISMM | Pore size distribution | - - - | - - - |
| 6 | IFMM | Enthalpy of adsorption | - - - | - - - |
| 7 | IFMM | Diffusivity | - - - | - - - |
| 8 | IFMM | Henry Selectivity | $\beta_{1,2} = \frac{K_{H,1}}{K_{H,2}}$ | Bae and Snurr, 2011 [74] |
| 9 | IFMM | Adsorption selectivity | $\propto_{1,2} = \frac{q_1^{ads}}{q_2^{ads}} \times \frac{C_2}{C_1}$ | Bae and Snurr, 2011 [74] |
| 10 | IFMM | Working Capacity | $WC = q_{ads,1} - q_{des,1}$ | Bae and Snurr, 2011 [74] |
| 11 | IFMM | Regenerability | $R = \frac{WC_1}{q_1^{ads}} \times 100\%$ | Bae and Snurr, 2011 [74] |
| 12 | HMM | Adsorbent figure of merit | $AFM = WC_1 \frac{(\propto_{1,2\,ads})^2}{\propto_{1,2\,des}}$ | Baksh & Notaro, 1998 [59] |
| 13 | HMM | Sorbent selection parameter* | $SSP = \propto_{1,2} \frac{WC_1}{WC_2}$ | Rege and Yang, 2001 [60] |
| 14 | HMM | Separation Factor | $SF = \frac{WC_1.C_2}{WC_2.C_1}$ | Pirngruber *et al.*, 2012 [61] |
| 15 | HMM | Adsorbent Performance Indicator | $API = \frac{(\propto_{12} - 1)^A WC_1{}^B}{\lvert \Delta H_{ads,1} \rvert^C}$ | Wiersum *et al.*, 2013 [62] |
| 16 | HMM | Adsorbent Performance Score | $APS = WC_1 \times \propto_{1,2}$ | Chung *et al.*, 2016 [63] |
| 17 | HMM | Separation Performance Metric (SPP) | $SPP = \frac{\left(\frac{M_{CH4,raff}}{M_{CH4,feed}}\right)}{\left(\frac{M_{ads}}{M_{CH4,aff}}\right) \times \left(\frac{E}{M_{CH4,raff}}\right)}$ | Braun *et al.*, 2016 [64] |
| 18 | HMM | Parasitic Energy (PE) | $PE = (0.75\eta_{T_{final}} \times Q) + W_{comp}$ | Lin *et al.*, 2012 [65] |
| 19 | PLM | Purity in PSA/VSA | $Purity = \frac{Total\ moles\ of\ CO_2\ in\ the\ extract\ product}{Total\ moles\ of\ CO_2\ and\ N_2\ in\ the\ extract\ product} \times 100$ | Rajagopalan *et al.*, 2016 [16] |
| 20 | PLM | Recovery in PSA/VSA | $\text{Recovery} = \frac{\text{Total moles of } CO_2 \text{ in the extract product}}{\text{Total moles of } CO_2 \text{ fed into the cycle}} \times 100$ | Rajagopalan *et al.*, 2016 [16] |
| 21 | PLM | Specific energy in PSA/VSA | $Specific\ Energy = \frac{\text{total energy used}}{\text{Total moles of } CO_2 \text{ captured}}$ | Rajagopalan *et al.*, 2016 [16] |
| 22 | PLM | Productivity in PSA/VSA | $Productivity = \frac{\text{Total moles of } CO_2 \text{ in the extract product}}{(\text{Total volume of adsorbent}) \times (\text{Cycle time})}$ | Rajagopalan *et al.*, 2016 [16] |
| 23 | GEM | General evaluation metric | $GEM = \frac{WC_1}{WC_{2,mod}^{1.32} \times \propto_{1,2\ des}^{0.25} \times \lvert \Delta H_{N_2} \rvert^{0.97}}$ | Leperi *et al.*, 2019 [72] |

* For Langmuir isotherms. For non-Langmuir systems $SSP = \frac{(\propto_{1,2\ ads})^2}{\propto_{1,2\ des}} \times \frac{WC_1}{WC_2}$ [74]

**Note 1.** Subscripts 1 and 2 always denotes stronger and weaker adsorbing components respectively.

**Note 2.** For evaluation metrics 1 to 16, $WC$, $\propto$, β, $C$, $K_H$ and $\Delta H$ represents working capacity, adsorption selectivity, ideal selectivity, concentration, Henry's constant and enthalpy of adsorption. For SPP, $M_{ads}$, $M_{i,k}$ and $E$ denote mass of adsorbent, moles of species *i* in streams *k* and total energy required for separation [64]. For PE, $Q$, $\eta$ and $W_{comp}$ are the thermal energy requirement, Carnot efficiency and compressor work respectively. For GEM, $\Delta H_{N_2}$ and $WC_{mod}$ stand for enthalpy of adsorption for nitrogen and the modified working capacity as defined in the corresponding reference [72].

## 5. Computational Screening of Porous Materials: A Historical Perspective

In the previous section, we discussed what metrics are available for material screening in adsorption applications through the prism of metric hierarchy from very simple "intrinsic" metrics to process-level metrics. In this section, we take a different, historical perspective on the development of computational screening strategies. This perspective will allow us to review how this field has evolved over time towards current multiscale workflows that incorporate elements of different types of simulation techniques and performance indicators.

The first material screening studies can be tracked back to more than 10 years ago [75-77]. In a pioneering study published in 2010 [55], Krishna and van Baten employed the configurational-biased Monte Carlo (CBMC) and molecular dynamics (MD) simulations to examine adsorption, diffusion, and permeation selectivities for separation of $CO_2/H_2$, $CO_2/CH_4$, $CO_2/N_2$, $CH_4/N_2$ and $CH_4/H_2$ mixtures in a number of zeolite, MOF, ZIF and carbon nanotube (CNT) structures. Their studies provided useful guidelines to the optimum choice of microporous layers that should be used in membrane separations representing a compromise between selectivity permeability and the permeability itself. This study also emphasized the importance of correlations between pore space properties (pore volume, limiting pore diameter, *etc.*) and transport properties (*e.g.* diffusion and permeation) in these classes of porous materials.

Building on the importance of the pore structure characterization, Haldoupis *et al.* [54] analysed pore sizes of more than 250,000 hypothetical silica zeolites to compute the size of the largest adsorbing cavity and pore-limiting diameter for all zeolites. This information can be used to reveal the range of adsorbate molecules that can possibly diffuse through each zeolite. Additionally, the authors computed Henry's constant of adsorption and diffusion activation energy for $CH_4$ and $H_2$ for a subset of 8000 zeolites using a computational method reported in their earlier study [78]. From the diffusion activation energies, they were able to estimate diffusivity of each adsorbate using a simple formulation of the Transition State Theory (TST). The method presented in this study for estimation of diffusion was limited to adsorption at infinite dilution. Calculation of transport properties at higher loadings is much more time-consuming which may limit the ability of the employed method for screening of large group of porous materials. Nevertheless, within the limitation of the methods, Haldoupis *et al.* could successfully demonstrate that using a combination of molecular simulation techniques, one can reasonably assess adsorption properties of a large group of nanoporous crystalline materials for a particular separation application [54].

Application of computational materials screening approaches took another step forward in 2012 when two major studies were published. Namely, Snurr and co-workers used a library of 102 building blocks and a "tinker-toy" algorithm to assemble a database of 137,953 hypothetical MOFs [79]. Using geometric characterization tools and Monte Carlo simulations, they explored their database to identify the most promising structures for methane storage. From this perspective, this is the first example of a computational screening strategy applied to a large group of MOF materials. Later in the same year, Snurr and co-workers [80] simulated adsorption of $CO_2$, $CH_4$, and $N_2$ in more than 130,000 hypothetical MOFs from the same database and subsequently examined their potential for $CO_2$ capture using five different performance metrics including $CO_2$ uptake, working capacity, regenerability, adsorption selectivity and sorbent selection parameter (as defined in **Table 1**). They showed that although the resulting structure-property relationships between pore size, surface area, pore volume, and chemical functionality provide several leads for design of new porous materials, none of the above metrics is actually a perfect predictor of $CO_2$ separation performance. The studies of Snurr and co-workers introduced several concepts that are now central to the computational screening strategies of porous materials. The concepts can be formulated as follows:

i. The modular nature of MOFs allows the use of simple, tinker-toy algorithms to assemble new hypothetical structures simply by linking the building blocks along the appropriate topology. This idea can be extended to other new classes of materials (ZIFs, COFs, etc).

ii. Each material within the database can be explored in terms of structural properties and functional properties. These properties can be used to classify, compare, and organize materials within the database.

iii. Computational screening studies calculate properties mentioned above. Two or more properties correlated to each other form clouds of data points, which can be explored to reveal some promising structure-property relations. An example of structure-property relationships for $CO_2$ separation in more than 130,000 MOFs is shown in **Figure *8***.

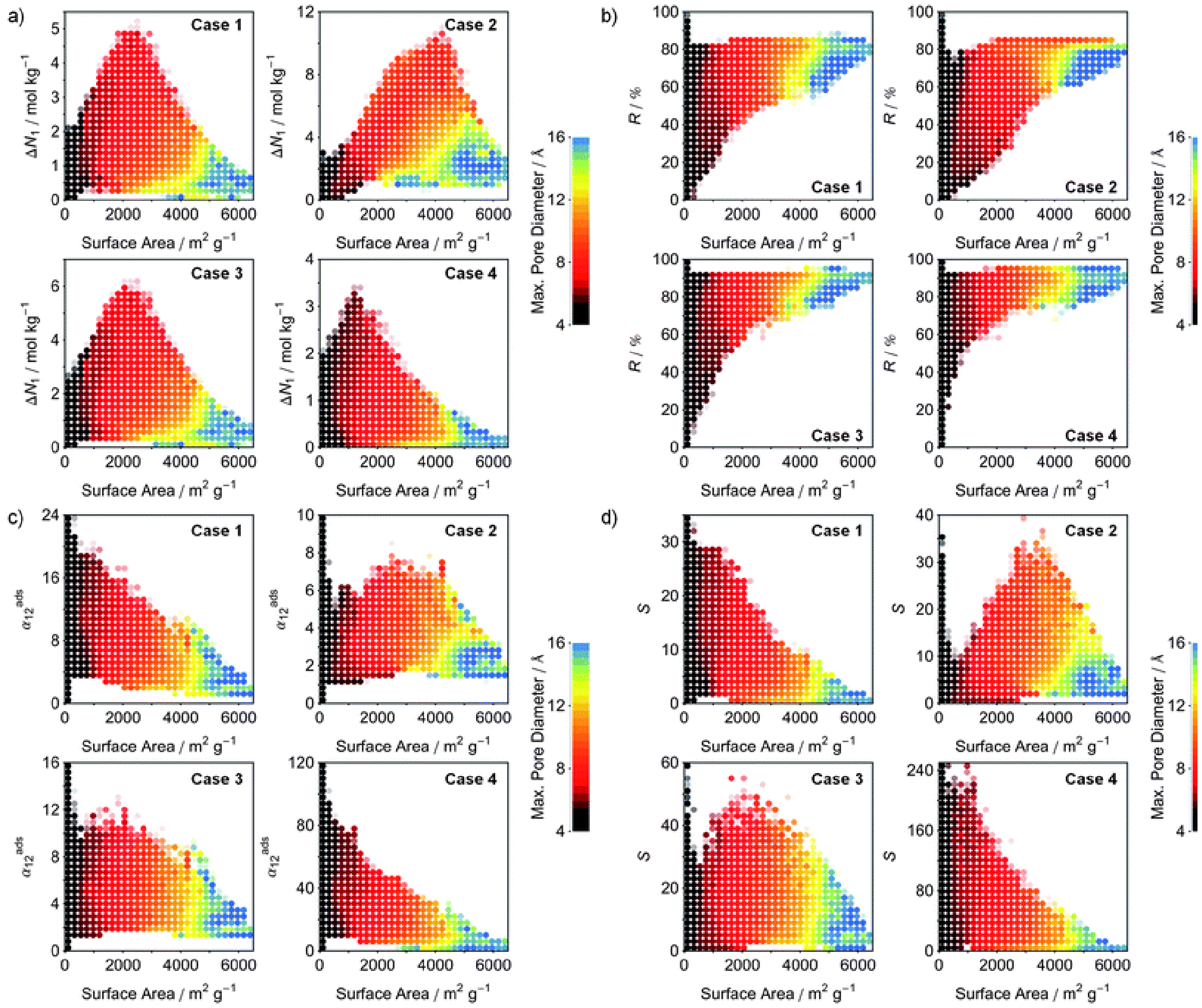

Figure 8. Structure-Property relationships of MOFs as obtained from molecular simulations for $CO_2$ separation. Figures show the relation between working capacity (a), regenerability (b), selectivity (c), and sorbent selection parameter (d) with surface area for four different cases. Each plot is divided into 30 × 30 regions that are represented by a filled circle, if more than 10 (or 25 for selectivity and sorbent selection parameters) structures exist within the region. The four separation cases include: Case (1) natural gas purification using PSA, Case (2) landfill gas separation using PSA, Case (3) landfill gas separation using VSA, and Case (4) flue gas separation using VSA. Reprinted with permission from Wilmer *et al.*, Energy & Environmental Science, 2012. 5(12): p. 9849-9856. Copyright (2012), Royal Society of Chemistry [80].

Further studies in this emerging field also identified some deficiencies of the original database of hypothetical MOFs constructed by Wilmer *et al.* [79] and, consequently, identified challenges and new directions of research. These can be summarized as follows:

i. Structures assembled in the tinker-toy algorithms require further accurate structure optimization using Quantum Mechanical (QM) methods to be more realistic.

ii. In general, we need systematic approaches to organize structures into databases that can be used in molecular simulations.

iii. Early computational and experimental studies in 2000-2010 very clearly demonstrated that the accurate molecular force fields are lacking for new classes of porous materials interacting with gases and liquids. A particularly striking manifestation of this was the failure of the conventional force fields to describe interaction of MOFs featuring open metal sites with carbon dioxide or unsaturated hydrocarbons. Interaction of adsorbents with water presents wholly separate and substantial challenge. This prompted the simulation community to put a significant effort into the development of a new generation of force fields based on the accurate QM potential energy surface. Despite some initial and significant advances, molecular force fields remain mainly specialized and non-transferable; and this is still very much a remaining challenge and an ongoing area of research.

iv. Early studies would use several simple, well-known algorithms to obtain structural characteristics of the porous material. Later, a number of comprehensive and versatile tools were developed (Zeo++ [81], Poreblazer [82], ZEOMICS/MOFomics [83, 84]) to calculate geometric descriptors of porous materials. These tools vary in the algorithms, characteristics they calculate and in the philosophy of use and access.

v. Development of Machine Learning algorithms is needed to establish structure-property relationships within the databases and drive the discovery of new materials with desired functionalities.

At the same time, Smit and co-workers [65] also published a new study on screening of hundreds of thousands of zeolite and ZIF structures using the parasitic energy (PE) as a promising metric for evaluation of materials performance in the context of post-combustion carbon capture. At the molecular simulation level, a combination of grand canonical Monte Carlo (GCMC) simulation, energy grid construction method and Widom test particle insertion technique were employed to obtain equilibrium adsorption characteristics of materials. The PE metric was then used to search for materials that have the potential to reduce the parasitic energy by 30–40% compared to the conventional amine-based absorption technologies [65]. This study proposed a theoretical limit for the minimal parasitic energy that can be achieved for a particular class of porous materials.

A series of articles by Sholl and co-workers [77, 85], and Keskin and colleagues [86-88] had laid the foundation of the computational screening methods for membrane gas separations. This was followed with Kim *et al*. [89] publishing a major study on screening of over 87,000 different zeolite structures for permeation separations. In this publication, the authors estimated the diffusion coefficients of $CO_2$, $N_2$ and $CH_4$ using free energy calculations and the Transition-State Theory (TST) and identified general characteristics of the best-performing structures for $CO_2/CH_4$ and $CO_2/N_2$ membrane separations. For $CO_2/CH_4$ separation, they predicted a structure that outperformed the best known zeolite structure by a factor of 4–7 based on the required area of an ideal membrane which is shown to be mainly dominated by and inversely proportional to the $CO_2$ permeability in the system [89]. In comparison with the results of Haldoupis *et al.* [54], Kim *et al.* demonstrated that screening of porous materials based on purely geometric approaches may deviate from what is predicted from a more advanced energy-based analysis [89].

The above study Kim *et al.* [89] was followed by two other publications with a greater emphasis on MOFs as an emerging group of porous solids for adsorption separation applications. The first study was published in 2014 by Sun *et al.* [56] where 12 materials including six MOFs, two ZIFs, and four zeolites were studied for removal of $SO_2$, $NO_x$ and $CO_2$ from the flue gas mixtures. They used grand canonical Monte Carlo (GCMC) simulations to predict mixture adsorption isotherms and selectivity of the candidate materials for separation of $SO_2$, $NO_x$ and $CO_2$ in a mixture containing $N_2$, $CO_2$, $O_2$, $SO_2$, $NO_2$ and NO. They compared the working capacity, absolute adsorption and adsorption selectivity as three different performance indicators to select the best performing materials. It was concluded that Cu-BTC and MIL-47 were the best adsorbents for separation of $SO_2$ from the flue gas mixture. For the removal of NOx however, Cu-BTC was identified as the best performing material. Finally, for the simultaneous removal of $SO_2$, $NO_x$ and $CO_2$, Mg-MOF74 was found to be the best candidate. The three performance indicators (namely the working capacity, absolute adsorption, and adsorption selectivity) used for evaluation of materials performance in this study only focus on the ability of materials to adsorb different gases at equilibrium. It does not take into account the role of transport which will be important in real dynamic processes. It also neglects the energy penalty associated with the regeneration of the bed.

The second study from this group was published by Huck *et al.* [66] in 2014 focusing on screening of 60 different synthesized and hypothetical materials including MOFs, zeolites and porous polymer networks (PPNs) for post-combustion carbon capture. Acknowledging that several evaluation criteria have been already introduced, this publication emphasized the use of parasitic energy as a more realistic metric for materials screening. This is because parasitic energy takes into account the energy penalty associated with the compression process (needed for $CO_2$ storage), as well as several essential thermodynamic properties such as the thermal energy required for heating up the adsorption bed, and the heat required to regenerate it [66]. **Figure *9*** demonstrates correlation of parasitic energy with other performance indicators calculated for separation of $CO_2$ using a temperature-pressure swing adsorption (TPSA) process from a coal-fired power plant.

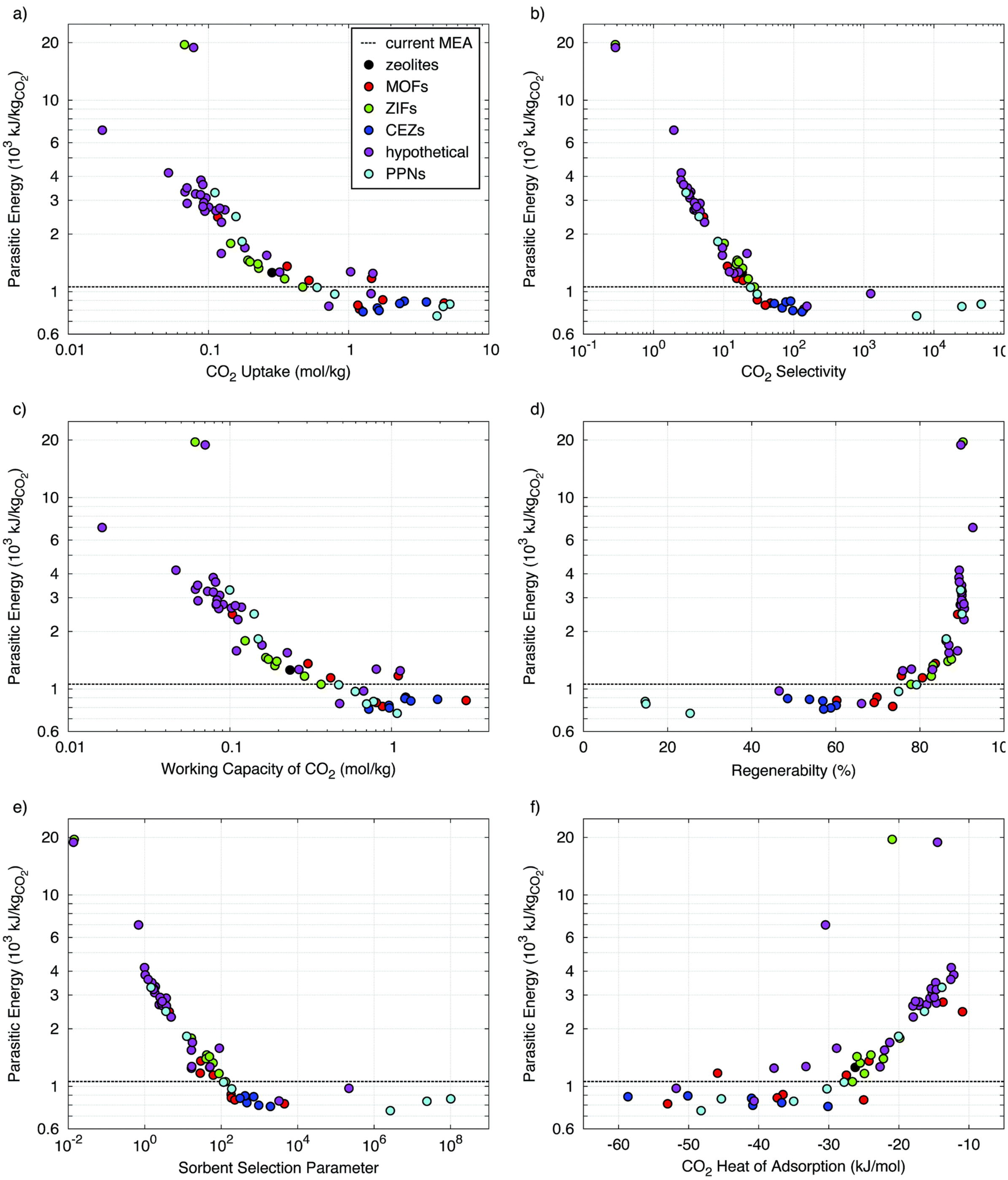

Figure 9. Correlation of parasitic energy with other performance indicators in a PTSA process for $CO_2$ capture. Reprinted with permission from Huck *et al.*, Energy & Environmental Science, 2014. 7(12): p. 4132-4146. Copyright (2014), Royal Society of Chemistry [66].

Using parasitic energy as the evaluation metric, the authors identified Mg-MOF-74, PPN-6-CH2TETA, and PPN-6-CH2DETA as the most promising materials for CCS in coal and natural gas fired power plants, and for direct air capture, respectively.

In a more recent study focused on membrane separation, Qiao *et al.* [57] screened 137,953 MOFs in an attempt to identify the best performing candidates for separation of $CH_4$, $N_2$ and $CO_2$. In a four-stages strategy, the authors employed a combination of geometric pore characterization metrics (*e.g.* pore limiting diameter (PLD), pore size distribution), equilibrium (Henry's constant) and transport properties (diffusivity and permeability) for materials screening showing that the PLD and pore size distribution are the two key factors governing diffusion and permeation of different gases in MOFs [57].

In early 2016, Braun *et al.* [64] published a new study to explore performance of all-silica zeolites for $CO_2$ capture from natural gas where for the first time, the inadequacy of some of the above mentioned adsorbent metrics for materials screening was highlighted. This study suggested that selectivity and working capacity are not necessarily representative of the economic drivers that chemical process designers actually consider [64]. The authors further argued that the use of these metrics can even be deceptive. As has been discussed in the previous section, as an alternative, the authors developed a new metric called separation performance parameter (SPP), which was designed to represent the economic drivers behind $CH_4/CO_2$ separation, and applied this metric to explore separation performance and structure–property relationship of tens of thousands of all-silica zeolites recorded in the International Zeolite Association (IZA) database [90] and the Predicted Crystallography Open Database (PCOD) of hypothetical zeolites [91].

The year 2016 also witnessed publication of more advanced screening studies. In particular, Snurr and co-workers reported on high-throughput screening of MOFs for $CO_2$ capture in the presence of water [58]. The article focused on the competitive co-adsorption of water as potentially an adverse issue in the deployment of adsorption-based $CO_2$ capture technologies. Here, the computational screening was conducted to search for MOFs with high $CO_2/H_2O$ selectivity. The screening workflow consisted of several steps as described below: initially, the framework charges were computed for 5109 MOFs using the extended charge equilibration method (EQeq) [92] which is an approximate, but computationally affordable technique for this purpose. In the next step, the Henry's constants of all MOFs were calculated using the Widom particle insertion. Following this step, the 15 most selective MOFs were identified based on the ratio of Henry's constant for $CO_2$ and $H_2O$. The resulting pre-screened materials were investigated further using more rigorous simulation techniques. For these materials, partial atomic charges were computed using the Repeating Electrostatic Potential Extracted ATomic (REPEAT) method [93], which is more accurate compared to the EQeq technique and is based on electron density distributions obtained from QM DFT calculations. Further, GCMC simulations were carried out to calculate the binary and ternary adsorption of $CO_2/H_2O$ and $CO_2/H_2O/N_2$ mixtures for the 15 pre-selected MOFs. GCMC simulated-adsorption isotherms were then used to identify MOFs with the highest $CO_2$ selectivity over both water and nitrogen. This study highlights the importance of electrostatic interactions in describing the $H_2O$-MOF interactions. On this basis, the authors suggested that accurate charge calculation methods are required to conduct similar screening studies. They also demonstrated a correlation between small pore sizes and strong binding of $CO_2$ which can limit adsorption of water at high humidity by preventing the formation of water clusters inside these pores [58].

Later in 2017, Li *et al.* [94] published a new screening study to explore multivariate metal–organic frameworks (MTV-MOFs). The authors constructed a new database of ~10,000 MTV-MOFs with mixed linkers and functional groups. A GCMC-based high-throughput computational screening method was employed to identify the high-performing candidates for $CO_2$ capture. They showed that compared to their parent MOFs, functionalized structures consistently exhibit better $CO_2/N_2$ selectivity; and in most cases even $CO_2$ capacity is improved. This work is particularly interesting as it demonstrated that arrangements of mixed linkers containing different functional groups can result in a combinatorial explosion in the number of possible structures which can then be mined to increase structural diversity and surface heterogeneity of materials space. This extended search space may contain candidate materials with higher potential for $CO_2$ capture.

Almost the entire studies reviewed up to this point had focused on the use of simple performance indicators (class of ISMM, IFMM and HMM) which are associated with structure or microscale function of adsorbents. These metrics normally consider simple properties such as the pore limiting diameter (PLD), pore size distribution (PSD), Henry's constant of adsorption ($K_H$), adsorption working capacity (WC), selectivity and micropore diffusion. As discussed earlier in Section 4, one can use these performance indicators to reveal correlations between materials structure and functions in microscale level which is important for fundamental understanding of the system, however these metrics fail to realistically predict separation performance of materials at the process level for dynamic adsorption processes such as PSA/VSA. The above realization gradually gave rise to the wider use of process-level metrics (PLM) for materials screening leading to design of multiscale screening workflows which combine various molecular simulation methods with process modelling and optimization.

The idea of constructing a multiscale simulation workflow through combining molecular simulations and process optimization for the purpose of materials screening has been originally presented by Hasan *et al.* in 2013 [30]. They had already used this method for cost-effective capture of $CO_2$ using zeolites as adsorbents. A similar multiscale approach was also adopted by Banu *et al.* [31] for hydrogen purification using MOFs. The studies of Farooq and co-workers [67, 95] brought into light the importance of multiscale performance-based methods for realistic materials screening especially in the context of post-combustion carbon capture. In their main screening study, Khurana and Farooq [67] evaluated the performance of 74 real and hypothetical adsorbents in a 4-step VSA process with light product pressurization (LPP). Process optimizations were carried out to minimize overall energy penalty of the process and maximize its productivity while simultaneously meeting the 95% $CO_2$ purity and 90% $CO_2$ recovery criteria for post-combustion carbon capture. As a result of this study, the authors identified several adsorbents with superior performance over Zeolite 13X, the current benchmark and the most studied adsorbent for post-combustion carbon capture.

This new development also provided additional evidence that process-level metrics (PLM) such as process productivity, overall energy consumption, and product purity do not directly correlate with the intrinsic properties of adsorbent materials [16, 67, 95-97] that have been widely used by scientists for materials screening over the last decade. The multiscale performance-based screening method discussed above addresses several important pitfalls associated with the traditional techniques where materials screening is performed solely based on intrinsic evaluation metrics:

(1) This approach can confirm whether the important $CO_2$ purity−recovery requirement can be met. ISMM, IFMM and HMM classes of evaluation metrics do not take this requirement into account. (2) It can identify the best performance for each adsorbent across a wide range of operating conditions while simultaneously satisfying the purity−recovery constraint. In contrast, adsorbent-based screening methods usually rank materials for a fixed set of operating conditions. (3) The process-level metrics (*e.g.* energy consumption and productivity) can be directly related to economic drivers of commercialized carbon capture plants (*e.g.* capital and operation cost).

The above approach for process-based screening of porous materials is particularly important in light of the available experimental evidence which support the predictions of the proposed screening platform. In a pilot plant study, Krishnamurthy *et al.* [98] demonstrated that the 95% $CO_2$ purity and 90% $CO_2$ recovery targets for post-combustion carbon capture can be achieved in experiment using the same 4-step VSA cycle with light product pressurization which was investigated by Khurana and Farooq [67, 98]. In a separate study, Perez *et al*. [99] also verified the ability of multi-objective optimization techniques to guide the design of PSA/VSA processes. In this study, it was shown that purity-recovery Pareto fronts of $CO_2$ as predicted by process modelling of the 4-step VSA-LPP cycle reasonably reproduce the experimental results [99]. These promising observations attracted more attention to the newly proposed process-based materials screening approaches and their combination with molecular simulation techniques, which would allow computational screening of porous materials. Recent examples of these new group of studies are outlined below:

In 2018, Farmahini *et al.* [68] used a similar multiscale platform by combining GCMC simulation with process modelling and optimization of the 4-step VSA-LPP cycle to explore the challenges associated with the interface between molecular and process levels of description. In this study, they identified several sources of inconsistency for accurate implementation of the multiscale screening workflow which can largely affect prediction of performance of materials at the process level. This includes the numerical procedures adopted to feed the equilibrium adsorption data into the process simulation, and the role of structural characteristics of adsorbent pellets including pellet porosity and pellet size.

In 2019, Balashankar and Rajendran [100] employed a two-stage approach to screen 119,661 hypothetical zeolites, 1,031 zeolitic imidazolate frameworks, and 156 zeolites catalogued by the International Zeolite Association [90]. In their study, the first stage was dedicated to the rapid screening of all materials under investigation using a computationally inexpensive batch adsorber analogue model to filter adsorbents which can meet 95% $CO_2$ purity and 90% $CO_2$ recovery targets. This stage was then followed by detailed process modelling of 15 top-performing candidates from the previous stage in addition to 24 synthesizable zeolites using the widely used 4-step VSA-LPP cycle to estimate the process level performance indicators more accurately. Out of the 39 adsorbents screened in the second stage, 16 material candidates outperformed Zeolite 13X both in terms of productivity and energy consumption [100].

A new generation of material screening studies based on process performance metrics appeared in 2020. In this year, Farmahini *et al.* [101] explored the role of the pellet morphology on materials performance. Pellet morphology belongs to the category of properties that cannot be evaluated at the molecular level and yet can greatly alter separation performance at the process level. The authors demonstrated that a series of competing mechanisms associated with diffusion into adsorbent pellets, convective mass transfer through adsorption column, and pressure drop across the bed can be tuned through optimization of pellet size and pellet porosity to maximize separation performance of different classes of porous materials including zeolites and MOFs [101].

Later, Park et al. [73] assessed separation performance of selected MOFs for subambient temperatures post-combustion carbon capture based on (i) a selection of simple adsorbent metrics (*e.g.* $CO_2$ swing capacity, selectivity, regenerability*, etc*); (ii) performance in an idealized 2-step PSA model (adopted from Ga *et al.* [102]) consisting of adsorption and desorption steps, (iii) performance in a rigorous model of 4-step Skarstrom cycle with light product pressurization. The results from this study showed that the order of high performing materials are different for the idealized 2-step model and the 4-step Skarstrom cycle. Moreover, it was illustrated that the simple adsorbent metrics that are strongly correlated with the predictions of the idealized model are not the same as those that are closely correlated with the predictions of the rigorous 4-step process model. This is an important observation, as it clearly demonstrates that the separation performance of porous materials is strongly influenced by the design of cycle configuration at the process level, and that materials ranking based on simple adsorbent metrics are not directly correlated with materials performance at the process-level.

Burns *et al.* [103] screened 1,632 experimentally characterized MOFs using a multiscale platform which combines molecular simulations with process optimization and machine learning models. In their screening study, they also employed the well-established 4-step VSA-LPP cycle and found that a total of 482 materials can achieve the 95% $CO_2$ purity and 90% $CO_2$ recovery targets out of which 365 materials have parasitic energies below that of commercial solvent-based $CO_2$ capture technologies [103]. Consistent with Danaci *et al.* [104], this study also highlighted the fact that nitrogen adsorption behaviour is also an important metric for the prediction of materials ability to separate $CO_2$ with very high purity and recovery in post-combustion $CO_2$ capture.

Another screening study from 2020 was published by Yancy-Caballero *et al.* [105] who compared process level performance of 15 promising MOFs with Zeolite 13X as a benchmark using three different process configurations including a modified Skarstrom cycle, a five-step cycle, and a fractionated vacuum swing adsorption cycle. The results from this study suggests that UTSA-16 and Cu-TDPAT perform equally well or even better than Zeolite 13X in all the three process configurations mentioned above. The authors also compared process-level ranking of these MOFs with other rankings obtained based on simplified HMM and GEM metrics. They showed that the rankings suggested by these metrics may differ significantly from the one predicted by detailed process optimizations [105] which is evident by various hierarchies of top-performing materials shown in **Figure *10***.

| | WC | α | S | $API_1$ | $API_2$ | $AFM_1$ | $AFM_2$ | SF | GEM | Mod. Skars. | 5-Step Cycle | FVSA |
|---|---|---|---|---|---|---|---|---|---|---|---|---|
| 1 | Ni-MOF-74 | SIFSIX-3-Ni | SIFSIX-3-Ni | Mg-MOF-74 | Mg-MOF-74 | SIFSIX-3-Ni | SIFSIX-3-Ni | SIFSIX-3-Ni | UTSA-16 | Zeolite 13X | UTSA-16 | Cu-TDPAT |
| 2 | Mg-MOF-74 | Zeolite 13X | Zeolite 13X | Zeolite 13X | Ni-MOF-74 | Zeolite 13X | Zeolite 13X | UTSA-16 | Cu-TDPAT | Cu-TDPAT | Zeolite 13X | UTSA-16 |
| 3 | Co-MOF-74 | Mg-MOF-74 | UTSA-16 | SIFSIX-3-Ni | UTSA-16 | Mg-MOF-74 | UTSA-16 | Zeolite 13X | Zeolite 13X | UTSA-16 | Cu-TDPAT | Zeolite 13X |
| 4 | UiO-66(OH)$_2$ | UTSA-16 | Cu-TDPAT | Ni-MOF-74 | Zeolite 13X | Ni-MOF-74 | Cu-TDPAT | Ni-MOF-74 | SIFSIX-3-Ni | Ni-MOF-74 | Ti-MIL-91 | Zn-MOF-74 |
| 5 | UTSA-16 | Cu-TDPAT | Ni-MOF-74 | UTSA-16 | SIFSIX-2-Cu-i | UTSA-16 | Ni-MOF-74 | Cu-TDPAT | Ti-MIL-91 | Mg-MOF-74 | Zn-MOF-74 | Ti-MIL-91 |
| 6 | SIFSIX-2-Cu-i | Ni-MOF-74 | Mg-MOF-74 | Cu-TDPAT | Co-MOF-74 | Cu-TDPAT | Mg-MOF-74 | Ti-MIL-91 | Zn-MOF-74 | SIFSIX-2-Cu-i | SIFSIX-3-Ni | SIFSIX-3-Ni |
| 7 | Zn-MOF-74 | Ti-MIL-91 | Ti-MIL-91 | SIFSIX-2-Cu-i | UiO-66(OH)$_2$ | SIFSIX-2-Cu-i | Ti-MIL-91 | Mg-MOF-74 | SIFSIX-2-Cu-i | SIFSIX-3-Ni | Mg-MOF-74 | |
| 8 | Zeolite 13X | SIFSIX-2-Cu-i | SIFSIX-2-Cu-i | Zn-MOF-74 | Zn-MOF-74 | Zn-MOF-74 | SIFSIX-2-Cu-i | SIFSIX-2-Cu-i | Ni-MOF-74 | Zn-MOF-74 | | |
| 9 | Cu-TDPAT | Zn-MOF-74 | Zn-MOF-74 | UiO-66(OH)$_2$ | Cu-TDPAT | Ti-MIL-91 | Zn-MOF-74 | Zn-MOF-74 | Sc2BDC3 | Ti-MIL-91 | | |
| 10 | Ti-MIL-91 | Sc2BDC3 | Sc2BDC3 | Ti-MIL-91 | Ti-MIL-91 | UiO-66(OH)$_2$ | Sc2BDC3 | Sc2BDC3 | Mg-MOF-74 | | | |
| 11 | Cu-BTTRi | UiO-66(OH)$_2$ | UiO-66(OH)$_2$ | Co-MOF-74 | SIFSIX-3-Ni | Co-MOF-74 | UiO-66(OH)$_2$ | UiO-66(OH)$_2$ | Cu-BTTRi | | | |
| 12 | NTU-105 | Co-MOF-74 | Co-MOF-74 | Cu-BTTRi | Cu-BTTRi | Cu-BTTRi | Co-MOF-74 | Co-MOF-74 | UiO-66(OH)$_2$ | | | |
| 13 | SIFSIX-3-Ni | Cu-BTTRi | Cu-BTTRi | NTU-105 | NTU-105 | NTU-105 | Cu-BTTRi | Cu-BTTRi | ZIF-8 | | | |
| 14 | MOF-177 | NTU-105 | NTU-105 | Sc2BDC3 | MOF-177 | Sc2BDC3 | NTU-105 | NTU-105 | Co-MOF-74 | | | |
| 15 | ZIF-8 | ZIF-8 | ZIF-8 | MOF-177 | Sc2BDC3 | MOF-177 | ZIF-8 | ZIF-8 | NTU-105 | | | |
| 16 | Sc2BDC3 | MOF-177 | MOF-177 | ZIF-8 | ZIF-8 | ZIF-8 | MOF-177 | MOF-177 | MOF-177 | | | |

Figure 10. Hierarchies of top-performing materials based various adsorbent-based performance indicators as compared with detailed process modelling and optimization for three cycle configurations namely modified Skarstrom, 5-step PSA, and fractionated vacuum swing adsorption (FVSA) cycle. Adsorbent metrics from left to right include $CO_2$ working capacity, selectivity, sorbent selection parameter, adsorbent performance indicators (APS 1 and 2), adsorbent figure of merits (AFM 1 and 2), separation factor, and general evaluation metric (GEM). Reprinted with permission from Yancy-Caballero *et al.*, Molecular Systems Design & Engineering, 2020. 5(7): p. 1205-1218. Copyright (2020), Royal Society of Chemistry [105].

As an example, Cu-TDPAT and UTSA-16 are the two top performing materials according to FVSA cycle, however based working capacity they are the 9th and 5th in the list of top-performing materials. Based on selectivity, these materials are the 5th and 4th materials from the top. Interestingly, GEM seems to provide a closer estimation of materials performance when compared with detailed process modelling for all three cycles. Another important observation here is the fact that the order of top-performing materials is a function of process configurations as shown by the first three columns from the right.

A recent study was published by Pai *et al.* [106] in 2020 who developed a generalized and data-driven surrogate model which can reproduce operation of PSA/VSA processes at cyclic steady state with high accuracy. The multiscale screening framework developed here simultaneously optimizes adsorption isotherm properties and process operating conditions in order to estimate performance indicators of the process. The framework makes use of a dense feed forward neural network trained with a Bayesian regularization technique and is able to significantly reduce the simulation and optimization time required for multiscale screening of porous materials for post-combustion carbon capture [106]. Development of such material-agnostic machine-learning models is particularly useful considering they can be employed for prediction of any arbitrary or hypothetical adsorbent as long as the equilibrium adsorption isotherms of $CO_2$ and $N_2$ for that material can be described by the implemented numerical adsorption model (*e.g.* a single-site Langmuir model in the case of this study).

Finally in 2021, Subraveti *et al.* [71] reported on a new attempt for integration of techno-economic analyses with detailed modelling and optimization of adsorption process for post-combustion carbon capture. They estimated the capture cost of $CO_2$ using Zeolite 13X, UTSA-16 and IISERP MOF2 as adsorbent in a 4-step VSA-LPP cycle. Their study showed that application of IISERP MOF2 in the above process leads to the lowest capture cost, while still being higher than the cost of carbon capture in MEA-based absorption process as the current industrial benchmark. According to this study, Zeolite 13X and UTSA-16 are the 2nd and 3rd in terms of the overall cost. An important message conveyed by the authors in this study was that the minimum cost configuration obtained from techno-economic analyses does not necessarily correspond to the most optimum configurations obtained by minimizing energy penalty and maximizing productivity of the VSA process, if complexities associated with scale-up of the process was not taken into consideration. This essentially means that realistic assessment of materials performance for industrial applications must go beyond optimization of the process, and that the multiscale screening workflows should encompass consideration of techno-economic analyses for materials screening.

This section was meant to provide the reader with a historical perspective of the topic without going into technical details of the screening methods. At the end of this section, it is useful to reflect on some of the key observations from our overview. It is clear that multiscale materials screening strategies have advanced significantly over the past decade evolving from screening of porous materials based on simple microscale properties, towards development of more realistic approaches based on process modelling and optimization for evaluation of materials performance, and finally to incorporating techno-economic assessment of the whole separation plant into the screening workflows. Overview of the studies discussed in this section, reveals lack of consistency among the hierarchies of top performing materials that are reported by different studies. This means that the screening studies conducted so far have not been able to propose a consistent set of materials as top performing candidates for post-combustion carbon capture. This is associated with the lack of consistency in calculation of a series of parameters that are used in performance-based materials screening workflows including but not limited to the force fields used in molecular simulations for prediction of equilibrium adsorption data; the numerical methods used for fitting adsorption isotherms; various model assumptions applied in describing kinetic of the process; application of different process and cycle configurations; and selection of the materials to screen. Addressing the issue of consistency in ranking of porous materials requires detailed knowledge about implementation of all the modelling modules and simulation techniques that are used as part of the materials screening workflow which is the topic of following section in this review.

# 6. Multiscale Screening Workflow

In the previous sections, we briefly discussed why materials screening is important in the context of PSA/VSA technologies for the post-combustion carbon capture. We also provided a historical perspective on the evolution of materials performance metrics and screening methods, which have been used so far. The main objective of these sections was to illustrate to the reader the importance and the gradual evolution of the research community towards adopting more complex multiscale screening workflows as the emerging way to evaluate separation performance of porous materials.

The objective of the current section is to introduce in an accessible, tutorial-style fashion the key elements and methods involved in the multiscale screening workflows. Some of these elements, such as molecular simulations and process modelling, have been also comprehensively covered in the several authoritative textbooks. The intention here is not to replace or replicate these sources, but to highlight only the essential aspects of the methods while focusing on the data they require, information they produce and on the gaps at the interfaces between different elements. To achieve this objective, the structure of this section logically follows the multiscale workflow diagram, shown in **Figure *11***. The starting point of this workflow is a database of porous materials. In Section 6.1, we review the currently existing databases and the computational tools required to characterise structural properties of the porous materials in these databases. Molecular simulations are used to obtain equilibrium and transport properties at a molecular level. These methods are introduced in Section 6.2. Finally, following the workflow we pass the information from molecular simulations to the process level modelling and optimization. Models, methods, and data required for this stage are reviewed in Section 6.3.

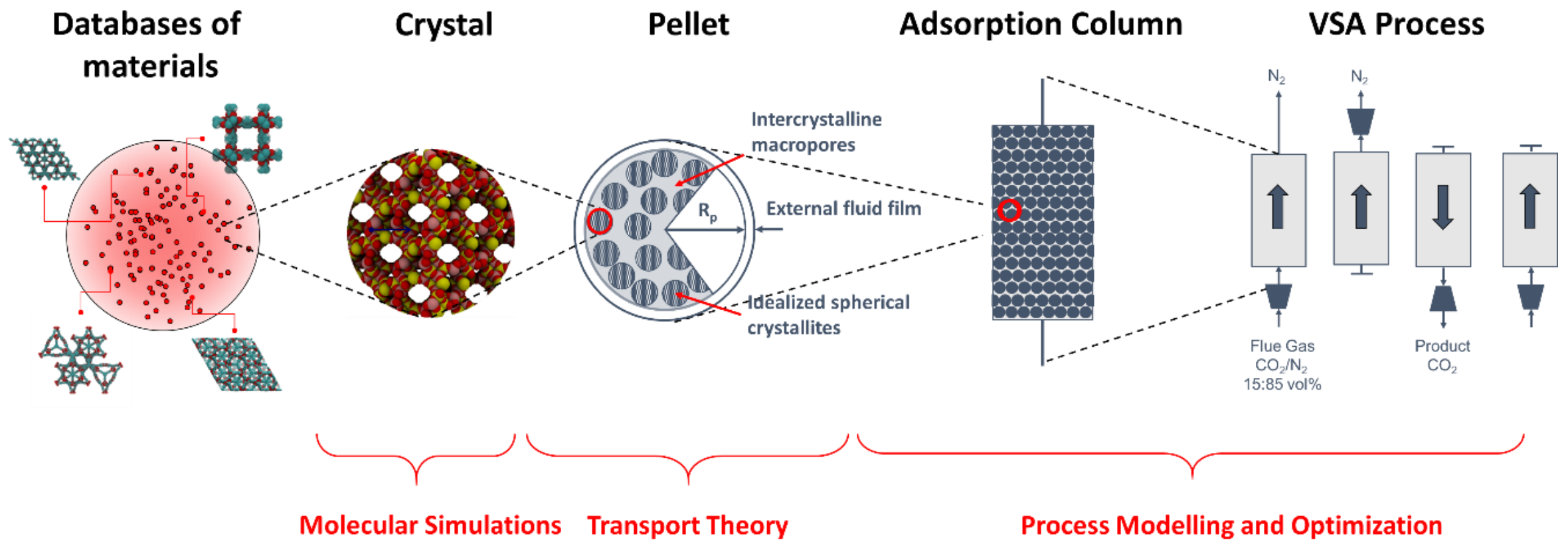

**Figure 11.** General structure of the multiscale screening workflow for materials screening

## 6.1. Material Databases and Characterization Tools

This section corresponds to the first step in the multiscale material screening workflow. The aim here is to provide concise and practical reference to the reader on what databases are currently available, what materials and data they contain and what tools are available to build geometric descriptors for materials in these databases.

### 6.1.1. Databases of Porous Materials

MOFs are the primary and most prominent example of the emerging families of materials [6-8] and it is useful to briefly review what these materials are. Although the origins of MOFs can be traced as far back as the late fifties, they were given their current name, Metal-Organic Frameworks, in the seminal paper by Yaghi and Li in 1995 [107]. To prepare a MOF one uses two types of building blocks: metal centers and organic molecules capable of forming strong coordination bonds with these centers. In the synthesis process, the building blocks form a crystalline framework where metal complexes comprise the vertices of the framework, connected by the organic linkers. Several papers that followed in the late nineties discovered few more examples of these frameworks, however, most importantly they demonstrated that these structures possessed permanent stable porosity, high surface area and new materials could be designed simply by variation of the building blocks, leading to the concept of isoreticular material design [108-110]. Since then, tens of thousands of new MOFs have been discovered: the most current assessment of the Cambridge Structural Database suggests *ca.* 100,000 reported structures that can be qualified as MOFs [111], with *ca.* 12,000 of them being porous, while the modular nature of MOFs implies that in principle infinite variation of structures is possible (if we assume that the diversity of MOFs can approach the diversity of the organic chemical space).

ZIFs, discovered a few years later [9, 112, 113], is a subclass of MOF materials that have zeolite framework topologies in which silicon atoms are replaced by transition metals and the bridging oxygens are substituted by imidazolate building units [114]. Currently, there are about 300 ZIFs reported in the CSD and potential application of these materials in the context of chemical separations has been recently reviewed by Pimentel *et al*. [115]. In contrast to materials based on coordinative assembly and coordination bonds, Covalent Organic Frameworks (COFs) do not feature metal complexes and are based on covalent bonds [10, 116]. Since their discovery in 2005, a substantial number of 2D and 3D COFs have been reported with diverse structural and chemical properties [116].

Crystalline materials, such as MOFs, ZIFs, COFs can be contrasted with several traditional and emerging classes of amorphous porous polymers, such as activated carbons [117], carbide-derived carbons [118] and Polymers with Intrinsic Microporosity (PIMs) [14, 15, 119]. Porous Aromatic Frameworks (PAFs) is another class of porous materials with rigid aromatic open-framework structure constructed by covalent bonds [12]. Although, PAFs are not crystalline they are ordered with regular and high porosity [120].

This wealth of new materials should not overshadow more traditional classes of porous materials such as zeolites, which, due to their stability, attractive cost, commercial availability and maturity in industrial applications, will likely remain the primary adsorptive materials for years to come. There are currently ~200 zeolite topologies recognized by the International Zeolite Association. Using computational methods, millions of new structures have also been hypothesized [91, 121]. An ongoing research is to understand the magnitude and diversity of the materials landscape for adsorption science [122, 123] and to evaluate what portion of this structural space is realizable in experiments [124]. Combined, these classes of materials provide an enormous chemical and structural diversity, collectively described as the materials genome [125]. Several efforts have been made to assemble databases of the experimentally synthesized or computationally constructed MOFs, ZIFs or porous polymer networks (PPN) [65, 79, 126-130]. Next, we review the most prominent examples of these databases:

**Databases of Hypothetical MOFs:** Hitherto, three main databases of hypothetical MOFs have been created which are discussed below:

***(a) The database by Wilmer et al.*:** This database contains 137,953 structures and is generated by recombining a library of 102 building blocks including secondary building units (SBU) and organic linkers from crystallographic data of already synthesized MOFs using a "tinker-toy" algorithm [79]. The resulting hypothetical database is, however, composed of only a few underlying framework topologies [131]. By testing a very limited set of MOFs including HKUST-1, IRMOF-1, PCN-14 and MIL-47, the authors suggested that their method can closely reconstruct molecular structures of the experimentally synthesized materials [79]. Nevertheless, generalization of this finding is subject to more comprehensive validations, considering no energy minimization has been performed for any of the constructed structures in this database. Also, the data associated with this database do not include partial electrostatic charges on atoms of MOFs, hence its application is limited to very few adsorption cases where electrostatic interactions are not important (*e.g.* $CH_4$ adsorption). In fact, Wimer *et al*. did use this database to search for MOF materials suitable for methane storage. They identified more than 300 MOFs with a predicted methane-storage capacity larger than that of any previously known material [79].

***(b) The database by Boyd and Woo*:** This new database of hypothetical MOFs was constructed using the topology-based algorithm of Boyd and Woo [132] and contains 324,426 structures which are generated by assembling a set of secondary building units containing 8 inorganic and 94 organic SBUs resulting in 12 different topologies [133]. The set was further diversified by chemical modification of MOFs, in which available hydrogens were replaced by functional groups. All MOFs in this database are structurally optimized using classical force fields. Framework charges for all structures included in this database were also computed using the charge equilibration method (Qeq) [134] and the MOF electrostatic potential optimized (MEPO) parameters [135].

**(c) ToBaCCo Database:** This database constructed using the Topologically Based Crystal Constructor (ToBaCCo) algorithm and contains 13,512 MOF structures with 41 different edge-transitive topologies [128, 136]. The database makes use of a top-down construction algorithm which uses topological blueprints and molecular building blocks as input to assemble MOF structures. The algorithm does not check for atom overlaps as part of the construction process therefore the geometry of the resulting structures must be optimized before being used in molecular simulations [136]. The database also includes partial atomic charges.

**Cambridge Structural Database (CSD):** The Cambridge Structural Database (CSD) contains more than a million of organic and metal-organic small-molecule crystal structures which are obtained from X-ray or neutron diffraction analyses [126]. The MOF structures deposited in this database are experimentally realized, nevertheless the use of CSD entries for high-throughput screening of porous materials is not straightforward. Checks must be performed to make sure that the candidate structures obtained from CSD are adequately porous and are free from residual substances that are leftover of the synthesis processes. As such, the first step in performing high-throughput screening of experimental MOFs is to construct curated subsets of CSD that can fulfil the above criteria (see more on the CSD-MOF Subset later in this section).

**Goldsmith Database of Experimental MOFs:** In 2013, Goldsmith *et al.* [137] constructed a MOF database containing 22,700 computation-ready structures which was derived from the CSD after the removal of unbounded guest molecules (*e.g.* residual solvents). By excluding disorder compounds and those with missing atoms, the total number of MOF structures were reduced to 4,000 [137] which do not include those with interpenetrated frameworks and charge-balancing ions [127]. The materials included in the database were subsequently characterized by calculating porosity, surface area and total theoretical $H_2$ uptake [137]. Goldsmith *et al*. used their MOF database to estimate the maximum theoretical uptake of hydrogen based on the so-called "Chahine rule" (see Ref [138] for further reading) known for hydrogen adsorption in microporous carbons but also shown to be valid across a wide range of other porous materials including MOFs [137].

**CoRE-MOF Database:** Construction of the Computation-Ready, Experimental Metal–Organic Frameworks (CoRE-MOF) was a major attempt in development of a MOF database that can be directly used in molecular simulations. The first version of CoRE-MOF [127] contains 5,109 3D MOF structures with pore-limiting diameter greater than 2.4 Å which are derived from CSD. The MOF structures were screened to make sure that all MOFs included in the database are crystalline (no disorder) and solvent-free. The database also reports helium void fractions of all MOFs in addition to their surface area, accessible volume, largest cavity diameter (LCD) and pore-limiting diameter (PLD). In the original version of the database, the structures were not optimized (except for very few MOFs that were manually edited) [127]. Following the initial release of CoRE-MOF, two modified subsets of this database were released in 2016 and 2017. The first subset contains 2,932 experimental MOFs whose partial atomic point charges were calculated using planewave DFT and the DDEC charge partitioning methods [139]. The second subset focuses on the geometry optimization of 879 experimentally synthesized MOFs using a periodic density functional theory (DFT) method [140]. The latter publication demonstrated that although the majority of MOF structures undergo less than 10% change in their structural parameters (*e.g.* pore size, lattice parameters, unit cell volume, helium void fraction) upon DFT optimization, many other MOF structures change significantly after geometry optimization especially those materials whose crystalline structure were cleaned from solvent residue molecules. More importantly, it was shown that the DFT optimization had a large impact on simulated gas adsorption in some cases, even for materials whose crystalline structure did not change significantly [140]. This study has important implication for high-throughput materials screening approaches that rely on databases of experimentally synthesized materials such as CSD [126] or the original CoRE-MOF [127]. The CoRE-MOF database was recently expanded to include approximately 14,000 structures (CoRE MOF 2019). The updated database includes additional structures that were contributed by CoRE-MOF users, obtained from updates of the CSD database and a Web of Science search [141]. CoRE MOF 2019 was released in two different sets: (1) free solvent removed (FSR) database for which only the free solvent molecules have been removed from the structures, (2) all solvent removed (ASR) database for which both bounded and free solvent molecules have been removed from the structures. CoRE MOF 2019 also summarizes a list of MOF structures that contain open-metal sites [141].

**CSD-MOF Subset:** In 2017, Moghadam *et al.* [130] constructed a new subset of CSD for solvent-free MOFs in which 69,666 1D, 2D and 3D MOFs were listed out of which 54,808 structures are non-disordered. These materials were characterized using the Zeo++ code [81] based on the Voronoi decomposition technique to calculate the accessible surface area, accessible pore volume, LCD, and PLD. It was found that 46,420 structures have gravimetric surface area equal to zero which essentially means that $N_2$ size molecular probes cannot access their pore spaces for geometric surface area calculations [130]. It is shown that the remaining 8,388 MOFs have PLD values larger than 3.7 Å which is approximately 3,600 structures more than what was previously published by Chung *et al.* [127] in the initial version of the CoRE-MOF database. Currently, the MOF subset of CSD database contains approximately 100,000 MOFs [111]. The main advantage of the CSD-MOF subset is that it is integrated into the Cambridge Crystallographic Data Centre's (CCDC) structure search program. This not only allows for tailored structural queries (*e.g.* generation of MOF subsets based on secondary building units or selection of non-disordered materials), but it can also be used to automatically update the database with subsequent addition of new MOFs to CSD [130].

**Hypothetical Zeolites Database (hZeo-DB):** hZeo is a database of computationally predicted zeolite-like structures which are generated by systematically exploring 230 space groups, unit cell dimensions between 3 Å to 30 Å, and T-atom densities from 10 to 20 per 1000 $Å^3$ [91, 142]. A computational procedure based on Monte Carlo search was employed to produce 3.3 million zeolite-like structures out of which 2.6 million topologically distinct structures were identified after energy minimization [142]. Roughly 10% of this number are the structures deemed to be thermodynamically accessible as aluminosilicates based on energy stability of the structures [91].

**Database of Zeolite Structures (IZA-DB):** IZA-DB provides information about the structure of all the zeolite framework types that have been approved by the Structure Commission of the International Zeolite Association (IZA-SC). The database currently contains 241 ordered and 11 partially disordered topologies [90].

**Database of Hypothetical Porous Polymer Networks (hPPN-DB):** The hypothetical PPN databse constructed by Martin *et al*. [129] contains almost 18,000 hypothetical structures of porous polymer networks which are predicted *in silico* using commercially available chemical fragments and two experimentally known synthetic routes; hence aiming to provide a database of synthetically realistic PPNs [129]. All structures from this database have their structure optimized using semiempirical electronic structure methods [129]. The structures are also characterized for their topological properties and methane adsorption characteristics [129].

**Hypothetical COF Database (hCOF-DB):** This database is a collection of 69,840 hypothetical covalent organic frameworks (COFs) which are assembled from 666 distinct organic linkers and four established synthetic routes [143]. It contains 18,813 interpenetrated 3D structures, 42,386 non-interpenetrated 3D structures and 8,641 2D-layered structures. All materials are structurally relaxed using classical force fields. The database does not include partial atomic charges for the deposited COFs.

**CoRE-COF Database:** In 2017, Tong *et al.* [144] compiled a computation-ready database of experimental covalent organic frameworks (COFs) containing 187 structures. The original version of the database contains 19 3D-COFs and 168 2D-COFs. The structures collected in this database are reported to be disorder-free and solvent-free which make them ready for computational studies. Although most of the structures available in CoRE-COF database are cleaned versions of the experimentally reported CIF files, some of the COFs collected in the database are constructed computationally based on the information reported in the literature where synthesis of the corresponding COFs had been reported without any CIF file. CoRE-COF materials are structurally optimized using a two-steps procedure [144] where optimization was initially performed using classical force fields and then later refined using the dispersion-corrected DFT method of Grimme (DFT-D2) [145]. The database also reports on structural features of each COF including their largest cavity diameter, pore-limiting diameter, accessible surface area and free volume. Since its first release, the CoRE-COF database has been updated regularly so that its most recent version (CoRE-COF Ver. 4.0)[1] contains 449 structures with the framework charges obtained from the charge equilibration (Qeq) method.

**CURATED COF Database:** Clean, Uniform, Refined with Automatic Tracking from Experimental Database (CURATED) of covalent organic frameworks (COFs) is another database of experimentally realized COFs [146]. The initial version of the database includes 324 structures, however the database has been updated recently so that its most recent version (Feb 2020) contains 482 structures. All structures collected in the CURATED COFs are cleaned from solvent molecules and have no partial

[1] CoRE-COF database: https://github.com/core-cof/CoRE-COF-Database (accessed on 25/04/2020)

occupation or structural disorder. They are structurally optimized using DFT with the DDEC framework partial charges included [146].

**Hypothetical ZIFs Database (hZIF-DB):** In 2012, Lin *et al.* [65] published a paper on computational screening of large number of zeolites and zeolitic imidazolate frameworks (ZIFs) for carbon capture. In this study, ZIF structures were generated computationally by using zeolite topologies of the International Zeolite Association (IZA) database. In doing so, the distance between zinc atoms and the centre of imidazolate rings was set to be 1.95 times larger than the silicon-oxygen distance in zeolites. ZIF frameworks were then generated by scaling the corresponding zeolite structures by the same factor and replacing every oxygen atom with an imidazolate group and substituting every silicon atom with a zinc atom. The resulting ZIF geometries were validated by comparing against geometries of two experimentally known ZIF structures (i.e. ZIF-3 and ZIF-10) [65]. This structure is not, however, available online or in a depository to further comment on its characteristics.

**Nanoporous Explorer Database (NE-DB):**

Nanoporous explorer is an aggregated database of nanoporous materials including CoRE-MOF [127], hypothetical MOFs [79], and hypothetical PPNs [129]. The database is part of a larger database developed under the Materials Project program [2] which is designed to provide a large collection of computed data for experimentally known and computationally predicted materials including nanoporous materials [147]. The NE-DB provides information about pore descriptors (e.g. PLD, LCD), adsorption properties (e.g. Henry's constant, adsorption isotherm, heat of adsorption), and simulated powder X-ray diffraction of many porous material. At the time of writing this article the Nanoporous Explorer database contains 530,243 entries.

**Nanoporous Materials Genome Database (NMG-DB):** NMG [125, 148] is a collection of a growing number of materials databases which currently encompasses more than 3 million hypothetical and synthesized porous materials. Most prominent examples of these databases are already discussed in this review. For the sake of completeness, we provide a full list of the constituting databases for NMG which includes hypothetical MOFs database [79, 133], computation-ready experimental MOFs database (CoRE-MOFs) [127, 141], hypothetical Zeolites [91, 142], ideal silica Zeolites obtained from the International Zeolite Association (IZA) database [90], hypothetical covalent organic frameworks (COFs) [143, 149], computation-ready experimental COF database (CoRE-COFs) [144, 146], hypothetical zeolitic imidazolate frameworks (ZIFs) [65], and hypothetical porous polymer networks (PPNs) [129].

**Database of Porous Rigid Amorphous Materials (PRAM-DB):**

So far, the databases we reviewed comprised of crystalline and ordered porous materials. In an important development, Thyagarajan and Sholl [150] have recently collected 205 atomistic models of amorphous nanoporous materials which had been previously published by various groups. This new database of porous rigid amorphous materials (PRAM-DB) contains several classes of materials with disordered porous structures including amorphous zeolite imidazolate frameworks (a-ZIFs) [151], activated carbons [152], carbide-derived carbons [153-157], polymers with intrinsic microporosity (PIMs) [158-161], hyper-cross-linked polymers (HCPs) [162-164], kerogens [165] and cement [166] which all have important applications in adsorption separation technologies. The database contains partial atomic charges for most of the materials. It also reports on a wide range of physical properties for each material. This includes pore limiting diameter (PLD), the largest cavity diameter (LCD), the accessible surface area and pore volume, pore size distribution (PSD), ray-tracing histograms, PXRD patterns and radial pair distribution functions (RDF) [150]. The new study also reports single-component and binary adsorption isotherms of several gases for these materials [150].

**Table 2. Structural database of crystalline porous materials**

| index | Database | Number of entries | Origin | Cleaned | Optimized | Charges included |
|---|---|---|---|---|---|---|
| 1 | Wilmer *et al*., [79] | 137,953 | Simulation | Yes | No | No |
| 2 | Boyd and Woo [133] | 324,426 | Simulation | Yes | Yes | Yes |
| 3 | ToBaCCo [128, 136] | 13,512 | Simulation | Yes | No | No |
| 4 | CSD [126] | >1M | Experiment | No | No | No |
| 5 | Goldsmit [137] | 4,000 | Experiment | Yes | No | No |
| 6 | CoRE-MOF 2019 [141] | ~14,000 | Experiment | Yes | Partially* | Partially** |
| 7 | CSD-MOF subset*** [130] | 96,000 | Experiment | Yes | No | No |
| 8 | hZeo [91, 142] | 2.6 M | Simulation | Yes | Yes | No |
| 9 | IZA [90] | 252 | Experiment | Yes | Yes | No |
| 10 | hPPN [129] | 18,000 | Simulation | Yes | Yes | No |
| 11 | hCOF [143] | 69,840 | Simulation | Yes | Yes | No |
| 12 | CoRE-COF [144] | 449 | Experiment | Yes | Yes | Yes |
| 13 | CURATED COFs [146] | 482 | Experiment | Yes | Yes | Yes |
| 14 | hZIFs [65] | - | Simulation | - | - | - |
| 15 | NE-DB [2] | 530,243 | Simulation & experiment | - | - | - |
| 16 | NMG [125, 148] | >3M | Simulation & experiment | - | - | - |
| 17 | **PRAM-DB** [150] | 205 | Simulation & experiment | - | - | Yes |

* 879 MOFs undergone geometry optimization which were released as part of CoRE MOF-DFT optimized 2017 [140].

** Partial atomic charges of 2,932 MOFs were computed which were released as part of CoRE MOF-DDEC 2016 [139]

*** As of Aug 2019 [111, 130].

#### 6.1.2. Computational Tools for Structural Characterization of Porous Solids

As can be seen from the reviewed studies, classification of materials within the databases and early efforts in computational screenings are based on the geometric descriptors of porous materials, such as the accessible surface area, pore limiting diameter and pore volume. As this is a practice-oriented review, we believe it is useful to mention the material characterization software available to obtain these geometric properties for crystalline and amorphous porous structures. To begin with, we refer the reader to several articles describing what properties of porous materials can be calculated and how they are related to the properties that can be measured and to the physical process of adsorption in porous materials [81, 82, 167, 168]. In principle, calculation of selected properties, such as the solvent-accessible surface area (in application to porous materials often called simply the accessible surface area), is available within many commercial and free software packages. Three packages available for a more comprehensive assessment of the materials are Poreblazer [82, 167, 169], Zeo++ [81], and PorosityPlus [170]. From this list, Poreblazer developed by Sarkisov and Harrison [82, 169] and PorosityPlus developed by Opletal [170] are written in Fortran and are available as open-source packages. Zeo++, developed by Haranczyk and co-workers [81] is a C++ package based on the Voronoi tessellation methods [81]. With Voronoi network being a dual graph of Delaunay network, the approach employed by Zeo++ is closely related to that of Foster *et al.* [171]. The program is downloadable from the website of the developers, with the source code available upon request only.

All three codes mentioned above are able to calculate accessible surface area (equivalent to the area of the surface formed by the nitrogen probe rolling on the surface of the atoms of the structure), pore volume (using several alternative definitions of this property) and pore size distribution. Poreblazer and Zeo++ can also calculate pore limiting diameter (PLD) of the porous frameworks, while PorosityPlus is also able to compute radial distribution function (RDF) of the adsorbed phase in the system. One important feature of Zeo++ software is its ability to read framework structures in CIF format, while the other two programs can only use XYZ format as their input for the porous framework. A detailed comparison of Poreblazer, Zeo++ and RASPA [172] has been recently provided by Sarkisov *et al*. [169] for structural characterization of CSD-MOF Subset database [130]. Here, we note that RASPA is a molecular simulation software which is mainly known for its capabilities for Monte Carlo simulations. This program is presented in the following section where we discuss grand canonical Monte Carlo (GCMC) technique for simulation of equilibrium adsorption isotherms.

**Table 3.** Computer software available for pore structure characterizations

| Item | Software | Surface area | Pore volume | PSD | PLD | RDF | Cif format supported | Code repository |
|---|---|---|---|---|---|---|---|---|
| 1 | Poreblazer [82] | Yes | Yes | Yes | Yes | No | No | https://github.com/SarkisovGroup/PoreBlazer |
| 2 | PorosityPlus [170] | Yes | Yes | Yes | No | Yes | No | https://data.csiro.au/collections/collection/CIcsiro:34838v1 |
| 3 | Zeo++ [81] | Yes | Yes | Yes | Yes | No | Yes | http://zeoplusplus.org/ |

## 6.2. Molecular Simulation

The purpose of this section is to briefly introduce the two main and most widely used molecular simulation techniques, grand canonical Monte Carlo (GCMC) and molecular dynamics (MD) simulations which are respectively used for simulation of adsorption and transport properties on a microscopic level. In the context of adsorption problems, comprehensive reviews on molecular simulations for Metal-Organic Frameworks have been provided by Yang and co-workers [173] and for zeolites by Smit and Maesen [174]. In this section however, we will discuss these techniques as two important elements of the multiscale screening workflows. In particular, we would like our intended reader to appreciate what parameters are required for these simulations, how they can be calculated, and what open-source software are available to researchers to perform these simulations.

In Section 6.2.1, we introduce fundamentals of GCMC method followed by Section 6.2.2 which presents the main publically available simulation software for performing this type of simulation. Next, in Section 6.2.3, fundamentals of molecular dynamics will be discussed, which will be also followed by a section related to the open-source programs that can be used to run MD (Section 6.2.4). Finally in Section 6.2.5, we will briefly introduce molecular force fields which are central to accurate simulation of molecular systems. The issues associated with the current gaps in the field of force field development and comment on their implications for multiscale materials screening studies will be reviewed later in a Section 8.1.

### 6.2.1. Grand Canonical Monte Carlo Simulation

In this section, we briefly review the Grand canonical Monte Carlo (GCMC) simulation method, which is widely used for calculation of equilibrium adsorption data. For a more comprehensive review of Monte Carlo methods, we would refer the reader to reference books [175-177] and several excellent articles by Dubbeldam and co-workers on the Monte Carlo methods and the organization of computer codes associated with them [172, 178].

The problem of interest here is the adsorption of small molecules (carbon dioxide, nitrogen, methane, hydrogen) in crystalline porous materials. The volume, $V$, and temperature, $T$, of the system are fixed, and the specified value of the chemical potential, $\mu$, establishes thermodynamic equilibrium between the system and the bulk reservoir, serving as a source and sink of adsorbate molecules. From the statistical-mechanical point of view, the system corresponds to the grand-canonical ensemble ($\mu$ V T), for which the Metropolis Monte Carlo is a widely-used method. This approach is suitable for rigid porous materials, which do not exhibit significant changes in the volume upon adsorption and desorption of guest molecules. For flexible materials, which represent a significant and interesting class of structures within the MOF family, more advanced simulation methods exist (such as the Osmotic Ensemble and Gibbs Ensemble Monte Carlo [178]).

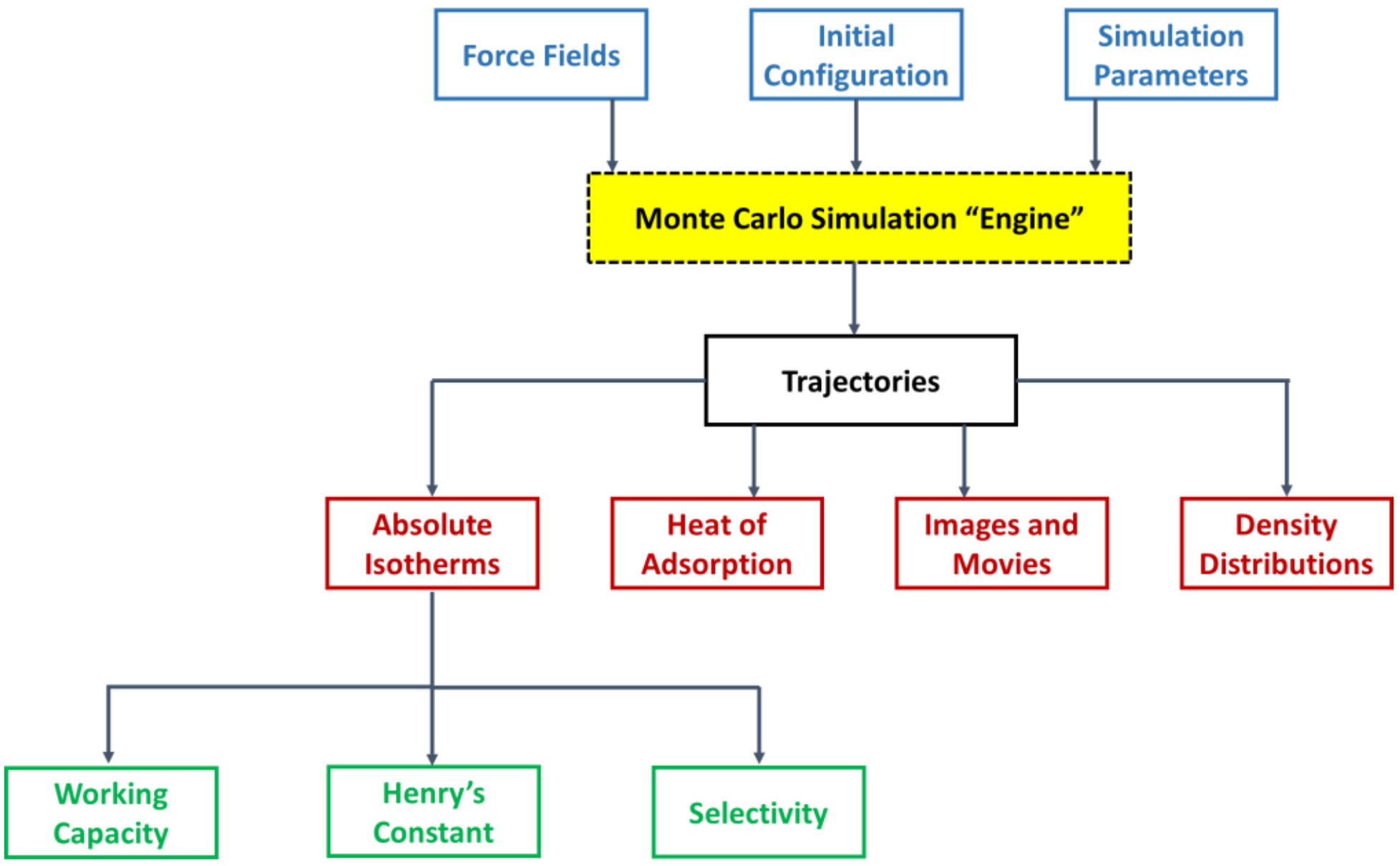

**Figure 12.** Schematic depiction of the workflow in the grand canonical Monte Carlo simulations. The blue boxes indicate the required input data and parameters for the simulations. In the most general terms, a simulation run generates a trajectory: a set of microstates of the system, corresponding to the particular ensemble. The red boxes indicate the primary properties that are directly calculated from the Monte Carlo trajectory. The green boxes are the secondary properties that can be calculated from the primary properties.

The schematic diagram of the GCMC workflow is shown in **Figure *12***. According to this scheme, a Monte Carlo simulation of adsorption requires the following inputs:

- Force field parameters: these parameters define what atoms and molecules are present in the system and describe how they interact with one another. This includes parameters associated with non-bonded dispersion interactions, partial charges on the atoms of the structure and molecules, geometry of the adsorbing molecules (distances and relative positions of the atoms within the molecule).

- Initial configurations of the species present in the system: this includes positions of the atoms of the porous structure and positions of any already adsorbed molecules.

- Simulation parameters, including details of the Monte Carlo protocol; number of steps allocated for the equilibration of the system; parameters associated with the statistical analysis of the simulation (*i.e.* number and size of blocks in the block-average analysis), temperature and fugacities of the adsorbing components. This input data category may also prescribe particular specialized methods to calculate electrostatic interactions between partial charges on the atoms.

Let us consider what happens within the Monte Carlo simulation engine. A configuration of the system with a particular number of molecules (in case of GCMC) and their positions is called a *microstate*. In the actual physical system these microstates occur according to the Boltzmann probability distribution. In a Monte Carlo simulation these microstates are generated by stochastically perturbing the state of the system: we can add molecules to the system or remove them, change their position and orientation. These different ways to change the state of the system are called *Monte Carlo moves*. To ensure Boltzmann distribution of the microstates, the probability to accept a move is calculated according to the equations (*1*) - (*4*):

Translation: $P_{ACC}(S \to S')$ (1)

$$P_{ACC}(S \to S') = \min\{1, \exp(-\beta\Delta U)\}$$

Rotation: $P_{ACC}(S \to S')$ (2)

$$P_{ACC}(S \to S') = \min\left\{1, \exp\left(-\beta\Delta U \frac{\sin\theta_S}{\sin\theta_{S'}}\right)\right\}$$

Insertion: $P_{ACC}(N_a \to N_a + 1)$ (3)

$$P_{ACC}(N_a \to N_a + 1) = min\left\{1, \frac{\beta f V}{N_a + 1}\exp(-\beta\Delta U)\right\}$$

Deletion: $P_{ACC}(N_a \to N_a - 1)$ (4)

$$P_{ACC}(N_a \to N_a - 1) = min\left\{1, \frac{N_a}{\beta f V}\exp(-\beta\Delta U)\right\}$$

where *U* represents the potential energy of interaction, $N_a$, and *V* are the number of molecules and volume respectively, $\beta$ is the reciprocal thermodynamic temperature, $1/k_BT$, with $k_B$ being the Boltzmann constant; $\theta$ is an Euler angle of the rigid body rotation. Here, *f* is the fugacity of the adsorbing species, which is related to the chemical potential as:

$$f = \frac{q_{rot}}{\beta\Lambda^3} exp(\beta\mu) \quad (5)$$

where $q_{rot}$ is the rotational partition function for a single rigid molecule, equal to 1 for a single particle molecule, and $\Lambda$ is the thermal de Broglie wavelength:

$$\Lambda = \left(\frac{\beta h^2}{2\pi m}\right)^{\frac{1}{2}} \quad (6)$$

where $h$ is the Planck's constant and $m$ is the molecule mass.

These probability factors depend on the potential energy of the system before and after the attempted move, $\Delta U$; in the case of the insertion and deletion moves, they also depend on the chemical potential or fugacity of the adsorbing species. Therefore, it is clear that at a fixed volume, temperature and chemical potential of the system, and given the molecular structure of the porous solid, the state of the system is governed by the interaction energy, based on the employed force field. Hence, the key message of this section is that in Monte Carlo simulations (as in Molecular Dynamics) the force field is the main input information required to setup the physical description of the system, whereas everything else can be treated as technical details.

As the simulation progresses, the positions of the molecules change and the number of the molecules fluctuates, producing a set of microstates over which the average properties of the system can be calculated. This set of microstates is called a trajectory and it is a common outcome of both Monte Carlo and molecular dynamic simulations (in a sense that it reflects the position of the system in the phase space), with the difference that the Monte Carlo trajectory is not a function of physical time and does not contain information about the velocities of the molecules.

The ensemble of microstates within the trajectory can be used to produce the relevant output properties of the system. In the context of the adsorption studies, the most important property is the average number of molecules present in the system. For a single value of the chemical potential or fugacity the simulation will produce an average adsorbed density. A series of simulations at increasing chemical potentials will produce an adsorption isotherm.

An important distinction has to be made between the absolute and excess amount adsorbed. The absolute amount adsorbed is the actual number of molecules present in the micropores at a particular fugacity. The excess amount is the difference between the absolute amount adsorbed and the number of molecules that would be present in the micropore volume according to the bulk gas density at the pressure and temperature of adsorption. The distinction between different definitions of adsorption and their connection to the experimental measurements has been discussed by Brandani *et al.* [179]. Monte Carlo simulations report absolute amount adsorbed, whereas experimental measurements are more often presented as the excess amount. The process simulations discussed in the next section take as an input analytical models for the absolute amount adsorbed.

Another important property that can be obtained from GCMC simulation is the enthalpy of adsorption. As will be discussed in the process modelling section of this review, in real processes and in process models based on adiabatic considerations, heat effects may play a role in the performance of the cycle. In molecular simulations, this property can be calculated either using the expression based on the result from the statistical-mechanical fluctuation theorem [180] or, in a direct analogy to the experimental methods, using the Clausius-Clapeyron equation. In the first case, a single isotherm is sufficient to calculate the heat of adsorption at each adsorption pressure. However, at high loading the reliability of this method deteriorates: it relies on the fluctuation of the number of adsorbed molecules in the system, and since at high loading the acceptance ratio for the insertion and deletion Monte Carlo moves is low, convergence of the method becomes problematic. This is not an issue for the approach based on the Clausius-Clapeyron equation [180], however, this method requires adsorption isotherms at several temperatures. Finally, simplified expressions are available if one is interested in the heat of adsorption in the Henry's law (zero loading) regime.

In addition to the properties directly required by the process simulation data (adsorption equilibria, heats of adsorption), molecular simulations also generate a wealth of information by visualizing the adsorption process on a molecular level (*e.g.* visualizations and density maps). These properties help to elucidate, for example, presence of specific binding sites and distribution of the molecules in the structure, which in turn can be used to construct new analytical models for adsorption.

So far, this brief introduction to the grand canonical Monte Carlo methods for adsorption problems implicitly assumed rigid crystal structures and rigid adsorbate molecules (with small gas molecules, such as nitrogen, carbon dioxide, methane being adequately described by this approximation). Extension of GCMC simulations to larger flexible molecules (*i.e.* alkanes) requires more advanced techniques, such as the Configurational-Bias GCMC [174]. Adsorption behaviour in flexible MOFs has also attracted significant attention over the years. To capture these phenomena, simulation in the Osmotic ensemble is required as well as advanced force fields to correctly represent the internal degrees of freedom within the framework [181].

### 6.2.2. Monte Carlo Simulation Codes

To make the review a practical reference, here we briefly introduce the open-source Monte Carlo codes for simulation of equilibrium adsorption isotherms in porous materials. These codes are listed in **Table 4**. We note here that a special issue of Molecular Simulation journal invited the community to reflect on the codes and algorithms available for the Monte Carlo simulations, their accessibility and applicability, efficiency and challenges [182]. In a recent study, we tasked ourselves with exploring the consistency of some of the most commonly used MC codes as listed in **Table 4** and examined their relative efficiency [183]. For this, we concentrated on a specific case study of carbon dioxide adsorption in IRMOF-1 material at conditions for which previous simulation results and experimental data were available [184]. It was a significant reassurance for us to observe that the codes were indeed consistent with each other. To assess their relative efficiency, we employed analysis based on the statistical inefficiency of sampling to compare trajectories from different codes on a consistent basis of the rate with which they were generating a statistically novel configuration. Our analyses revealed some differences in the overall performance of various MC codes, nevertheless this variation was found to be relatively negligible [183]. RASPA, MuSiC and DL_MONTE were overall the top performing programs in the analysis. Within the same article, we also generated consistent setups and scripts for all the codes for the above test case, which can be used by the molecular simulation community as a template for consistency tests and validation of future MC codes. These materials are available from our online github repository[2] [183]. Consistency and efficiency of MC codes are particularly important in the context of materials screening and multiscale simulation workflows.

**Table 4.** Monte Carlo simulation codes

| Software | Reference | Website |
|---|---|---|
| Cassandra | Shah et al [185] | https://cassandra.nd.edu/ |
| DL Monte | Purton *et al.* [186] | https://www.ccp5.ac.uk/DL_MONTE |
| MuSiC | Gupta *et al.* [187] | http://zeolites.cqe.northwestern.edu/Music/ |
| RASPA | Dubbeldam *et al.* [172] | https://www.iraspa.org/RASPA/index.html |
| Towhee | Martin [188] | http://towhee.sourceforge.net/ |

Here, we briefly introduce the codes listed in **Table 4**:

[2] GCMC Benchmark: https:// https://github.com/SarkisovGroup/gcmcbenchmarks

**Cassandra:** is a MC program developed in the group of Maginn at the University of Notre Dame. It is an effective package for simulation of the thermodynamic properties of fluids and solids [185]. Cassandra supports canonical (NVT), isothermal-isobaric (NPT), grand canonical (μVT), osmotic (μpT), Gibbs (NVT and NPT versions) and reactive (RxMC) ensembles. The code can be compiled to run in parallel using OpenMP [185].

**DL_Monte:** is another Monte Carlo simulation software that can be run in parallel [186]. It is originally developed by Purton and co-workers at Daresbury Laboratory in the UK with special emphasis at materials science. It is now being developed as a multi-purpose simulation package in collaboration with Wilding (University of Bristol) and Parker (University of Bath) research groups. The code can simulate systems in canonical (NVT), isobaric-isothermal (NPT), grand canonical (μVT), semi-Grand canonical, and Gibbs ensembles [186]. DL_MONTE is a twin sister code of DL_POLY package, a molecular dynamics simulation software that will be introduced later in this review. With regard to parallelization of MC codes such as DL_Monte and Cassandra, Gowers *et al.* [183] have demonstrated that the measured performances of existing implementations show poor efficiency due to various reasons. At least in the context of adsorption simulations and computational screening of porous materials, parallel execution of multiple MC runs offers higher efficiency and larger overall speed up as compared to parallelization of MC codes [183].

**MuSiC:** The Multipurpose Simulation Code (MuSiC) is an object-oriented software developed in Snurr's research group from Northwestern University[187]. The code supports grand canonical (μVT), canonical (NVT), and isobaric-isothermal (NPT) ensembles. It can also be used to perform hybrid MC and molecular dynamics (MD) simulations [189].

**RASPA:** is a molecular simulation code designed for simulation of adsorption and diffusion processes in nanoporous materials, including flexible structures [172]. The code was developed in Snurr's group at the Northwestern University in collaboration with several other scientists in the field of molecular simulations [172]. RASPA supports a variety of ensembles including micro canonical (NVE), canonical (NVT), isobaric-isothermal (NPT), isoenthalpic-isobaric (NPH), Gibbs (NVT and NPT versions), and isobaric-isothermal ensemble with a fully flexible simulation cell (NPTPR) [172]. It can be used to perform both Monte Carlo and molecular dynamics simulations, however it is best known for its capability as a MC software. The code also supports configurational bias Monte Carlo (CBMC), and continuous fractional component Monte Carlo (CFMC) for rigid and flexible molecules [172, 178].

**MCCCS Towhee:** The Monte Carlo for Complex Chemical Systems (MCCCS) program was originally developed in Siepmann's research group at the University of Minnesota. It is currently being developed and maintained by Martin [188, 190]. The code was initially designed for the prediction of fluid-phase equilibria, however it has been extended later to simulate different systems including porous materials. Towhee supports a variety of ensembles including NVT, NPT, μVT and Gibbs ensemble [188].

### 6.2.3. Molecular Dynamics Simulation

In this section, we turn our focus to molecular dynamics, which is widely employed for calculation of time-dependent phenomena across different fields from gas separation to materials science, geological sequestration of gases, biomolecular science, and drug discovery [173, 191-196]. The brief description provided here solely concerns molecular diffusion of simple gases in crystalline porous materials, which is relevant to the topic of this review. This section is meant to serve as an introductory material for non-expert readers. For more in-depth discussion of this technique, the reader is referred to numerous resources available in the literature [175, 196-200].

In contrast to Monte Carlo method where the microstates of the system are generated stochastically, in MD, we consider evolution of the system in space and time by numerically solving Newton's classical mechanics equations of motion [199]. In a system of particles interacting with each other and their environment, the total force exerted on each particle is given by [199, 201]

$$F_i(t) = m_i a_i(t) = m_i \frac{dv_i}{dt} = m_i \frac{d^2 r_i}{dt^2} = -\nabla_i U(r_i) \qquad i = 1, 2, 3, \dots, N \tag{7}$$

where $F_i$, $v_i$, $m_i$, $a_i$ and $r_i$ denote the force, velocity, mass, acceleration and position of the $i^{th}$ particle respectively, and $U$ and $t$ stand for the potential energy of interaction and time. The above equation is normally solved from a Taylor series expansion about initial position and velocity of particles in the system [199, 202]. There are several algorithms in the literature for time integration of equation (7) such as the Leapfrog [203] and Verlet [201]. In the latter one, which is not only one of the simplest methods but also one of the most widespread algorithms [197], the position of the particles at each time step is calculated by:

$$r_i(t + \Delta t) \approx 2r_i(t) - r_i(t - \Delta t) + \frac{F_i(t)}{m_i} \Delta t^2 \tag{8}$$

The above estimate of the new position of particle *i* contains an error that is of order $\Delta t^4$, where $\Delta t$ is the time step in the MD simulations [197].

In the context of gas adsorption where diffusion of particles in porous materials is monitored, MD simulations are normally carried out in the canonical (NVT) ensemble where volume, *V*, temperature, *T*, and the number of particles in the system, *N*, are conserved. This approach is suitable for molecular diffusion in materials with rigid porous frameworks. For materials with large framework flexibility, simulations can also be performed in NPT ensemble where pressure, *P*, is constant instead of the system volume, *V* [204, 205]. This would allow volume of the system to change under constant pressure, which is often the case in diffusion experiments.

**Figure *13*** depicts the schematic diagram of the MD workflow and the properties that can be calculated from typical MD simulations. In MD, we need to define a set of starting (*i.e.* initial) configurations for the system which are often obtained from GCMC simulation. Similar to the MC method, interatomic interactions of all particles must be defined using an appropriate set of force fields along with other simulation parameters that are normally supplied to an MD program as input data (*e.g.* time step, temperature, pressure, *etc*). MD generates time trajectory of the system containing positions of all particles and their associated potential energies. From these data, a number of transport [174, 196, 206], and thermal properties [191, 207-209] can be calculated. Similarly to the key message in the discussion of the GCMC method, here we again emphasise that given a set physical constraints (*e.g.* NVT) these properties are a function of how molecules interact with each other, whereas all other parameters can be treated as technical details of the protocol. These technical details may influence the efficiency of sampling and convergence of the results but not the physical properties of the system. Hence, the force field is the main input property that defines the physics and the behaviour of the system of interest.

From the perspective of the multiscale workflows, the key data we are interested in obtaining using MD are transport properties of multicomponent mixtures. Indeed, obtaining information on multicomponent diffusion from experiments is not trivial and requires advanced techniques. Similarly to the GCMC simulation, extension of simulation from a single component system to multicomponent mixtures does not make the MD simulations significantly more complicated and this is the main advantage of molecular simulations. It is also important to recognize that "transport properties" is an umbrella term for several distinct diffusion phenomena and frameworks of description associated with them. Below we consider these phenomena using the single component and multicomponent cases. In the process we comment on what properties associated with these phenomena can be obtained from MD and what properties are required in process modelling.

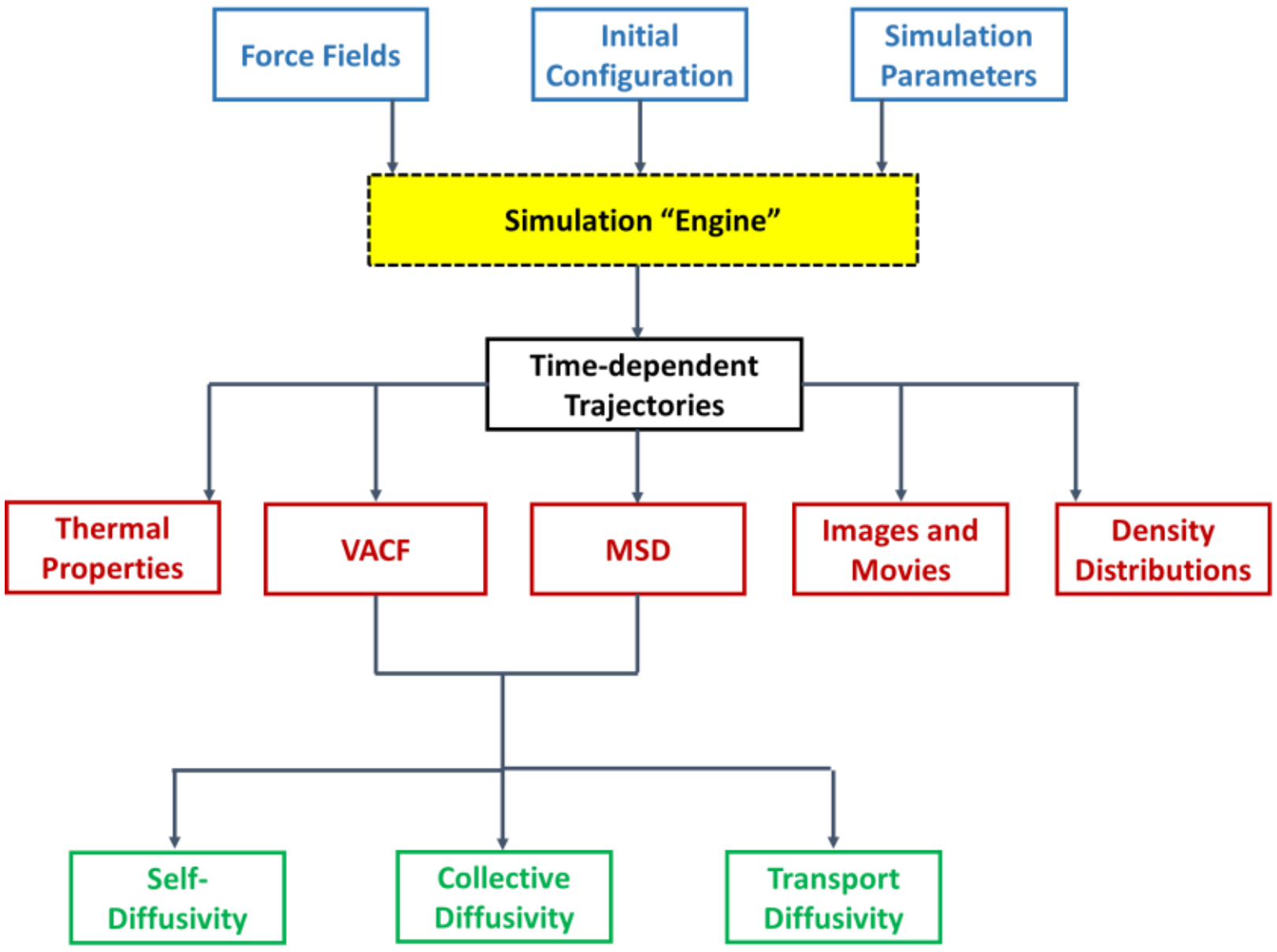

**Figure 13**. Schematic depiction of the workflow in the molecular dynamic simulations. The blue boxes indicate the required input data and parameters for the simulations. MD simulation generates a time-dependent trajectory from which the primary properties (red squares) such as mean-squared displacement (MSD) and velocity auto correlation function (VACF) are calculated. The green boxes are the secondary properties that can be calculated from the primary properties.

*Diffusion in single-component systems:*

Self-diffusivity, collective diffusivity, and transport diffusivity are three types of diffusion phenomena that are commonly studied by molecular simulations [174, 196, 200, 206, 210, 211]. Self-diffusivity ($D_s$) describes the motion of individual labelled molecules through a fluid in the absence of the chemical potential or concentration gradients. In experiments, this property is measured using tracer diffusivity techniques, such as pulsed field gradient (PFG) NMR. In simulation, equilibrium molecular dynamics (EMD) is extensively used to calculate self-diffusivity of adsorbate molecules in different types of porous frameworks [212-217]. Self-diffusivity can be conveniently computed from the mean-squared displacement of particles using the Einstein relationship given by:

$$D_s = \frac{1}{2Nd} \lim_{t\to\infty} \frac{1}{t} \langle \sum_{i=1}^{N} |r_i(t) - r_i(0)|^2 \rangle \tag{9}$$

Where $d$ is dimensionality of the system. $D_s$ can also be computed from the time integral of the velocity autocorrelation function (VACF) defined by:

$$D_s = \frac{1}{Nd} \int_0^\infty \langle \sum_{i=1}^{N} v_i(t) \cdot v_i(0) \rangle \, dt \tag{10}$$

here, $v_i\ (t)$ is the centre of mass velocity vector of molecule $i$. The brackets in (9) and (10) indicate an ensemble average taken over the simulation run time. As diffusion in porous materials is an activated process, temperature dependence of $D_s$ is typically captured in the well-known Arrhenius relation $D_s = D_0 \exp(-\frac{E_a}{k_B T})$, where $D_0$ is the pre-exponential constant and $E_a$ is the activation energy.

In contrast to self-diffusivity, the transport ($D_t$) and collective, or corrected, ($D_c$) diffusivities are associated with the macroscopic flux of molecules arising from the spatial concentration gradient in the fluid [173, 199]. The transport diffusivity also referred to as the Fickian or chemical diffusivity is related to net flux in the system, which is described by the Fick's first law:

$$J(q) = -D_t(q)\nabla q \tag{11}$$

here, $J$ and $\nabla q$ are the flux and concentration gradient in the adsorbed phase, respectively.

Equation (11) can also be described in terms of the chemical potential gradient $\nabla\mu$ [199]:

$$J(q) = -L(q)\nabla\mu = -\frac{q}{k_B T} D_c(q)\nabla\mu \tag{12}$$

where, $L$ is the Onsager transport coefficient, and $D_c$ is the corrected, or collective [199].

The transport diffusivity ($D_t$) is related to the collective diffusivity ($D_c$) through a term associated with curvature in the adsorption isotherm [174, 206]. This parameter is called the thermodynamic or Darken correction factor, $\Gamma$, described by:

$$\Gamma \equiv \frac{1}{k_B T}\left(\frac{\partial \mu}{\partial \ln q}\right)_T \tag{13}$$

Given the relation of fugacity with chemical potential $\Delta\mu \equiv k_B T \ln\left(\frac{f}{f_o}\right)$, one can rewrite equation (13) in the following form [174, 199].

$$\Gamma \equiv \left(\frac{\partial \ln f}{\partial \ln q}\right)_T \tag{14}$$

here, $f$ represents the fugacity of the bulk fluid in equilibrium with the adsorbed phase and $q$ denotes the concentration of the adsorbed phase. The thermodynamic correction factor can be calculated from the adsorption isotherm, which itself is obtained from GCMC simulation as explained in Section 6.2.1.

Therefore, the relation between $D_t$and $D_c$can be rewritten as [199]

$$D_t(q) = \frac{k_B T}{q} L(q)\Gamma = D_c(q)\Gamma \tag{15}$$

The collective and transport diffusivities can be calculated from both equilibrium molecular dynamics (EMD) and non-equilibrium molecular dynamics (NEMD) simulations. In the latter approach, the chemical potential gradient is the driving force for transport which is imposed on the system in the dual control volume grand canonical molecular dynamics (DCV-GCMD) [218, 219].

In EMD, the collective diffusivity can be computed from either of the following equations [173, 199]

$$D_c = \frac{1}{2Nd} \lim_{t\to\infty} \frac{1}{t} \langle \left| \sum_{i=1}^{N} [r_i(t) - r_i(0)] \right|^2 \rangle \tag{16}$$

$$D_c = \frac{1}{Nd} \int_0^\infty \langle \left( \sum_{i=1}^{N} v_i(t) \right) . \left( \sum_{i=1}^{N} v_i(0) \right) \rangle \, dt \tag{17}$$

In process modelling, the mass balance equations are formulated using the Fick's description of transport phenomena and therefore, it is the data and models for the transport diffusion coefficient, $D_t$, that are required to set up a process simulation.

*Diffusion of multi-component systems:*

To this point, we have discussed methods required for the calculation of different types of diffusion in single-component systems. Diffusion in multicomponent systems is generally an advanced topic with extensive literature available on the fundamentals and practical applications [220]. Here, we mention only essential concepts to illustrate what properties can be obtained from molecular simulations and challenges associated with the incorporation in the process models.

Several equivalently rigorous formulations of multicomponent diffusion exist: *e.g.*, Onsager, Maxwell-Stefan, and the generalized Fick's approach [221, 222]. Briefly, for an n-component system, the generalized Fick's law can be formulated as:

$$[J] = [D_t][\nabla q] \tag{18}$$

here, [J] is the column vector of diffusion fluxes of the components in the system and $[\nabla q]$ is the column vector of the diffusion gradients in the adsorbed phase. The mutual diffusion matrix, $[D_t]$, is given by:

$$[D_t] = [B]^{-1}[\Gamma] \tag{19}$$

where

$$B_{ii} = \frac{x_i}{Đ_{in}} + \sum_{\substack{k=1 \\ k\neq i}}^{n} \frac{x_k}{Đ_{ik}} \;, \quad B_{ij\,(i\neq j)} = -x_i \left( \frac{1}{Đ_{ij}} - \frac{1}{Đ_{in}} \right) \tag{20}$$

with $\delta_{ij}$ being the Kronecker delta and $Đ_{ij}$ is the Maxwell-Stefan diffusion coefficients, and $[\Gamma]$ is a matrix of thermodynamic correction coefficients. Equivalently, equation (20) could be formulated using a matrix of Onsager coefficients $[L]$, which can be shown to be related to $[B]^{-1}$ [220, 223].

In principle, all properties in equation (19) can be obtained from molecular simulations. Mutual diffusion coefficients and the components of the Onsager matrix can be obtained using expressions, similar to equation (16) for a multicomponent system:

$$L_{ij} = \frac{1}{2N_j d} \lim_{t\to\infty} \frac{d}{dt} \langle \left( \sum_{l=1}^{N_i} [r_l^i(t) - r_l^i(0)] \right) \times \left( \sum_{k=1}^{N_j} [r_k^j(t) - r_k^j(0)] \right) \rangle \tag{21}$$

whereas elements of $[\Gamma]$ could be obtained from GCMC simulations of multicomponent systems. This immediately points to two challenges. Firstly, construction of the comprehensive data for multicomponent diffusion requires a substantially larger number of simulations, with properties, such as $L_{ij}$ difficult to converge. The complete matrix of thermodynamic correction factors also requires GCMC simulation of multicomponent systems, which may be associated with substantial parameter space (*i.e.* the variation of the composition of the gas and adsorbed phases). Secondly, the process simulations require a continuous analytical model of the transport and equilibrium properties. Hence, the data obtained from molecular simulations for the properties above would need to be fitted to some simplified models (*e.g.* the Darken approximation of Maxwell-Stefan coefficients) or be amiable to interpolation within the process model. This will be further complicated, if one wants to incorporate temperature dependence of the diffusion coefficients, since in the micropores it is an activated process.

*What data on transport properties is required in process simulations?*

The general theoretical framework for multicomponent transport phenomena may require substantial number of parameters that are difficult to obtain both in experiments and simulation. However, to construct a process model such a level of description may not be actually needed. To understand this, it is useful to broadly identify three regions of the process where transport of the components of the mixture take place: the space between the pellets of the porous material in the adsorption column, the macropores within the pellets and the micropores in the small crystal grains (crystallites) constituting the pellets.

In the gas phase of the interstitial space between the pellets and in the macropores, concentration dependent diffusion coefficients would be required for the cases when the number of components is more than two and when the system is expected to significantly deviate from the ideal gas. This is not the case for low pressure binary mixtures of $N_2$ and $CO_2$. Hence, as we will see in the process modelling Section 6.3, for the diffusion in these regions we have a range of classical models, such as the Chapman-Enskog model for molecular diffusivity, that provide concentration independent Fickian diffusion coefficient.

What about the micropores? In the same section on process modelling, we will also explain why in the commonly adopted process models for PSA post-combustion carbon capture, the diffusion in micropores of the crystal is not considered at all. The assumption is that for micropores larger than the size of adsorbing molecules (for species such as $CO_2$ and $N_2$, the micropores should be larger than 4Å), the micropores are in instant equilibrium with the gas phase in the macropores of the pellet and we will provide a comment on why it is a reasonable assumption.

Hence, the remaining domain of processes and applications where the multicomponent data are indeed required in sufficient detail is associated with kinetic separations, such as for example the separation of oxygen and argon in molecular sieves, or propane-propylene separation using 4A zeolites. However, even in the kinetically controlled systems, single component diffusivities coupled with the gradients of the chemical potential will provide a reasonably good model for process simulations in the most cases. Molecular simulations, however, could be useful to probe under what conditions these assumptions are correct; to test when the models of additional, intermediate complexity may be required and identify reliably approaches to calibrate them. In summary, we are not aware of process modelling studies that incorporated the description of the multicomponent diffusion in its full complexity, although some studies employed simple models for micropore diffusion based on concentration independent single component data [31, 224, 225].

### 6.2.4. Molecular Dynamics Codes

In this section, we briefly introduce some of the most widely-used open-source molecular dynamics simulation software. There are numerous MD codes developed by various research groups amd commercial developers [178, 226], some of which are purpose-built software that are developed with particular applications in mind such as, for example, large biological systems (*e.g.* NAMD [227], CHARMM [228]). In this section however, we only focus on MD packages that offer many useful features for simulation of fluid transport in nanoporous materials. These softwares are listed in **Table 5** and are briefly described here:

**LAMMPS:** the Large-scale Atomic-Molecular Massively Parallel Simulator (LAMMPS) is a highly efficient and scalable classical molecular dynamics simulation code developed by the US Sandia National Laboratories with a focus on materials modelling [229]. It can be used for simulation of solid-state materials (metals, semiconductors), soft matter (biomolecules, polymers), coarse-grained, and mesoscopic systems [229]. LAMMPS can be employed as a parallel particle simulator at the atomic, meso, or continuum scales [229]. LAMMPS is written in C++. Many features of the code support accelerated performance on CPUs, GPUs, Intel Xeon Phis, and OpenMP [229] .

**GROMACS:** The Groningen Machine for Chemical Simulations is a MD simulation software primarily designed for simulation of biochemical molecules [230], however due to its computational efficiency it is also highly popular in the domain of materials modelling and simulation of transport processes in porous media. The code is written in C/C++. It was originally developed at the Department of Biophysical Chemistry in the University of Groningen. Since 2001, two teams at the Royal Institute of Technology (KTH) and the Uppsala University in Sweden have been responsible for development and maintenance of the GROMACS software.

**DL_POLY:** is another classical MD simulation software, which was developed at Daresbury Laboratory in the UK. It is a massively parallel code written in Fortran 90 which is suitable for simulation of macromolecules, polymers, ionic systems, solutions, and transport in porous media [178, 231].

**Table 5.** Molecular dynamics simulation codes

| Software | Reference | Website |
|---|---|---|
| LAMMPS | Plimpton [229] | https://lammps.sandia.gov |
| GROMACS | Abraham *et al.* [230] | http://www.gromacs.org |
| DL_POLY | Todorov *et al.* [231] | http://www.ccp5.ac.uk/DL_POLY |

### 6.2.5. Force fields

A comprehensive review of the current state-of-the art in force fields for adsorption phenomena in nanoporous materials has been recently provided by Dubbeldam and co-workers [232]. Here, we mention only essential elements required in the context of the multiscale workflows. A force field is a set of equations and parameters that describe how molecules interact with each other and with their environment, and governing the thermophysical properties of the system of interest.

Let us consider adsorption of $CO_2$ in a rigid porous material. Small molecules such as $CO_2$ can be also treated with reasonable accuracy as rigid structures. The total energy of interaction in this case is associated only with non-bonded (not involving a chemical bond) contributions and can be seen as composed of two terms: molecules of the gas interacting with each other (we call this for simplicity *fluid-fluid* interactions) and with the atoms of the porous structure (*fluid-solid* interactions):

$$U_{non-bonded} = U_{fluid-fluid} + U_{fluid-solid} \tag{22}$$

In their turn, each of these terms can be seen as composed of the dispersion or the van der Waals interactions and the polar interactions, due to the inhomogeneous distribution of the electron density within the molecule and the porous material. This inhomogeneous distribution of the electron density can be permanent, as in polar molecules, or induced. The commonly adopted model to describe dispersion interactions between two atoms is the Lennard-Jones potential:

$$u_{LJ,ij} = 4\varepsilon_{ij}\left[\left(\frac{\sigma_{ij}}{r_{ij}}\right)^{12} - \left(\frac{\sigma_{ij}}{r_{ij}}\right)^{6}\right] \tag{23}$$

where $\varepsilon$, $\sigma$, and *r* are the potential well-depth, collision diameter and distance, respectively and the indices *ij* indicate that these properties are obtained for a pair of atoms *i* and *j*.

The polar interactions are usually captured by placing partial charges on the specific atoms within the molecule. The interaction between the two partial charges is then obtained using the usual Coulomb equation:

$$u_{Coul,ij} = \frac{q_i q_j}{4\pi\varepsilon_o r_{ij}} \tag{24}$$

where $q_i$ and $q_j$ are individual partial charges on atoms *i* and *j*, and $\varepsilon_o$ the vacuum electrical permittivity. The total dispersion and electrostatic interaction energy in the system are then simply a sum of all pairwise terms according to equations (23) and (24) between atoms and charges in the system. In practice, these calculations are performed within a particular cut-off distance around each individual atom. For short range interactions such as the Lennard-Jones potential, the calculated value quickly converges as a function of distance, leading to a small error if the cut-off is equal to a few atom diameters. This is not the case for the long-range Coulombic interactions and advanced techniques such as the Ewald summations have to be employed to account for this.

For rigid porous materials and small, rigid adsorbate molecules, the collection of all Lennard-Jones parameters of the atoms in the system, partial charges assigned to them and the particular rules associated with the calculation of the cross-term for unlike atoms constitute the simplest force field.

If one wants to consider more complex systems featuring, for example, flexible molecules or flexible porous structures, additional energy terms to describe internal degrees of freedom (bond and angle vibrations, dihedral rotations, *etc*) will be required as defined by equation (*25*) [178, 232]:

$$\begin{aligned} U_{total} = &\sum_{bonds} u_b(r) + \sum_{bends} u_\theta(\theta) + \sum_{tortion} u_\phi(\phi) + \sum_{out\ of\ plane\ bends} u_\chi(\chi) \\ &+ \sum_{non-bonded} u_{nb}(r) + \sum_{bond-bond} u_{bb'}(r,r') + \sum_{bond-bend} u_{b\theta'}(r,\theta) \\ &+ \sum_{bend-bend} u_{\theta\theta'}(\theta,\theta') + \sum_{bond-torsion} u_{r\phi}(r,\phi,r') \\ &+ \sum_{bond-torsion} u_{\theta\phi}(\theta,\phi,\theta') + \cdots \end{aligned} \tag{25}$$

In this case, it is the collection of all functions and parameters involved in equation (*25*) that constitute a complete force field.

Over the years a substantial number of force fields have been developed. They differ in the functional forms employed in equation (*25*), target properties they reproduce, specialization and numerical procedures used to optimize the force field parameters to capture the target properties. Important characteristics of the force fields are:

(1) Availability: the force field has parameters available for a particular group of molecules and species of our interest.

(2) Accuracy: the force field is able to reproduce particular properties of the system.

(3) Transferability: the same force field can be applied to another class of molecules, while retaining its accuracy.

From this perspective it is useful to distinguish very specialized force fields, such as AMBER [233-239] for biological systems, which are very accurate for specific properties within a specific group of chemical species, but may not be available for other classes of chemicals or have limited transferability and generic force fields such as the Universal Force Field (UFF) [240-242] and DREIDING [243] that are based on a small number of basic elements, are able to describe a broad range of chemicals and species, but also for the same reason may lack consistent accuracy in description of the properties of interest.

A special comment should be also made on the assignment of the point charges within a particular force field. There is currently no universally accepted system of point charge assignment, because point charges are not experimentally observable properties. As a result, different force fields adopt different strategies on how to assign partial charges on the molecules under considerations. For example, the UFF model was originally calibrated to work with no charges assigned or charges obtained using the Qeq charge equilibration method [240]. For porous materials, the common practice is to assign charges in a separate step as these charges are not readily available from the standard force fields. For this, again, many algorithms were developed over the years, including empirical approaches, based on fitting some target properties, and a wide range of methods based on information from quantum-mechanical (QM) calculations, including the Mulliken Population Analysis [244], Density Derived Electrostatic and Chemical (DDEC) charges [245], and ChelpG [246] to name a few. What is important to recognize here is that this large variety of methods differ in their fundamental principles, in the level of theory they use and the system they consider to calibrate the charges (periodic systems, fragments). The complexity of charge assignment for materials such as MOFs has been recently explored by Sladekova *et al.* [247], who also provided a useful introduction to the previous studies investigating the influence of the choice of the charge assignment scheme on the adsorption properties of the material.

In the context of adsorption in porous materials, a number of force fields have been developed for zeolites. In particular, accurate force fields have been developed to describe adsorption of hydrocarbons in all-silica zeolites [248]. These force fields stem from the Transferable Potentials for Phase Equilibria Force Field (TraPPE) model that has been developed to accurately capture phase equilibria of alkanes and other organic species [249-254]. Accurate force fields for $CO_2$, $N_2$, and some other small gases in zeolites are also available from Gacia-Sanchez *et al.* [255] and from Martin-Calvo *et al*. [256].

In case of MOFs the situation is more complex due to significant chemical heterogeneity of these materials. Early molecular simulation studies adopted generic force fields such as UFF and DREIDING for the sole reason that these force fields contained some parameters for metal atoms, required to describe MOFs [257]. These force fields in fact proved quite reasonable in description of adsorption of simple non-polar molecules, such as methane, and noble gases [53].

The situation became more difficult as the focus of the research community shifted to adsorption of polar molecules, such as carbon dioxide and water. Adsorption of these molecules in MOFs and ZIFs requires assignment of partial charges on the atoms of the structure. As we discussed above, the number of possible methods to assign these charges is significant, and there is not yet a single, agreed procedure for this step.

An additional challenge is posed by MOFs with the open metal sites. The exposed metal sites interact quite strongly with molecules such as $CO_2$, water and unsaturated carbons, and this is where generic force fields fail [258]. Accurate description of interactions of these molecules with open-metal site MOFs has been subject of intensive investigation in recent years [259-264]. The employed approaches involved accurate QM calibration of the functional forms of the potentials and associated parameters and led to several specialized force field for certain groups of MOF materials, such as the MOF-74 family [261, 262, 264, 265]. These specialized force fields have, however, low transferability to other MOFs and so far have been focused on specific adsorbate molecules, such as $CO_2$, whereas the comprehensive implementation of the multiscale frameworks requires accurate description of adsorption of all components in the multicomponent mixture, including nitrogen. As the flue gas also contains some water, modelling of water adsorption in addition to $CO_2$ and $N_2$ would also allow to construct more accurate and realistic process models. However, accurate molecular simulation of water adsorption in all materials, regardless their nature, is still a very challenging problem.

Finally, we note that although the developed force field may be reliable in the prediction of the equilibrium adsorption properties, it does not necessarily implies accurate prediction of transport properties using the same force field.

In summary, even within the constraints of rigid structures and small rigid gas molecules, accurate force fields for $CO_2$ and $N_2$ adsorption are available and have been validated only for a handful of materials. Later in this review, we will discuss this challenge and its implications on the computational screening workflows.

**Beyond rigid structures: Force fields for prediction of structural transitions and lattice vibrations**

Molecular simulations typically assume adsorbent materials to have rigid frameworks. Recently, novel porous materials have been discovered that exhibit structural flexibility [266, 267]. Development of force fields that can correctly capture this behaviour is an ongoing area of research. This is particularly important for the studies of MOFs, as all MOFs exhibit *some* forms of structural flexibility [181, 232, 266] ranging from lattice vibrations at equilibrium to large-scale structural transformations upon external stimuli [232], such as temperature [268], guest adsorption [269] and electric field [270]. Among different types of structural flexibilities, structural vibrations and phonon properties of the lattice determine specific heat capacity of porous solids [207, 271, 272], whose importance for performance-based materials screening has been recently demonstrated [101, 104].

As elucidated by Kapil *et al.*, thermal properties of the lattice can be described by a quantum harmonic treatment [272]. However, the heat capacity of loaded porous frameworks requires a combination of quantum and anharmonic treatment [272]. Analysis of phonon properties for estimation of thermal properties of materials requires costly quantum mechanical calculations [273] which are not affordable for routine screening of large numbers of porous materials. To address this limitation, development of purpose-built and computationally affordable force fields has been recently undertaken by several groups [273-276], nevertheless further developments for improved accuracy and transferability of these force fields are required [277].

## 6.3. Process Modelling and Optimization

The main objective of this section is to give an accessible guide on the PSA/VSA process modelling from fundamentals to practical implementation. We begin with the basics of the mass, energy and momentum balances in the adsorption column packed with pellets of adsorbent material (Section 6.3.1). We will introduce the hierarchy of models, differing in the levels of details in their description and in the assumptions involved. We will briefly review the commonly involved methods in the solution of the introduced balance equations, under the appropriate boundary conditions.

Setting up a process model requires a number of parameters and properties. For a non-practitioner, it can be overwhelming to see the process model in its full complexity, and hence in Section 6.3.2 we tasked ourselves with explaining what parameters are required and how their values can be obtained.

A pressure swing adsorption process involves several columns, each of them going through a cyclic sequence of steps. In Section 6.3.3, we will use a simple 4-step cycle to introduce the PSA process and the key concepts associated with its cycle, such as cyclic steady state (CSS), performance of the cycle in terms of purity, recovery, productivity and energy consumption. Furthermore, using this example of the 4-step process, we will briefly explore the concentration profiles during different steps at CSS and how to interpret them.

A specific cycle configuration may not operate at the optimal conditions. Hence, a significant part of the process modelling research is focused the cycle optimization. In Section 6.3.4, we introduce currently used optimization methods, such as genetic algorithms, and essential concepts associated with the process optimization.

As has been already discussed in the section on the process metrics, in general process simulations are time consuming. This prompted a significant research effort into more efficient alternatives for process performance evaluation that work in tandem with detailed process simulations. These developments are reviewed in Section 6.3.5.

Finally, following the spirit of the review, we conclude the section on process modelling with a brief overview of the available codes for this type of modelling, their capabilities and access (Section 6.3.6).

### 6.3.1. Fundamentals

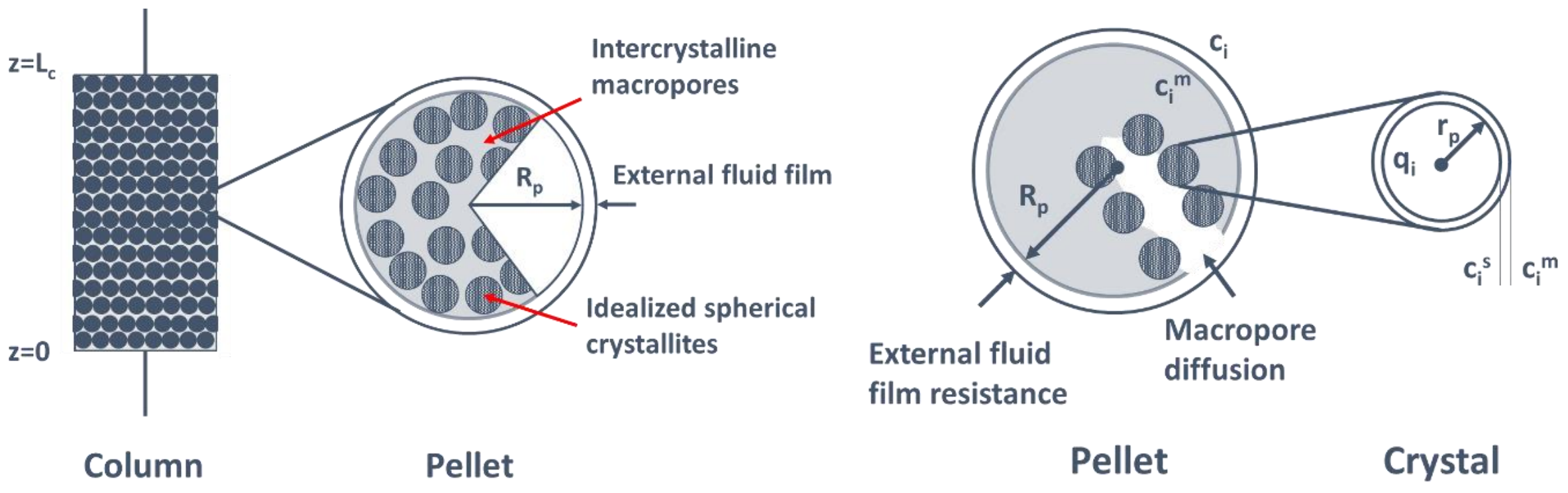

**Figure 14.** Schematic depiction of the adsorption system under consideration. The column is treated as a vessel filled with pellets of porous materials (on the left). Each pellet can be seen as an agglomerate of crystallites held together by inert binder. Other properties and processes are explained in the text.

An adsorption column is the basic unit of the adsorption process. In this section, we provide a brief summary of the mass, energy and moment balances around this unit, which are either solved numerically in the process simulations or serve as starting points for simplified analytical models. For a more comprehensive analysis we refer the reader to the seminal books by Ruthven *et al*. [278, 279] on fundamentals of adsorption and PSA processes.

Consider a schematic of a packed column in **Figure *14*** . The column has length of $L_c$, $z$ is used as the position within the column in the axial direction and the feed is introduced to the column from the bottom at $z = 0$. The column is packed with pellets of adsorbent material. The pellet consists of microporous crystallites, which are held together by an inert binder. Thus, the pellet has intercrystalline macropores and intra-crystalline micropores. In the description of the various transport processes, we adopt the following convention: *macropore* refers to the pore space between the crystallites and *micropore* refers to the pores inside the crystallites. On the right of **Figure *14***, we show an idealized spherical pellet of radius $R_p$ and volume $V_p$. In the model, we can also assume that crystallites are spherical particle of radius $r_p$. The pellet volume consists of the macropore volume $V_{macro}$ and crystal volume $V_{cr}$, which in turn consists of the micropore volume $V_{micro}$ and the skeletal volume $V_{skel}$:

$$V_p = V_{macro} + V_{cr} \tag{26}$$
$$V_{cr} = V_{micro} + V_{skel} \tag{27}$$

The bulk density $\rho_{bulk}$, pellet density $\rho_p$, crystal density $\rho_{cr}$ and skeletal density $\rho_{skel}$ are defined as follows:

$$\rho_{bulk} = \frac{m_p}{V_C} = \frac{m_p}{V_{gas} + V_p} \tag{28}$$
$$\rho_p = \frac{m_p}{V_p} = \frac{m_p}{V_{macro} + V_{cr}} \tag{29}$$
$$\rho_{cr} = \frac{m_p}{V_{cr}} = \frac{m_p}{V_{micro} + V_{skel}} \tag{30}$$
$$\rho_{skel} = \frac{m_p}{V_{skel}} \tag{31}$$

Here, $V_{gas}$ is the volume of the gas phase in the column and $m_p$ is the total mass of the adsorbent pellets. This mass includes both the mass of the adsorbent crystals as well as the mass of the binder. Thus, it is assumed that the binder volume is part of the skeletal volume of the pellet. Therefore, the saturation capacity of the adsorbent has to be corrected for the mass of the binder if the adsorption isotherms were measured for the non-pelletized adsorbent crystals. The bed void fraction $\varepsilon$, pellet void fraction $\varepsilon_p$ and crystal void fraction $\varepsilon_{cr}$ are defined as follows:

$$\varepsilon = \frac{V_{gas}}{V_C} = 1 - \frac{V_p}{V_C} = 1 - \frac{\rho_{bulk}}{\rho_p} \tag{32}$$
$$\varepsilon_p = \frac{V_{macro}}{V_p} = 1 - \frac{V_{cr}}{V_p} = 1 - \frac{\rho_p}{\rho_{cr}} \tag{33}$$

$$\varepsilon_{cr} = \frac{V_{micro}}{V_{cr}} = 1 - \frac{V_{skel}}{V_{cr}} = 1 - \frac{\rho_{cr}}{\rho_{skel}} \tag{34}$$

A common starting point for many process modelling approaches is the material balance in the column based on the axial dispersed plug flow model (although more complex and complete formulations are also possible, *i.e.* including radial dispersion term, *etc*):

$$\frac{\partial c_i}{\partial t} + \frac{(1-\varepsilon)}{\varepsilon} \cdot \frac{\partial \bar{Q}_i}{\partial t} + \frac{\partial (c_i \cdot v)}{\partial z} + \frac{\partial J_i}{\partial z} = 0 \tag{35}$$

Here, $c_i$ is the gas phase concentration of component *i*, $c_i^m$ is the macropore concentration of component *i* in the adsorbent pellet, $v$ is the interstitial velocity, and $J_i$ is dispersive flux of component *i*. In this equation, the first and the second terms are the accumulation terms in the gas phase and in the pellets, respectively. The amount adsorbed in the pellet can be seen as the composite of the amount as gas in the macropores of the pellet $\varepsilon_p c_i^m$ and the absolute amount adsorbed in the micropores of the adsorbent material, $(1-\varepsilon_p)q_i$:

$$Q_i = \varepsilon_p c_i^m + (1-\varepsilon_p) q_i \tag{36}$$

where $q_i$ is the sorbate concentration of component *i* in the micropores of the adsorbent. In the column mass balance, the average amount adsorbed in the pellet is needed:

$$\bar{Q}_i = \frac{3}{R_p^3} \int_0^{R_p} Q_i r^2 dr \tag{37}$$

and, similarly the average adsorbed amount in a crystallite can be defined:

$$\bar{q}_i = \frac{3}{r_p^3} \int_0^{r_p} q_i r^2 dr \tag{38}$$

The third term in equation (35) describes the convective flow of the gas across the bed and the final term describes the dispersion process relative to the bulk flow. The dispersive flux is given by:

$$J_i = -D_i^L c_T \frac{\partial x_i}{\partial z} \tag{39}$$

where, $D_i^L$ is the axial dispersion coefficient, $c_T$ is the total gas concentration, and $x_i$ is the mole fraction of component *i*. For the axial dispersion coefficient ($D_i^L$) correlations are available [279], such as:

$$D_i^L = 20 \times \frac{D_i^m}{\varepsilon} + 0.5 V_0 \times \frac{2R_p}{\varepsilon} \tag{40}$$

here, $D_i^m$ is molecular diffusivity which is defined later in the section, and $V_0$ is the average superficial fluid velocity through the packed bed.

Although equation (35) provides the overall mass-balance in the column, it does not describe the actual process of diffusion into the pellets. For this, a separate set of material balance equations can be formulated around the pellet. In the most general case, the model will contain terms associated with the external film resistance at the pellet surface, macropore diffusion from the bulk gas phase into the pellet, barrier and film resistance at the adsorbent crystals boundary and micropore diffusion in the adsorbent crystals. A schematic of an adsorbent pellet with relevant properties is shown on the right of **Figure *14***. Let us consider these processes in more detail.

First, let us focus on the overall material balance for the pellet. The amount adsorbed the pellet is governed by the following mass-balance equation, based on the second Fick's law formulated for the spherical pellet geometry:

$$\varepsilon_p \frac{\partial c_i^m}{\partial t} + (1 - \varepsilon_p)\frac{\partial \bar{q}_i}{\partial t} - \frac{1}{r^2}\frac{\partial}{\partial r}\left(D_{macro,i}^e r^2 \frac{\partial c_i^m}{\partial r}\right) = 0, \qquad 0 < r < R_p \tag{41}$$

Here, $D_{macro,i}^e$ is the effective macropore diffusion coefficient. The first term of equation (41) represents the accumulation in the macropores, the second term describes the accumulation in the micropores and the last term describes diffusive mass-transport due to the concentration gradients inside the pellet (the second Fick's law).

At the surface of the pellet, diffusion from the bulk gas phase into the pellet can be described via mass-transfer process across the film at the surface:

$$D_{macro,i}^e \left.\frac{\partial c_i^m}{\partial r}\right|_{r=R_p} = k_{i,f}(c_i - c_i^m) \tag{42}$$

where $k_{i,f}$ is the external fluid film mass transfer coefficient. This equation sets the boundary condition at $R_p$, whereas at $r = 0$ the boundary condition is $\left.\frac{\partial c_i^m}{\partial r}\right|_{r=0} = 0$ which is required due to the assumption of spherical symmetry.

The effective diffusion coefficient reflects various mass-transfer mechanisms into the pellet and is obtained by combining the molecular diffusion $D_i^m$, Knudsen diffusion $D_i^K$, surface diffusion $D_i^S$ and the viscous diffusion coefficients $D_i^V$:

$$D_{macro,i}^e = \frac{\varepsilon_p}{\tau}\left[\left(\frac{1}{D_i^m} + \frac{1}{D_i^K}\right)^{-1} + D_i^S + D_i^V\right] \tag{43}$$

where, the individual diffusion coefficients are estimated using the well-known expressions:

$$D_i^m = 1.86 \times 10^{-7} \frac{\sqrt{T^3(MW_1^{-1} + MW_2^{-1})}}{P\sigma_{12}^2\Omega_{12}} \tag{44}$$

$$D_i^K = 97 r_{pore}\sqrt{\frac{T}{MW_i}} \tag{45}$$

$$D_i^S = \frac{1 - \varepsilon_p}{\varepsilon_p} K D_i^{S_0} exp\left(\frac{-E}{RT}\right) \tag{46}$$

$$D_i^V = 10^{-5}\frac{P r_{pore}^2}{8\mu_i} \tag{47}$$

here, $MW_i$ is the molecular weight in *g mol$^{-1}$*, $\sigma_{12}$ is the collision diameter from the Lennard-Jones potential in *Å*, $\Omega_{12}$ is a function depending on the Lennard-Jones force constant and temperature, $r_{pore}$ is the mean macropore radius in *m*, $K$ is the Henry's constant of adsorption, and $E$ is the diffusional activation energy. These expressions along with the theories behind them and the values of the parameters are discussed in the classical textbooks on transport phenomena [278, 280]. We further note, that typically in the process models the values are obtained at some fixed, representative conditions, while in reality the conditions change dynamically in the actual process, and hence, these properties would also vary in time in a more accurate model.

Similarly, for the diffusive process in the micropores inside the crystallites, modelled as spherical particles of size $r_p$, we can formulate a similar general mass-balance equation, based on the second Fick's law of diffusion:

$$\frac{\partial q_i}{\partial t} - \frac{1}{r^2}\frac{\partial}{\partial r}\left(D_i^{\mu} r^2 \frac{\partial q_i}{\partial r}\right) = 0, \qquad 0 < r < r_p \tag{48}$$

here, $D_i^{\mu}$ is the effective diffusion coefficient in micropores and other terms are described as before. Similar to the processes at the pellet surface, the diffusion into the crystallite particle from the surface can be described using transfer resistances across the surface:

$$D_i^{\mu} \frac{\partial q_i}{\partial r}\bigg|_{r=r_p} = k_{i,f}^{\mu}(c_i^m - c_i^s) = k_{i,b}\left(q_i^*(c_i^s) - q_i|_{r_p}\right) \tag{49}$$

Here, $q_i^*$is the adsorbed concentration of component *i* in equilibrium and $c_i^s$ is the concentration of component *i* at the crystal boundary. In equation (49) we equivalently consider fluxes across the external fluid film, governed by the mass-transfer coefficient $k_{i,f}^{\mu}$, or across the crystal boundary, governed by the mass-transfer coefficient $k_{i,b}$. Equation (49) defines a boundary condition for $r = r_p$, whereas at $r = 0$ it is $\frac{\partial q_i}{\partial r}\Big|_{r=0} = 0$ (again similar to the boundary condition of the pellet).

A similar hierarchy of equations can be formulated for the energy balance in the column. In the most general non-isothermal case, the following equations govern the heat-transfer processes:

$$\varepsilon \frac{\partial \breve{U}_f}{\partial t} + (1-\varepsilon)\frac{\partial \breve{U}_p}{\partial t} + \epsilon \frac{\partial(\breve{H}_f \cdot v)}{\partial z} + \frac{\partial J_T}{\partial z} + \varepsilon \sum_{i=1}^{N} \frac{\partial(J_i \widetilde{H}_i)}{\partial z} + h_w \frac{A_c}{V_c}(T_f - T_w) = 0 \tag{50}$$

here, $\breve{U}_f$ is the internal energy in the fluid phase per unit volume, $\breve{U}_p$ is the internal energy in the pellet per unit volume, $\breve{H}_f$ is the enthalpy in the fluid phase per unit volume, $J_T$ is the thermal diffusive flux, $\widetilde{H}_i$ is the partial molar enthalpy of component *i* in the fluid phase. $T_f$, $T_w$ are temperatures of the fluid and the wall, respectively, while $h_w$ is the heat transfer coefficient between the wall and the surroundings. $A_c$ and $V_c$ are the surface area and the volume of the column, respectively. The first two terms in equation (50) are accumulation terms for the gas phase and the solid phase respectively; the third term is associated with the convective flux of the fluid stream, with enthalpy $\breve{H}_f$. The next two terms are the axial dispersion terms. The first one, $\frac{\partial J_T}{\partial z}$, describes thermal flux due to the temperature gradients along the *z* axis, whereas the second term is associated with the diffusive fluxes due to the concentration gradients along *z* axis (and hence enthalpy fluxes coupled with them). The last term on the left in equation (50) describes heat transfer from fluid to the wall of the column.

The heat transfer across the wall of the column can be described as:

$$\rho_w \hat{c}_{P,w} \frac{\partial T_w}{\partial r} + h_w \frac{A_c}{V_c}(T_f - T_w) + U\alpha_{wl}(T_w - T_\infty) = 0 \tag{51}$$

where $T_\infty$ is the temperature of the surroundings. The column wall is defined by the column wall density $\rho_w$ and specific heat capacity, $\hat{c}_{P,w}$. The ratio of the logarithmic mean surface area to volume of the column wall $\alpha_{wl}$ is given by:

$$\alpha_{wl} = \left[(2R_c + \delta_w)\ln\left(\frac{2R_c + \delta_w}{2R_c}\right)\right]^{-1} \tag{52}$$

where $R_c$ and $\delta_w$ are the radius of the column and the thickness of the wall, respectively.

At the level of the pellet, uniform temperature profile is typically assumed across the pellet (no temperature gradients) and this has been shown to be consistent with the experimental observations [281].

The overall energy balance for the pellet can be then formulated as:

$$\frac{\partial \breve{U}_p}{\partial t} = \varepsilon_p \frac{\partial \breve{U}_{p,f}}{\partial t} + (1 - \varepsilon_p)\frac{\partial \breve{U}_{p,s}}{\partial t} = h_p(T_f - T_p)\frac{V_p}{A_p} + \sum_{i=1}^{N_c} \frac{\partial \bar{Q}_i}{\partial t}\bar{H}_{i,f} \tag{53}$$

Here, the first equality simply indicates that the total energy change in the pellet can be seen as a sum of the change in energy in the fluid phase in macropores and change in energy associated with the adsorbed phase (solid + micropores). The second equality links this change to the heat transfer across the pellet boundary with heat transfer coefficient $h_p$ and heat flux associated with the adsorption of the components in the system, where $\bar{H}_{i,f}$ is partial molar enthalpy of the component *i*.

Finally, the thermal axial dispersion flux $J_T$ is given by:

$$J_T = -\lambda_f^L \varepsilon \frac{\partial T_f}{\partial z} - \lambda_p^L (1 - \varepsilon)\frac{\partial T_p}{\partial z} \tag{54}$$

Here, the axial thermal conductivity in the fluid and pellet are given by $\lambda_f^L$ and $\lambda_p^L$, respectively. There are also alternative ways to formulate the energy balance, for an example of which we refer the reader to the article by Zhao *et al.* [282].

The momentum balance is described by the Ergun pressure drop equation:

$$-\frac{dP}{dz} = \frac{150\mu(1-\varepsilon)^2}{\varepsilon^2 4R_p^2} v + \frac{1.75(1-\varepsilon)^2 \rho_f}{\varepsilon 2R_p} v|v| \tag{55}$$

where, $\mu$ is the fluid viscosity and $\rho_f$ is the fluid density.

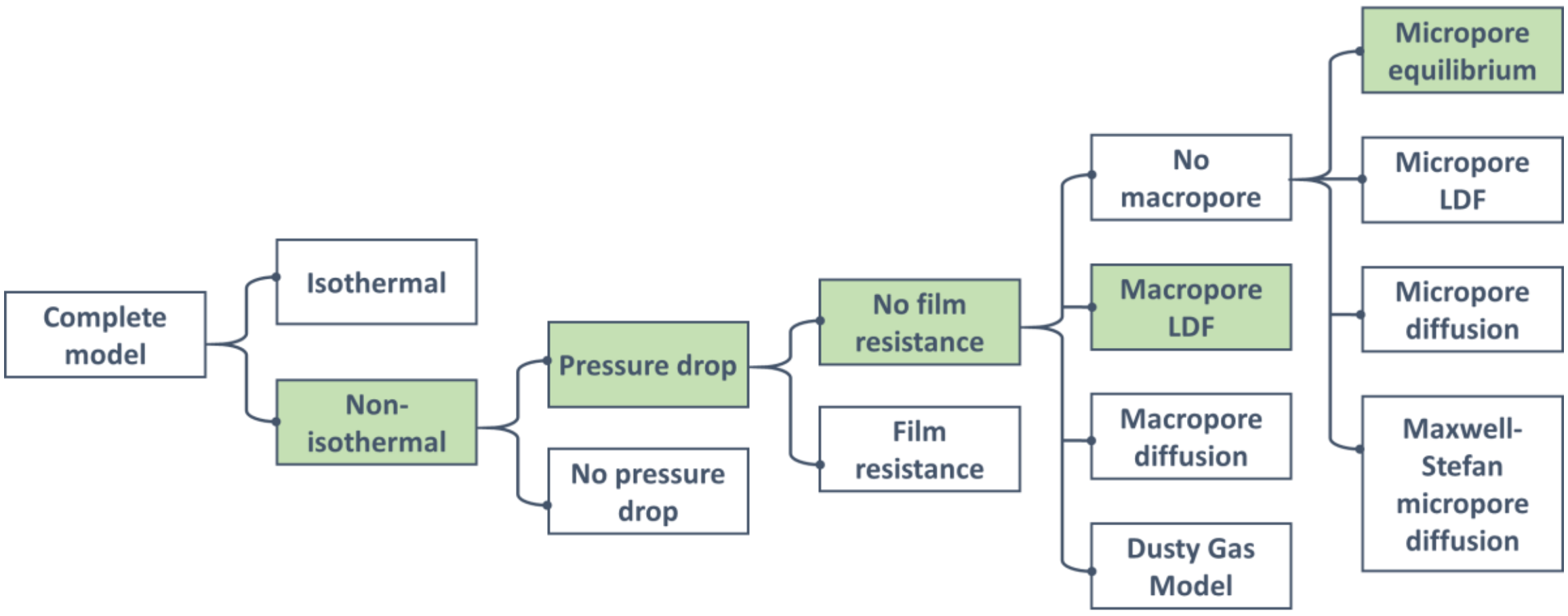

**Figure 15.** Hierarchy of the models available for the mass and energy balances in the adsorption column (not an exhaustive list). Squares shaded green reflect the combination of the models employed in the studies by Farmahini *et al.* [68, 101] and also commonly adopted by other practitioners in the field.

The equations above provide a complete and general description of the mass and energy balances in the column. These equations serve as a starting point for more simplified models. Indeed, **Figure *15*** illustrates the hierarchy of the models with each model based on its own set of assumptions and resulting simplifications of the governing equations. Reading this diagram from left to right, the system can be considered as isothermal (hence no energy balance equations are required) or non-isothermal. Then, within each branch, we can either include or ignore the pressure drop across the system. For each branch we can further consider whether we include film resistance at the surface of the pellet or not and so on. This hierarchy demonstrates that we can construct order of $10^2$ models depending on the combination of the assumptions we use. The boxes shaded green in **Figure *15*** represent the choice of the assumptions adopted in the studies of Farmahini *et al.* [68, 101] , as well as in many other previous studies [96, 283, 284]. In this case, the following assumptions are considered:

1. The system is modelled as non-isothermal with heat transfer allowed between the packed bed and its wall, but the pellets and gas phase are kept at the same temperature.

$$T_f = T_p \tag{56}$$

2. Pressure drop is considered across the bed. The pressure drop is modelled using the Ergun equation (55).

3. No external film resistance is considered. In this case, equation (42) vanishes and the following condition applies:

$$c_i^m(R_p\,, z, t) = c_i(z,t) \tag{57}$$

4. The macropore resistance is modelled using the Linear Driving Force (LDF) approximation. Effectively, all the resistances to diffusion are lumped into a single effective parameter, while the driving force of the process is simply the difference between the concentration of species *i* in the gas phase $c_i$ and in the macropores, $c_i^m$. As a result, equation (41) can be replaced with a simplified model:

$$\varepsilon_p \frac{\partial c_i^m}{\partial t} + (1-\varepsilon_p)\frac{\partial \bar{q}_i}{\partial t} = k_i^p \frac{A_p}{V_p}(c_i - c_i^m) \tag{58}$$

here, $k_i^p$ is the LDF coefficient for the pellet. This parameter can be calculated using the effective macropore diffusivity with the Glueckauf approximation, which is equivalent to assuming a parabolic concentration profile [285]:

$$k_i^p = \frac{5D_{macro,i}^e}{R_p} \tag{59}$$

5. Micropore equilibrium is assumed. This assumption implies that the crystallites are in instant equilibrium with the gas phase in the macropores of the pellet. This would be the case when the overall mass-transfer into the pellets is controlled by macropores, and not micropores. Although this seems counter intuitive, the validity of this assumption for materials with pore sizes that do not impose significant kinetic constrains on diffusion of small molecules (larger than 4 Å) has been discussed on several occasions [278, 286]. To illustrate this point, let us return to the equation (41) describing the mass balance around the pellet. If we make an assumption that the isotherm is linear ($\bar{q}_i = K_{H,i}c_i^m$ where $K_{H,i}$ is the Henry's constant for component *i*), equation (41) can be rearranged as:

$$\left[\varepsilon_p + K_{H,i}(1-\varepsilon_p)\right]\frac{\partial c_i^m}{\partial t} = \frac{1}{r^2}\frac{\partial}{\partial r}\left(D_{macro,i}^e r^2 \frac{\partial c_i^m}{\partial r}\right) \tag{60}$$

which can be further rearranged to obtain the Fick's diffusion equation and the effective pore diffusivity of component *i*, $D_{P,i}^e$:

$$\frac{\partial c_i^m}{\partial t} = \frac{1}{r^2}\frac{\partial}{\partial r}\left(D_{P,i}^e r^2 \frac{\partial c_i^m}{\partial r}\right) \tag{61}$$

$$D_{P,i}^e = \frac{D_{macro,i}^e}{\left[\varepsilon_p + K_{H,i}(1-\varepsilon_p)\right]} \tag{62}$$

While it is obvious that $D_i^\mu$ is always smaller than $D_{P,i}^e$, what is important in determining the controlling mass transfer mechanism is the comparison of the molar fluxes. In particular, the two diffusional time constants which should be compared to each other are then the macropore diffusion constant ${R_p}^2/D_{P,i}^e$ and the micropore diffusion time constant, ${r_p}^2/D_i^\mu$. Small crystals (small ${r_p}^2$), relatively large beads (large ${R_p}^2$) and large value of the effective Henry's constants lead to $\frac{{R_p}^2}{D_{P,i}^e} \gg \frac{{r_p}^2}{D_i^\mu}$ , or in other words mass transfer controlled by the macropore diffusion.

Hence, we assume that the micropores are in the instantaneous equilibrium with the gas phase in the macropores, described by the concentration $c_i^m$. This assumption is equivalent to the following condition:

$$q_i(z,t) = q_i^*(c_i^m) \tag{63}$$

Instead of the condition above, one may wish to include a more detailed model of micropore diffusion using the LDF approximation, then the following simplification can be employed to describe transport into the crystallites:

$$\frac{d\bar{q}_i}{dt} = k_i^{cr}\frac{3}{r_p}(q_i^* - \bar{q}_i) \tag{64}$$

The LDF coefficient $k_i^{cr}$ can be calculated from the effective micropore diffusivity by:

$$k_i^{cr} = \frac{5D_i^{\mu}}{r_p} \tag{65}$$

Regardless of the details of the model, the combined mass and energy balance equations form a system of Differential Algebraic Equations (DAEs). These equations are usually discretized in the spatial domain by an appropriate numerical method such as finite difference, finite element, orthogonal collocation, or finite volume method. This produces a system of Ordinary Differential Equations (ODEs) which can be solved using a number of approaches, such as the internal functions within the available simulation packages (*e.g.* MATLAB), or existing numerical solution libraries (*e.g.* SUNDIALS [287] ). Here, it is also useful to reflect on the simplified LDF-based models versus detailed diffusion equation models. The motivation to develop simplified LDF models is driven primarily by the numerical efficiency. Indeed the simplified LDF model with 30 axial volumes and 2 components corresponds to 120 DAEs, while the same system with the diffusion equation would be approximately 600 DAEs. Including also diffusion in the micropores would lead to *ca.* 3000 equations. The computational costs would be at least proportional to the total number of equations N if the code is well written and $N^2$ for a not-so-well written code.

### 6.3.2. Complete Hierarchy of Data Required for Multiscale Process Simulation

One of the primary aspirations of this review is to provide a useful guide on PSA/VSA process models for non-practitioners. Reading a standard research paper on process modelling of adsorption processes can often be overwhelming because of the number of parameters and properties one needs to specify, with their sources not necessarily being obvious. Here, we also emphasize that even after reading our review we do not expect a novice in process modelling to be able to setup their own simulations. However, we hope they will be able to understand the requirements for these simulations and be aware of the potential sources of data. Broadly, we can split the data required for setting PSA/VSA process simulations into the following categories. Column properties describe the geometric dimensions of the column, its length, diameter, and thickness of the walls. These properties either are taken to reflect the actual experimental unit, or given some specific, physically meaningful values. For example, certain parameters of the column have been used in several studies and they have now become commonly employed parameters for several groups to ensure consistent comparison of the process modelling results [67, 68, 72, 95, 99-101, 103]. The balance equations described in section 6.3.1, also imply that to solve these equations we need values of the properties associated with the thermophysical characteristics of the material of the column and how it interacts with its environment (*e.g.* heat capacity, heat transfer coefficient, etc).

In the next category, we have all properties associated with the pellet: pellet size, pellet porosity, and pellet tortuosity. In the same category, we also include properties associated with the transport in macropores of the pellets, such as different contributions to the overall macropore diffusivity (*e.g.* molecular diffusion, Knudsen diffusion, *etc*).

Further down in the hierarchy of scales shown in **Figure *15*** is crystallites, and hence the next category of properties is associated with the properties of the adsorbent material crystals: crystal density, crystal thermal and transport properties, etc. In principle, the pellet is made out of crystallites and binder and properties of the pellet, such as the specific heat capacity or thermal conductivity, are a composite property of the two materials, binder and crystallites. However, the common convention is to assume these properties of the binder to be equivalent to the properties of the adsorbent crystals.

Generally, equilibrium adsorption data should also belong to the category of the crystal properties. However, this requires a special consideration. Adsorption data, both in experiments and in simulations, are typically obtained as single component adsorption isotherms comprised of discrete data points. However, process simulations require an analytical expression describing adsorption equilibria in order to be able to solve the mass balance equations described in Section 6.3.1. Moreover, the accuracy of process modelling also depends on how well the supplied models describe the multicomponent equilibria – hence accurate interpolation of single component isotherms may not be sufficient for the correct behaviour of the model in the actual process simulations. A common approach is to use the dual-site Langmuir (DSL) adsorption model to obtain an analytical description of adsorption isotherms. For a single component system, the DSL isotherm for species *i* is defined by:

$$q_i^* = \sum_{j=1}^{2} \left[ q_{j,i}^s \frac{b_{j,i} \times P}{1 + b_{j,i} \times P} \right] \tag{66}$$

here, $q_{j,i}^s$ is saturation capacity of site *j* with respect to species *i*, $b_{j,i}$ is affinity of each site described by the van't Hoff equation: $b_{j,i} = b_{oj,i} \exp\left(\frac{-\Delta H_{j,i}}{RT}\right)$. In the van't Hoff equation, $\Delta H_{j,i}$ is the heat of adsorption at adsorption site *j* and $b_{oj,i}$ is the pre-exponential factor.

As seen here, there are six parameters ($q_{1,i}^s$, $q_{2,i}^s$, $b_{o1,i}$, $b_{o2,i}$, $\Delta H_{1,i}$ and $\Delta H_{2,i}$) for each gas component *i* which can be obtained. The thermodynamic consistency requires that the saturation capacity of each site is the same for all adsorbing species (for example, for the binary $CO_2/N_2$ adsorption this implies $q_{1,N_2}^s = q_{1,CO_2}^s$ and $q_{2,N_2}^s = q_{2,CO_2}^s$), unless adsorbing molecules differ significantly in size. In an early study, Myers showed that these conditions are essential for the accuracy of the multicomponent DSL model [288]. This poses additional constraints on the fitting of equation (66) to the reference adsorption data using non-linear least-square regression. Adsorption of species *A* from a binary gas mixture of *A* and *B* at fixed temperature is described by the extended version of the Dual-Site Langmuir model (extended DSL) which is given by:

$$q_A^* = \left[ q_{1,A}^s \frac{b_{1,A} \times P \times y_A}{1 + (b_{1,A} \times P \times y_A) + (b_{1,B} \times P \times y_B)} \right] + \left[ q_{2,A}^s \frac{b_{2,A} \times P \times y_A}{1 + (b_{2,A} \times P \times y_A) + (b_{2,B} \times P \times y_B)} \right] \tag{67}$$

where $y_A$ and $y_B$ are mole fractions of components *A* and *B* in the gas phase. To obtain physically meaningful parameters for the DSL model, normally the fitting algorithm is guided through a set of mathematical constraints, which also help the algorithm to converge [68]. The quality of the DSL model is ultimately tested by its ability to predict binary adsorption equilibria. This data may not be readily available from experiments, however in molecular simulations it is relatively easy to implement and carry out these tests. In our previous publications, we explored systematic ways to obtain parameters of the DSL model and we refer the reader to the original publication [68].

It should be noted that for many new materials such as phase-change adsorbents it is not easy to propose a suitable functional form that can properly describe equilibrium adsorption data [289, 290]. The alternative approach here is to describe the equilibrium relationship between adsorbed phase and fluid phase as a set of discrete points. Haghpanah *et al.* [291] have proposed a method to obtain discrete equilibrium data from single-component breakthrough experiments and include it into computer simulations so that a continuous functional form is no longer required. In this method, adsorbed phase concentration ($q$) is defined for a set of discrete values of the fluid phase concentration ($c$) within the range of the feed concentration. Within the dynamic breakthrough model, the adsorbed phase concentration of any point between two adjacent discrete points is then calculated by interpolation [291]. To extract discrete equilibrium data, a single-component breakthrough experiments are performed for different fluid phase concentrations. The actual values of the corresponding solid loadings are found by solving an optimization problem and reducing the error between the experimental breakthrough results and predictions of the process model [291]. The above computational technique has been further developed by other research groups [292, 293]. For example, Rajendran *et al.* [293] have extended this method by incorporating discrete single-component equilibrium data into the Ideal Adsorbed Solution Theory (IAST) [294] in order to describe binary equilibrium data.

The final category of parameters that are required for process modelling include properties of the feed such as its temperature and composition which are typically specified by the design problem at hand (*e.g.* post-combustion carbon capture). **Table 6** summarizes the full set of properties needed to set up a PSA/VSA process simulation along with their sources according to the categories provided above.

**Table 6.** Complete set of input parameters for process simulation

| **Column properties** | | |
|---|---|---|
| **Parameter** | **Symbol** | **Source** |
| Wall (ambient) temperature (K) | $T_w$ | Design specification |
| Column length (m) | $L_c$ | Design specification |
| Inner column radius (m) | $R_{c,i}$ | Design specification |
| Outer column radius (m) | $R_{c,o}$ | Design specification |
| Column void fraction | $\varepsilon$ | Heuristic values |
| Specific heat capacity of column wall (J/kg·K) | $\hat{c}_{P,w}$ | Literature data |
| Density of column wall (kg/m$^3$) | $\rho_w$ | Literature data |
| Wall heat transfer co-efficient (J/m$^2$·K·s) | $h_w$ | Literature data |
| Outside heat transfer co-coefficient (J/m$^2$.K.s) | $U$ | Heat-transfer engineering correlations, available from the literature |
| **Pellet properties** | | |
| Pellet porosity | $\varepsilon_p$ | Mercury porosimetry experiment |
| Pellet radius (m) | $R_p$ | Geometric measurement using conventional callipers |
| Pellet tortuosity (τ) | $\tau_p$ | Often heuristic values are used. However, dynamic tortuosity can be obtained from the measurement of the effective pellet diffusivity at different temperatures and pressures [295]. |
| Pellet heat transfer coefficient (J/m$^2$·K·s) | $h_p$ | Analytical correlations [278] |
| Average macropore diameter (m) | $r_{pore}$ | Mercury porosimetry experiments |
| | | |
| Molecular diffusivity (m$^2$/s) | $D^m$ | Predicted from kinetic theory of gases or measured in bulk gas mixtures.<br>Eq. (44) corresponds to the Chapman-Enskog theory. |
| Knudsen diffusivity (m$^2$/s) | $D^K$ | Predicted from the standard kinetic theories, *e.g.* Eq. (45) |
| Surface diffusivity (m$^2$/s) | $D^S$ | Measured experimentally, several methods exist [296], Eq. (46) |
| Viscous diffusivity (m$^2$/s) | $D^V$ | Eq. (47) |
| **Crystal properties** | | |
| Crystal density (kg/m$^3$) | $\rho_{cr}$ | Experimental crystallographic data |
| Microporosity (-) | $\varepsilon_{cr}$ | Helium pycnometry experiment on powder;, interpretation of nitrogen and argon adsorption isotherms at 77 K and 87 K, respectively; or $CO_2$ adsorption isotherm at 273 K |
| Crystal radius (m) | $r_p$ | Optical microscopy |
| Specific heat capacity (J/kg·K) | $\hat{c}_{P,cr}$ | Experimental calorimetry, empirical group contribution methods, ab initio simulation methods |
| Micropore diffusivity (m$^2$/s) | $D^\mu$ | Molecular dynamic simulation, NMR experiments, other experimental techniques [297] |
| Activation energy (kJ/mol) | $E_a$ | Molecular dynamics, NMR experiments, other experimental techniques [297] |
| **Properties of competitive adsorption isotherms (e.g. in case of the DSL model)** | | |
| Saturation capacity for site 1 of the DSL model | $q_{s1}$ (mol/m$^3$) | DSL fit to experimental adsorption or GCMC simulation data |
| Pre-exponential constant for site 1 of the DSL model | $b_{01}$ (bar$^{-1}$) | DSL fit to experimental adsorption or GCMC simulation data |
| Enthalpy of adsorption on site 1 for site 1 of the DSL model | $-\Delta H_1$ (J/mol) | DSL fit to experimental adsorption or GCMC simulation data |
| Saturation capacity for site 2 of the DSL model | $q_{s2}$ (mol/m$^3$) | DSL fit to experimental adsorption or GCMC simulation data |
| Pre-exponential constant for site 2 of the DSL model | $b_{02}$ (bar$^{-1}$) ( | DSL fit to experimental adsorption or GCMC simulation data |
| Enthalpy of adsorption on site 2 for site 1 of the DSL model | $-\Delta H_2$ (J/mol) | DSL fit to experimental adsorption or GCMC simulation data |
| **Fluid properties** | | |
| Viscosity (Pa.s) | $\mu$ | Literature data |
| Fluid thermal conductivity (J/m·K·s) | $\lambda_f^L$ | Literature data |
| Axial dispersion coefficient (m$^2$/s) | $D_i^L$ | Eq.(40) |
| **Feed Properties** | | |
| Feed composition (-) | $c_{F,i}, x_{F,i}$ | Design specifications |
| Feed temperature (K) | $T_F$ | Design specifications |

From the table above it is clear that setting up a model requires a combination of properties that can be measured experimentally (*e.g.* adsorption isotherms, properties of the pellet), or for which well-established thermophysical models exist (*e.g.* molecular diffusivity, Knudsen diffusivity). Some other properties have well-known literature values (*e.g.* heat conductivity of steel). In general, the large number of parameters required to set up the model in combination with the large number of potential models (hierarchies used as described in **Figure *15***) often makes comparison and reproduction of data between various research groups a challenging task and we can only advocate detailed disclosure of the sources, parameters and algorithms used for every simulation.

A separate challenge is the implementation of the complete *in silico* workflows. As can be seen from **Table 6**, only a limited set of properties can be obtained from molecular simulations (*e.g.* equilibrium adsorption data, micropore diffusivity, heat capacity and thermal conductivity of adsorbent crystals). For other properties, particularly those pertaining the morphology of the pellets, we can either adopt some conventional estimates based on what is known from previous experimental measurements or we can use these parameters as optimization variables within a specified range of known values. The former approach is however prone to inaccuracy and inconsistency, considering pellet morphology is not standardized and various manufacturers produce adsorbent materials with different characteristics (*e.g.* different size and porosity, various types of binder). Optimization of these parameters however have proved to be a more promising approach in some cases. In a recent study, Farmahini *et al.* [101] have demonstrated that size and porosity of pellets can be used as decision variables during process optimization not only to achieve maximum theoretical performance of adsorbent materials, but also for consistent comparison of different screening studies. To fully understand the impact of these two approaches, we advocate for sensitivity and error propagation analyses of the multiscale materials screening workflows for the parameters that cannot be calculated from molecular simulations or any other theoretical methods. The results of such analyses will show whether the use of estimated reference values for these properties has a significant impact on the overall predictions of the multiscale workflows.

### 6.3.3. PSA/VSA Process and Cycle Configuration

In Section 3, we briefly introduced the PSA/VSA process. In the previous sections, we also covered the mass, energy and momentum balance equations, governing the behaviour of the adsorption column and data needed to set up the process model. Here, we consider in more details a particular 4-step VSA cycle and essential elements of cycle configuration. For the sake of concreteness and consistency we continue with the same case study of the post-combustion feed, comprised of carbon dioxide (15%) and nitrogen (85%).

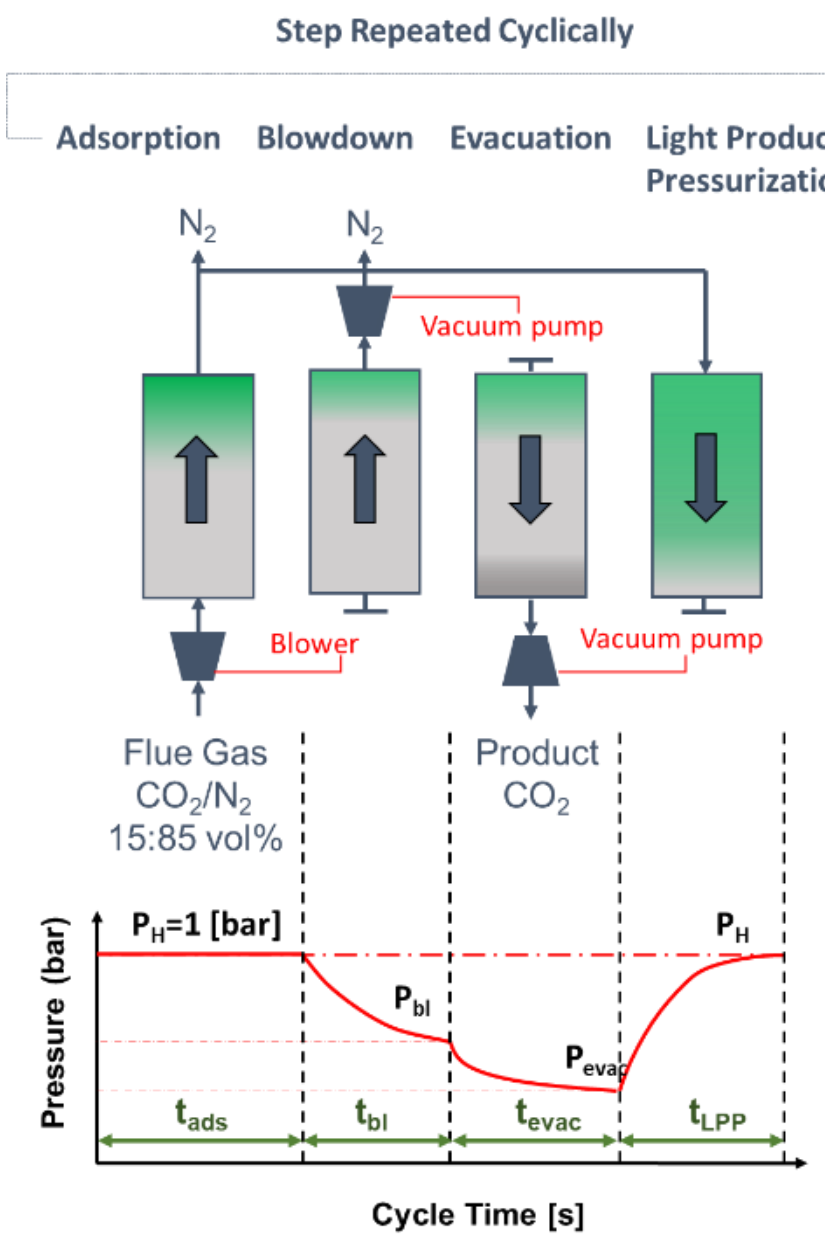

**Figure 16.** Schematic depiction of a 4-step process with Light Product Pressurization (LPP). From left to right the column goes through adsorption, co-current blowdown, counter-current evacuation and LPP steps. The bottom panel shows the pressure profiles during the steps and their duration within the cycle time. The green colour within the column unit schematically indicates distribution of nitrogen at the end of each step. The figure has been adapted from Burns *et al.* [103].

**Figure *16*** shows a 4-step cycle which first appeared in the work of Ko *et al*. [298], who referred to this as the fractionated vacuum swing adsorption cycle. The first step of the process is the adsorption step. The feed is introduced to the column at the pressure close to atmospheric. This is followed by a co-current blowdown step: the column is closed at the feed end, and the pressure is reduced to remove excess nitrogen present in the column in order to increase the purity of the product. Next is the counter-current evacuation step, where the pressure is reduced further, causing desorption of carbon dioxide. The product of this step is a carbon dioxide-rich stream. Finally, this step must be followed by bringing the pressure of the column back to the adsorption pressure, which is done in the repressurization step. In principle, repressurization can be done using the feed stream. However, previous studies demonstrated that counter-current repressurization with the light product stream, as schematically depicted in **Figure *16*** leads to a much better process performance [98, 299]. This effect stems from the counter-current repressurization helping to concentrate carbon dioxide closer to the feed end of the column as it will increase purity and recovery of carbon dioxide during the evacuation step later in the sequence. As can be seen from **Figure *16***, each step in this process is therefore associated with a particular pressure profile and duration. These parameters, namely time of the adsorption step $t_{ads}$, time of the desorption step $t_{bd}$, and time of the evacuation step $t_{evac}$, the blowdown pressure $P_{bd}$, and the evacuation pressure $P_{evac}$, along with feed pressure $P_H$, and feed flow rate are called cycle variables for this particular process and their specific values define the cycle configuration. The cycle variables are typically constrained by a number of considerations. For example, $P_H$ cannot be set too high otherwise the compression of the dilute gas makes the process not viable. $P_{evac}$ is another important example. In practical systems 0.2 - 0.3 bar would be a reasonable value for this parameter, but often much lower values are used in process simulations in order to achieve the required purity/recovery targets.

Adsorption processes operate at the Cyclic Steady State (CSS), and equations described in 6.3.1 can be solved iteratively to arrive to the CSS. Alternatively, time can be discretised and the CSS in this case is calculated directly, but this approach resulting in a large set of nonlinear equations is not necessarily faster [300]. Although the actual industrial process features several adsorption units in a different stage of the cycle at any given moment, as they all go through the same steps, it is possible to consider modelling of this process with only one unit. This so-called *unibed* approach has been originally described by Kumar *et al*. [301]. This in general allows to study multicolumn process at a similar numerical cost as a simple Skarstrom cycle. The numerical procedure starts with some initial conditions and solution of the balance equations in the adsorption step. This produces concentration profiles for each component of the system in the adsorbed phase and in the gas phase. These concentration profiles and the composition of the product stream serve as the initial conditions for the next step in the adsorption cycle (in this case, the blowdown step), and so on. The iterative process continues until the numerical CSS is reached: this happens when the state variables start to depend only on the spatial position in the system and the time relative to the start of the cycle. One can employ several mathematical criteria to establish whether the solution has reached the CSS [302, 303]. We note, however, that this is not a simple problem especially for non-isothermal systems and (or) with one very strongly adsorbed component (for example water): in this case convergence may require thousands of cycles. Complexity of the PSA/VSA processes is very well illustrated by the concentration, temperature and pressure profiles that are calculated for each process cycle. **Figure *17*** depicts concentration profiles of $CO_2$ at the end of each step for the 4-step VSA-LPP cycle shown in **Figure *16***. The corresponding cycle variables are provided in **Table 7**.

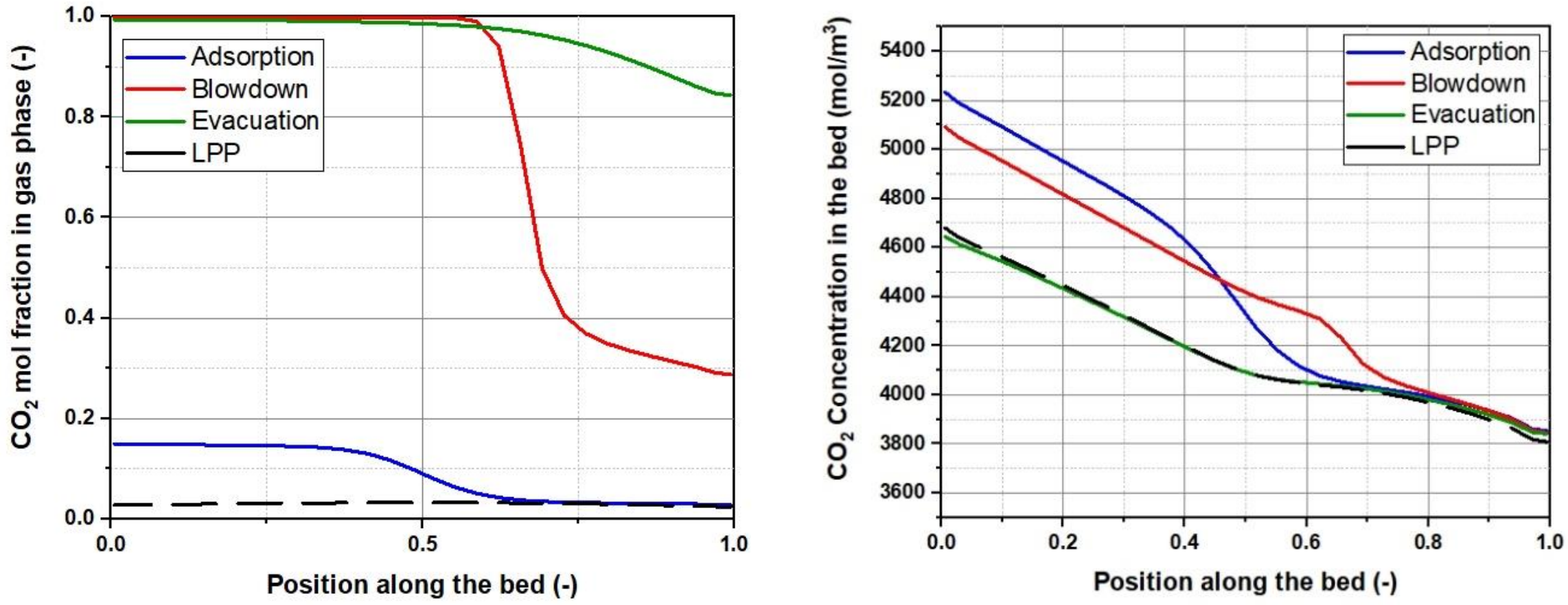

**Figure 17.** Examples of the concentration profiles for carbon dioxide in the column as a function of the dimensionless position along the bed. The profiles correspond to the end of adsorption, blowdown, evacuation and LPP steps in the gas phase (on the left) and in the adsorbed phase (on the right) at the cyclic steady state condition. The conditions and other parameters of the process are provided elsewhere [101].

**Table 7. Cycle variables used for simulation of Figure 17**

| Decision variable | Feed (mol/s) | $t_{ads}$ [s] | $t_{bd}$ [s] | $t_{evac}$ [s] | $P_{bd}$ [bar] | $P_{evac}$ [bar] |
|---|---|---|---|---|---|---|
| Value | 0.793 | 79.9 | 15.8 | 85.3 | 0.085 | 0.02 |

Correct interpretation of the bed profiles is vital for the analysis of the performance and efficiency of the PSA/VSA processes. The CSS implies that these profiles do not change anymore (within the numerical convergence criteria) as we continue with the numerical iterations and they will remain looking like this at the end of their respective steps.

Let us focus on these profiles in a step by step fashion. The LPP step prepares the bed for the next adsorption step and the LPP profile reflects the state of the column before the adsorption step is started. In the gas phase the concentration of carbon dioxide is very low. In the adsorbed phase, the concentration of carbon dioxide is also low, however some carbon dioxide remains in the adsorbed phase close to the feed end of the column (dimensionless bed position = 0). At the end of the adsorption step, the profile in the adsorbed phase reflects the higher amount of carbon dioxide now present in the solid. It starts with saturation values at the feed end slowly diminishing towards the light product end of the column (dimensionless bed position = 1). This reduction in saturation value is due to a non-uniform temperature distribution along the column: at the adsorption front the heat of adsorption increases the temperature which in turn reduces the saturation value; behind the adsorption front the temperature reduces slowly and, in turn, the saturation value increases towards the feed end. In the gas phase, the concentration of carbon dioxide is low at the end of the adsorption step. The main purpose of the blowdown step is to remove the remaining nitrogen in the gas phase. At the end of the blowdown, some of carbon dioxide is released from the porous material and it is concentrated at the feed end of the column in the gas phase. The available carbon dioxide at the end of the blowdown step will contribute to the heavy product, *i.e.* $CO_2$-rich stream, during the counter-current evacuation step. At the end of this step, the gas phase consists almost of pure $CO_2$, while in the adsorbent phase the concentration of $CO_2$ is lowered. It is important to note from the profiles discussed that the porous material is never fully regenerated – the amount of carbon dioxide it captures is represented by the difference between blue (adsorption) and the green (evacuation) lines, indicating that the working capacity of the material is only a fraction of the absolute capacity. From the same graph it is also clear that the adsorption step is stopped before the complete breakthrough occurs and the portion of the bed (between 0.75 and 1.00 in the dimensionless coordinates along the bed length) is never used.

To quantify performance of PSA/VSA processes, the following properties are normally evaluated:

1) Purity, $Pu_{CO_2}$: this property characterizes the composition of the final product. It is the ratio of the number of moles of carbon dioxide evacuated to the total number of moles of gas mixture evacuated during a single cycle:

$$Pu_{CO_2} = \frac{\text{Moles of CO}_2\text{ recovered in evacuation}}{\text{Total moles out in evacuation}} \tag{68}$$

2) Recovery, $Re_{CO_2}$: this property describes the amount of carbon dioxide recovered as part of the product stream compared to what is originally fed into the column.

$$Re_{CO_2} = \frac{\text{Moles of CO}_2 \text{ recovered in evacuation}}{\text{Total moles CO}_2 \text{ in the feed}} \quad (69)$$

The other two properties include energy penalty and productivity of the process, which have been already defined in **Table 1**, however we will explain them here again: :

3) Energy penalty: it is defined as the total amount of energy used for separation of one mole of $CO_2$ from the feed.

$$Specific\ energy\ penalty = \frac{\text{total energy used}}{\text{Moles of CO}_2 \text{ captured}} \quad (70)$$

4) Productivity: it is the amount of $CO_2$ captured in the product stream per unit volume of adsorbent per unit time.

$$Productivity = \frac{\text{Total moles of CO}_2 \text{ in product}}{(\text{Total volume of adsorbent}) \times (\text{Cycle time})} \quad (71)$$

Here, it is also instructive to reflect on the nature of the energy used in the process. In the PSA/VSA cycle this work is associated either with compression or pulling vacuum. In the 4-step process considered here, the most significant energy penalty comes from pulling vacuum during the evacuation step, however it may shift to other steps in more complex processes [304].

The complexity of this picture, its dynamic nature and the fact that it depends on a number of parameters, including the configuration variables of the cycle, explains why it is difficult to find some simplified metric which would comprehensively capture the efficiency of PSA/VSA separation processes.

### 6.3.4. Process Performance and Optimization

In the previous section, we considered a single cycle configuration with specific values ascribed $t_{ads}$, $t_{bd}$, $t_{evac}$, $P_{bd}$, $P_{ext}$, and flow rate of the feed, *F*. However, in reality, the resulting process may or may not be able to meet the design objectives to recover more than 90% of $CO_2$ with at least 95% purity. It also may not operate optimally; hence incurring additional energy penalties. The objective of the optimization process is to adjust the values of the cycle parameters in such a way that the process can meet its design constraints, while operating at the highest possible productivity and minimum energy penalty. In the optimization language, the cycle parameters described above become decision variables, while mathematically the optimization problem can be formulated as follows:

$$\Theta_{min} = \min_{t_{ads}, t_{bd}, t_{ads}, P_{bd}, P_{evac}, \mathrm{F}} \Theta_i \qquad i = 1, 2; \quad (72)$$

$$\Theta_1 = \text{Energy}/100 \quad (73)$$

$$\Theta_2 = -\text{Productivity} \quad (74)$$

Subject to: $Re_{CO_2} \geq 90\%;\ Pu_{CO_2} \geq 95\%$

The optimization conditions above form an optimisation problem with two objective functions and two constraints. Here, it is important to realize that the two optimization targets, minimal energy penalty and high productivity, are in competition with each other. Indeed, higher productivity may be achieved using higher flow rates given the same amount of the active adsorbent material in the column. However, this approach may require faster cycles and lower evacuation pressures, which will lead to higher energy penalties. In contrast, lower energy penalty can be achieved with more moderate vacuum during the evacuation step but it will be achieved at a cost of processing lower flow rates in the system or having to resort to longer individual steps, leading to lower productivity. As a result, the actual solution to the optimization problem is not a single set of values of the cycle parameters, but multiple combinations of these parameters, each of them associated with a particular combination of purity, recovery, energy penalty and productivity values.

From the mathematical perspective, the problem above corresponds to a multi-objective optimization. In general, this is a challenging problem as the search for the solution takes place in a multidimensional space of the decision variables, which can form clusters of feasible solutions, separated by non-feasible regions. The study of Fiandaca *et al*. [305], showed that the objective function is non-smooth and non-convex, and also that the design space is non-convex. Several approaches have been proposed to deal with this problem over the years, with Ref [305] briefly reviewing available approaches up to 2009. However, in recent years the conventional practice became to invoke the evolutionary Genetic Algorithms (GA), because of their ability to achieve global convergence, and a large number of tools available to implement them. In particular a set of methods associated with the second and third generation of non-dominated sorting genetic algorithm (NSGA-II,III) has been a popular choice. It has been implemented in many commercial packages such as MATLAB and also available as a set of free libraries [306-308].

The initial step in the optimization problem is to identify a range of values within which each decision variable can change. A number of initial operating conditions (so called, *population* in GA terms) is selected from this range (either randomly or using more sophisticated approaches such as Latin Hypercube Sampling). For each combination of the decision variables, the PSA process is simulated as described in Section 6.3.3. Promising candidates are identified, and their features are combined (using mutations and crossover moves) to give a new generation of operating conditions. As the optimization process evolves from generation to generation, the cloud of points representing the cycle configurations on the Energy Penalty-Productivity progresses towards higher values of productivity and lower values of energy (subject to purity and recovery constraints) until this process effectively stops (further progress of the cloud is not visible within the convergence criteria). At this point, the optimization simulation has converged to its final set of solutions.

This process can be illustrated with two useful graphs commonly employed in the process simulation and optimization studies. The first plot has purity and recovery as *X* and *Y* axes. It identifies the proportion of cycle configurations that are able to meet the 95%-90% constraints for purity and recovery. The second plot shows the evolution of the cycles in Energy Penalty – Productivity coordinates. **Figure *18*** illustrates typical examples of these graphs.

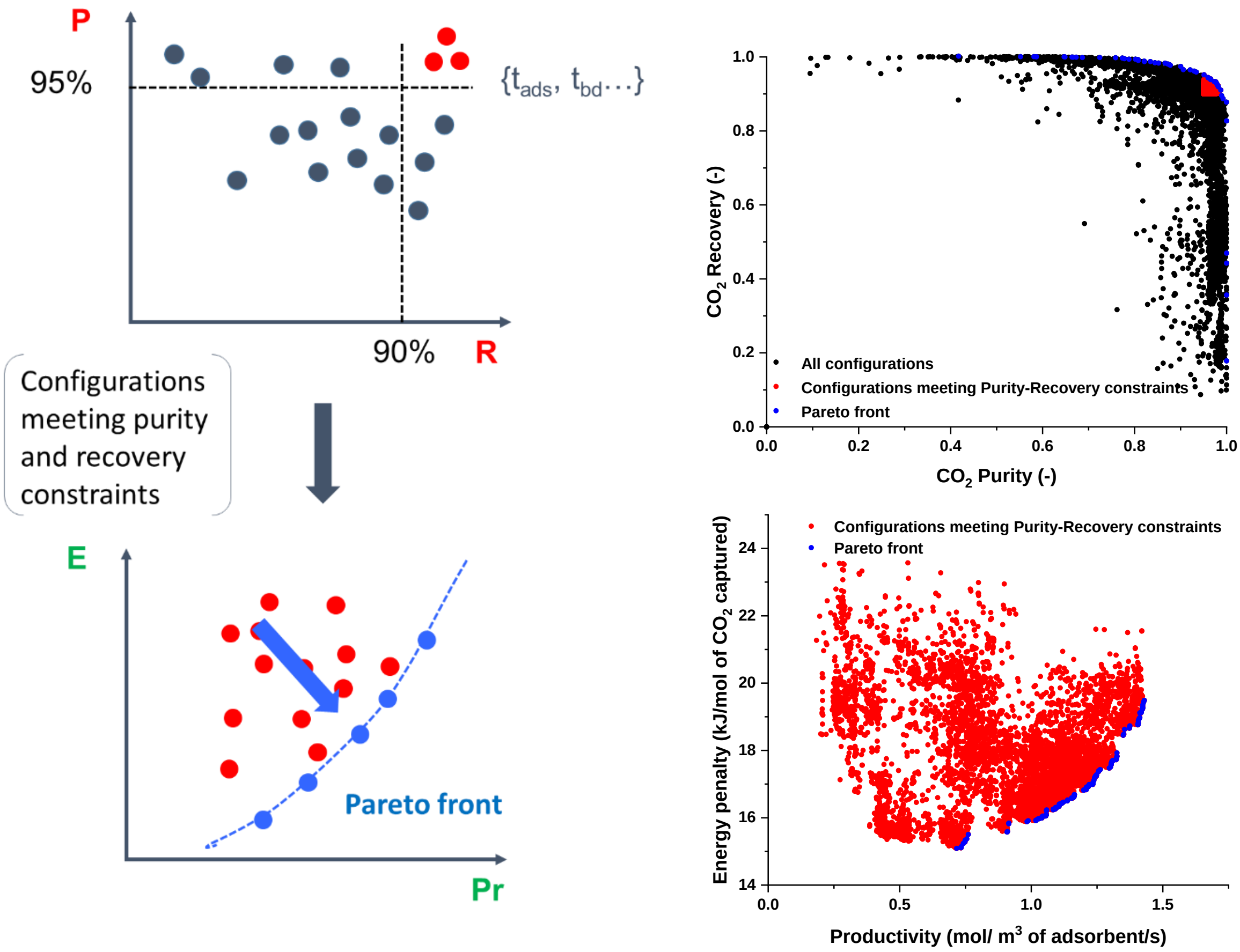

**Figure 18**. Process performance characterized in terms of Purity-Recovery coordinates (constraints, top graph) and Energy Penalty – Productivity coordinates (Pareto front, bottom graphs); graphs on the left is a schematic for the illustration; whereas graphs on the right correspond to a case studied in our recent publication [101].

The front edge of the clouds shown in the above figure are called the Pareto fronts. These are the set of cycle configurations that combine the highest purity-recovery and energy-productivity for a given process configuration subject to its pre-defined process constraints.

As already mentioned, this implies that for each material there is a number of possible operating conditions to choose from (points on the Pareto front). High productivity processes will incur higher energy cost, but lower footprint and capital cost of the plant. Low energy processes, on the other, hand will benefit from lower energy penalties, however, may incur larger capital costs due to larger required footprint of the plant.

Assessment of the performance of two materials then invariably becomes the comparison of their corresponding Pareto fronts. If two specific values are provided as the metric of the performance of the material (*e.g.* energy penalty value per unit of $CO_2$) it is important to specify to what conditions these values correspond to: to the lowest energy penalty on the Pareto front, or the highest productivity.

### 6.3.5. Emerging Numerical Techniques for Process Optimization

The process simulations we have covered so far are computationally expensive: a single process simulation for a given set of design variables takes minutes to complete. Process optimization to obtain a Pareto front as described in **Figure *14*** requires thousands and tens of thousands of simulations, leading to an overall cost of the process optimization exercise of $10^2$-$10^3$ CPU hours for a single material. Clearly, routine screening of tens, hundreds or thousands of materials at the process level is prohibitive.

This promoted the development of several strategies to reduce this cost. These strategies can be split into three main categories:

1. Reduce the pool of candidate materials by low cost, preliminary screening strategies
2. Reduce the computational complexity of the individual process simulations in the optimisation process:
    a. Accelerate the convergence to CSS
    b. Use a simpler model from the model hierarchy
    c. Replace the high-fidelity model with a surrogate model trained on the high-fidelity model
3. Reduce the computational effort of the optimisation process

The three approaches can be combined, but all have disadvantages and limitations, which can compromise the screening process so that the optimal material and cycle configuration is missed. Here we review studies with a focus on accelerating process optimization using strategies outlined above.

Strategies in the first category can use any of the previously published performance metrics to eliminate candidates from the candidate pool so that the expensive computational effort can be spent on the most promising candidates. As described in Section 4, simple performance metrics are not able to correctly and accurately rank materials for the complex and highly dynamic adsorption processes where the performance is given by a balance between the competing objectives of energy penalty and productivity as well as the competing constraints of purity and recovery. Thus, it is crucial to have very conservative exclusion criteria so that potentially promising candidates are not removed from the candidate pool. On the other hand, a number of these metrics can be computed very quickly so that the least promising candidates can be removed for a low computational cost.

Burns *et al.* [103] performed a detailed multi-objective process optimisation and ranking for a large range of materials for post-combustion capture. Afterwards, they trained Machine Learning (ML) classifiers to predict the objectives, *i.e.* purity, recovery, parasitic energy and productivity, based on 29 sorbent metrics such as working capacity, selectivity, and isotherm parameters. They showed that the $N_2$ adsorption behaviour is crucial for the correct classification of materials that meet the 95% purity 90% recovery constraints and achieved a prediction accuracy of 91% for this. However, the prediction of parasitic energy and productivity for materials that achieve the 95% purity 90% recovery constraints achieved only very low prediction accuracy. They concluded that full process simulations are required for accurate ranking of parasitic energy and productivity. An interesting approach is followed by Khurana and Farooq [67] who trained a classification neural network based on five equilibrium isotherm characteristics which cover the parameter space of the dual-site Langmuir isotherm. Their model can predict with 94% accuracy whether a material can meet the 95/90 purity/recovery constraints for post-combustion capture with the LPP-VSA cycle. In addition to the good accuracy, the false negatives, *i.e.* materials that can achieve the target but were wrongly classified, showed high energy penalties and low productivity. For the materials that met the 95/90 purity/recovery constraints, they developed a meta-model to predict the energy penalty and productivity and achieved $R^2$ values of around 0.9 for minimum energy penalty and maximum productivity.

The second category is split into three methods to reduce the computational complexity of the individual process simulations. First, instead of simulating cycle after cycle to reach CSS, so-called successive substitution, several studies have explored methods to accelerate the convergence to CSS. For example, Smith and Westerberg [309] and Ding and LeVan [310] used Newton and quasi-Newton steps to reduce the cyclic deviation. This method requires the calculation of the Jacobian and can achieve about an order of magnitude faster convergence. Alternatively, derivative-free extrapolation methods such as the epsilon extrapolation used by Friedrich *et al.* [311] can reduce the required number of cycles to CSS by a factor of 3. Pai *et al.* [312] used artificial neural networks to predict the bed profiles at CSS and use this to initialise the high-fidelity simulations. In their tests it reduces the average number of cycles which need to be simulated to reach CSS by a factor of 6.

Second, simpler but still physics-based models are used instead of the high-fidelity models. These simplified models should be fast to calculate while still capturing the main physics of the separation process. Balashankar *et al.,* [313] use a batch adsorber analogue model as a simplification for the full VSA model with spatial discretisation. The simplified model assumes that the system is isothermal, well-mixed and has no mass transfer resistance, but still captures part of the physics of the separation and can be solved in seconds. The authors compared the output from the simplified model with the detailed process optimisations and developed a classifier that achieved a Matthew correlation coefficient of 0.76 in the classification of materials that meet the 95% purity 90% recovery constraints. In addition, they calculated a linear regression for the energy penalty, which estimated the energy penalty with reasonably good accuracy, *i.e.* within 15% for 83% of the materials. However, Biegler *et al.* [314] evaluated the use of simplified models for process optimisation and concluded that it can lead to convergence failure and even to false optima.

Third, surrogate models are built based on the output of the high-fidelity models. These models are faster to evaluate and are usually embedded into optimisation methods. Agarwal *et al.* [24] used proper orthogonal decomposition (POD) to replace the detailed spatial discretisation with a reduced order model (ROM), leading to a system of Differential Algebraic Equations (DAE) of a significantly lower order. The ROM was trained on a number of bed profiles for different cycle conditions simulated to CSS. Because only the largest singular values are used for the ROM, the size of the discretised model is reduced by an order of magnitude. The ROM is accurate close to the training cycle conditions but loses accuracy further away from these points. This means that the ROM needs to be retrained if the optimisation moves away from the original training points.

In recent years, the focus has moved to directly using the optimisation objectives and constraints to build ROM instead of using ROM of the bed profiles. This approach replaces the process simulation (reduced order or high-fidelity) with fast-to-calculate surrogate models (also called ROM, meta models or emulators), which directly calculate the optimisation objectives based on the optimisation variables. These models are built from the input-output relations generated with the high-fidelity models and can be used with any black-box optimisation algorithm. This enables the interfacing with state-of-the-art multi-objective optimisation methods to handle the trade-off between competing objectives and constraints. We review the principles of these models and recent studies based on them below.

The process of a surrogate optimisation is shown in **Figure *19***. The process starts with an initial Design of Experiments (DoE) which should cover the entire design space. The high-fidelity model is used to simulate the responses for these initial designs. Then the optimisation loop starts by building a surrogate model based on these input-output relations. The optimisation method operates on this, fast-to-calculate, surrogate model to find promising design points. The choice of the next design points is a balance between exploring the design space and exploiting the best-predicted design or designs. The new design point is evaluated with the high-fidelity model, added to the input-output relations and a new iteration of the optimisation loop starts, *i.e.* we build a new surrogate model. The optimisation loop is stopped once a stopping criterion, which is often a computational budget, is fulfilled.

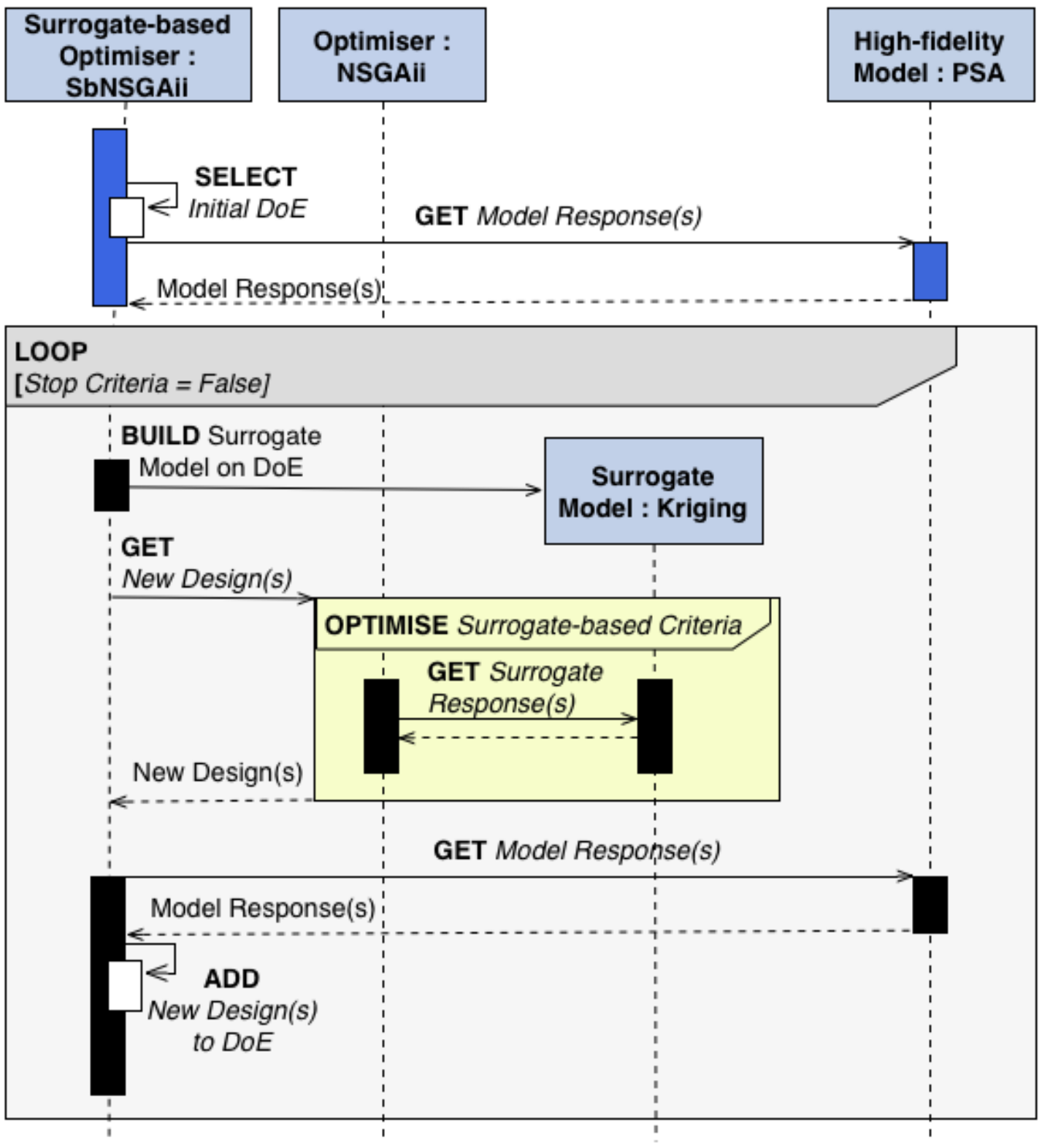

**Figure 19.** Sequence diagram for a surrogate-based optimiser showing the interplay between the optimiser, surrogate model and high-fidelity model.

Beck *et al.* [25, 284] used Kriging regression based surrogate models with the NSGA-II optimiser to simultaneously optimise the $CO_2$ purity and recovery for post-combustion capture. The Kriging regression models the input-output relation as a Gaussian process and gives the best linear unbiased prediction. In addition, it also provides confidence bands for the prediction, which can be used to explore the design space. They achieved a five-times reduction in computational effort and also investigated the specific energy penalty [284].

The rapid development of machine learning methods and, in particular, artificial neural networks (ANN) is mirrored in the application of machine learning to adsorption process optimisations. Sant Anna *et al.* [315] developed a three layer neural network (input layer, one hidden layer, and output layer) surrogate model for the separation of nitrogen and methane. They trained the neural network on around 500 training samples and performed a multi-objective optimisation of $N_2$ purity and recovery on the trained network without further updating the surrogate model. Comparing the optimal values with the high-fidelity simulations showed that the maximum relative difference was 1.4% for $N_2$ purity and 4% for $N_2$ recovery.

Instead of directly approximating the optimisation objectives and constraints, Leperi *et al.* [316] used ANN based surrogate models to approximate each basic step, *e.g.* counter-current pressurisation and co-current feed, of the PSA cycles. This approach enabled them to build arbitrary PSA cycles and to include cycle synthesis in the optimisation procedure without the need to retrain the ANN for each process configuration. They built 12 surrogate models for each step: one ANN for the state variables at 10 locations along the column and one for each end of the column to predict the inflow/outflows during the step. The ANNs are trained with high-fidelity simulations for 300 Latin hypercube samples and used to predict the column profiles as well as purity and recovery for three process configurations and two adsorbents for post-combustion capture. The predictions were used in an optimisation loop to find the purity/recovery Pareto front. The solutions on the Pareto front were used to test the accuracy of the ANN prediction and to retrain the ANN in case the prediction is too far from the high-fidelity simulation. After retraining, the relative errors for both purity and recovery were below 1.5% for all cases.

Subraveti *et al.* [317] used a surrogate model based on ANN in the multi-objective optimisation of purity and recovery of pre-combustion $CO_2$ separation and achieved a 10 fold reduction in computational effort. For the first five generations of the multi-objective NSGA-II algorithm they used the high-fidelity model. This generated training data for the ANN, which would already be biased towards the optimal region of the design space and should improve the prediction accuracy in the optimal region. The ANN with one hidden layer with 10 neurons was trained on the generated input-output data. The remaining 45 generations of the optimisation were performed on the ANN. The Pareto front was close to the one generated with the high-fidelity model, but had a relative error around 1% in both objectives. In a subsequent paper, the group compared a range of machine learning methods and showed that Gaussian process regression achieves an $R^2$ value above 0.98 for purity, recovery, energy penalty and productivity with a training set of 400 randomly sampled high-fidelity simulations [312]. Their optimisation on this surrogate model (without further refinement) was within 3% of the high-fidelity simulation for purity and recovery as well as for energy penalty and productivity. However, the latter was for a reduced 95/80 purity/recovery constraints.

Pai *et al.* [106] developed a material-agnostic surrogate model called MAPLE that fully emulates operation of the 4 steps VSA-LPP cycle at the cyclic steady state. The framework is based on a dense feedforward neural network trained with a Bayesian regularization technique. The framework accepts the adsorbent properties, the Langmuir adsorption isotherm parameters, and operating conditions as input. It predicts key performance indicators of the process including $CO_2$ purity and recovery in addition to productivity and overall energy consumption of the process as output. The model was trained by a set of data generated using detailed process modelling. In order to reduce computational time of the multi-objective optimization, MAPLE was used to calculate the CSS performance indicators and feed them back to the optimizer. The fully trained model, predicts each performance indicators with less than 2% error compared to the detailed process modelling. The computational time required for simulation and optimization of the process was also reduced from 1500 core-hours per adsorbent to ≤1 core-min for each adsorbent which shows a significant improvement for screening of large databases of porous materials [106].

Strategies in the third category include a range of methods to reduce the computational effort of the optimisation method itself, *i.e.* reduce the number of required iterations to reach an optimal value or Pareto front. The first strategy should be the reduction of the search space. This includes the removal of parameters, which have no or only a small impact on the performance and the reduction of design space, *i.e.* reduce the evacuation pressure range. For example, Balashankar *et al.* [313] removed the blowdown and evacuation times from the list of optimisation variables. This was acceptable in their optimisation because these variables have very limited impact on the purity and recovery. However, they have a large effect on productivity and energy penalty.

Yancy-Caballero *et al.* [105] performed a hierarchical, multi-objective optimisation with NSGA-II. They first optimised purity and recovery to screen for materials that achieve the 95/90 purity/recovery constraints and then optimised the promising materials for energy penalty and productivity. The energy penalty and productivity optimisation was seeded with the results from the initial optimisation and was performed in two steps: the first step used a low spatial resolution, which reduces the computational complexity, and the second step used a high spatial resolution and was pre-seeded with the low resolution results.

Finally, Ding *et al. [318]* and Jiang *et al.* [319] presented a strategy which combines the reduction of the computational complexity of individual simulations with a reduction of the computational complexity of the optimisation. They included the CSS condition as a constraint in the constrained single-objective optimisation problem so that both the objective and approach to CSS were optimised simultaneously. This approach, called the simultaneous tailored approach by Biegler and co-workers [17], removes the expensive calculation of CSS for each iteration and has reduced the computational time by a factor of 10 for single-objective optimisations of air separation VSA cycles [319].

### 6.3.6. Available Tools and Software for Process Modelling and Optimization

The objective of this section is to introduce the reader to several software packages and libraries that are available for PSA/VSA process simulations. Broadly, these can be divided into three categories: codes developed by the academic groups, codes developed within various companies and commercial software packages, with built-in adsorption process simulators. From this classification we can also identify the most significant challenge in a consistent description of these tools: they are not open source software (with one exception discussed below) and we do not have direct access to the organization, functionality, implemented models or capabilities of these codes to make the comparison consistent. Hence, from the onset we admit that this section is likely to be incomplete, however our main objective is to provide the reader with an overview of the options available for PSA/VSA simulations.

***Commercial software:***

**gPROMS:** The process builder developed by Process Systems Enterprise (PSE) has an adsorption process library which has been used for simulation of pressure and temperature swing adsorption processes [320]. In the adsorption process library, it is possible to use the dispersed plug flow or the plug flow model. The adsorption isotherms of Langmuir, dual-site Langmuir and virial isotherms can be used in gPROMS. Here, the flow sheet can be built by joining individual units such as valves, headers mass flow controllers, sources and sinks and adsorption columns. The adsorption process model is a system of partial differential equations that are discretized in the spatial domain using either finite difference (backward, forward and central), finite element, finite volume or orthogonal collocation with finite element. The flow controllers supply constant amount of gas, while the sources and sinks are used to specify initial and final operating conditions. gPROMS also has the facility to account for column headers to distribute flow and these are modelled as continuous stirred tank reactors. Building a flow sheet enables one to schedule various steps operating in multiple columns. In principle, within gPROMS it is also possible to perform optimization and scheduling of VSA process using in house libraries [321, 322]. PSA processes have been optimized in gPROMS for the maximization of $CO_2$ product purity and recovery, with the number of beds, process configuration, feed pressure, particle diameter, length to diameter ratio and feed flowrate as the decision variables by Nikolaidis *et al*. [323]. The same approach was used in an earlier study by Nikolic *et al*. [324] for $H_2$ recovery from steam methane reformer off-gas. However, it should be noted that these studies have not reported any Pareto fronts.

**Aspen Adsorption:** It is a flow sheet simulator that can design, simulate and optimize adsorption processes [325, 326]. Few studies exist in literature that have simulated PSA and dual-reflux PSA processes for $CO_2$ capture using this program [327-330]. In Aspen Adsorption, it is possible to simulate multi-bed PSA processes with isothermal or non-isothermal model and use non-ideal gas equations of states. In most publications with Aspen Adsorption, a Langmuir model has been used. Moreover, it is also possible to use finite difference or finite volume numerical schemes to solve the model equations. To the best of our knowledge no cycle optimization studies have been published using Aspen Adsorption software, although it is possible to couple Aspen products and MATLAB [331].

**ProSim DAC**: ProSim DAC is a dynamic simulation software from ProSim [332]. It is capable of simulating adsorption and desorption steps using TSA, PSA and VTSA processes. From the model hierarchy point of view, the process model can be isothermal and non-isothermal, it can further include a pressure drop, while transport in macropores is modelled using the LDF approach. Data for a wide variety of adsorbents is available (e.g., activated carbon and zeolites) and is accompanied by many different models for equilibrium data (adsorption isotherms) and mass transfer models. DAC is a relatively new addition to the ProSim family of process simulation tools so far employed predominantly in solvent recovery and in adsorption of volatile organic compounds. We are not aware of any academic article on carbon capture simulation and optimization that have used ProSim DAC.

***Academic codes:***

Several research groups have been developing codes for adsorption process modelling starting from the 1980ties, including SAXS (Swing Adsorption X=Pressure, Temperature Software) from Da Silva and Rodrigues, Dynamic Adsorption Process Simulator (DAPS) by Ebner and Ritter, PSA SW from Mazzotti and co-workers, MINSA Webley and co-workers, and CySim by Brandani and co-workers. The key challenge in the discussion of the codes used by the academic groups is similar to the issues associated with the industrial software: the codes are usually not open source, and full details of the algorithms employed, implementation and capabilities are not readily available. Below we briefly review the information available on MINSA code, our own simulator CySim and the recently published open-source code by Yancy-Caballero *et al.*

**MINSA** (Monash Integrated Numerical Simulator for Adsorption): MINSA is a generalized cycle simulator which has been developed by Webley and co-workers for PSA simulations using the DVODE integration scheme of Brown, Bryne and Hindmarsh [333] written in FORTRAN [334-337]. This simulation package solves mass and energy balance equations that have been discretized by the finite volume method [336, 337]. The software has been used extensively for various adsorption processes and verified against experimental data over the past two decades [335, 338-340]

**CySim (Cycle Simulator):** CySim is a modular computer program for simulation of adsorption processes which has been developed by Brandani and co-workers [311, 341] in University of Edinburgh. CySim can be used to simulate breakthrough curves, ZLC experiments [341], dual piston PSA [342] and other PSA processes. The user defined structure is translated into a system of differential algebraic equations, which are solved with the SUNDIALS library. This can be interfaced with either MATLAB or Python's genetic algorithm packages such as *gamultiobj* [343, 344], *inspyred* [345] and *Platypus* [306] to perform process optimization. Recently Farmahini *et al.* have used CySim to simulate and optimize the 4-step VSA process with LPP for post-combustion carbon capture [68, 101]. CySim is regularly updated with new models and applications, for example for monolithic adsorbents to include inlet and flow maldistributions [303, 346].

**Code by Yancy-Caballero *et al.*:** Recently, Yancy-Caballero and co-workers developed a MatLab code to simulate PSA/VSA processes [105]. In particular, the code uses the finite volume method with the weighted essentially nonoscillatory (WENO) scheme to discretize the PSA model; and the ode15s solver within MATLAB to solve the resulting ODEs. NSGA-II algorithms within MATLAB is employed for process optimization. In the most recent study, this code has been applied for performance ranking of several MOF materials in post-combustion carbon capture processes, using Skarstrom, a fractionated vacuum swing adsorption (FVSA), and a five-step cycles. A notable feature of the code is that it is open source and is publically available from the github depository of the academic group which has developed the code. **Table 8** provides a list of process simulation software that have been discussed in this section.

**Table 8.** List of academic and commercial software for PSA/VSA simulation

| Software | Reference | Website |
|---|---|---|
| **Commercial software** | | |
| gPROMS | PSE [320] | https://www.psenterprise.com/products/gproms |
| Aspen Adsorption | AspenTech [325] | https://www.aspentech.com/en/products/pages/aspen-adsorption |
| ProSim DAC | ProSim [332] | https://www.prosim.net/en/product/prosim-dac-dynamic-adsorption-column-simulation/ |
| **Academic software** | | |
| MINSA | Todd *et al.* [334, 335] | http://users.monash.edu.au/~webley/minsa.htm |
| CySim | Friedrich *et al.* [311] | https://www.carboncapture.eng.ed.ac.uk/lab/cysim |
| Code by Yancy-Caballero *et al.* | Yancy-Caballero *et al.* [105] | https://github.com/PEESEgroup/PSA |

## 7. Carbon Capture with Advanced Process Configurations

The main objective of this section is to introduce the reader to more complex PSA/VSA process configurations and review recent studies on application of process modelling to assess the viability of PSA/VSA technologies for carbon.

In the previous section, we used the 4-step VSA-LPP process to introduce several essential concepts and fundamentals of the PSA/VSA process and optimization. One of the issues associated with this specific process is that it can meet the required purity/recovery constraints only by going to very low evacuation pressures (*e.g.* 0.01 bar). Although from the Pareto front analysis this process is very competitive compared to other alternatives, in practice it is not viable, as the standard industrial pumps do not typically go below the range of 0.13-0.2 bar [347]. This necessitates a search for more complex process configurations. One option is to consider a two-stage process [286]. In this case, the first stage focuses on maximizing the recovery, while the second smaller polishing unit would aim to achieve the required purity. Indeed, Abanades *et al.,* summarized recent studies of 2-stage PSA processes [286]. According to their summary, it is clear that most process simulations arrive at VSA configurations that require approximately 0.5 – 0.75 MJ/kg $CO_2$, and that they can operate at evacuation pressures between 0.05 - 0.1 bar, which is more comparable to the industrial standards.

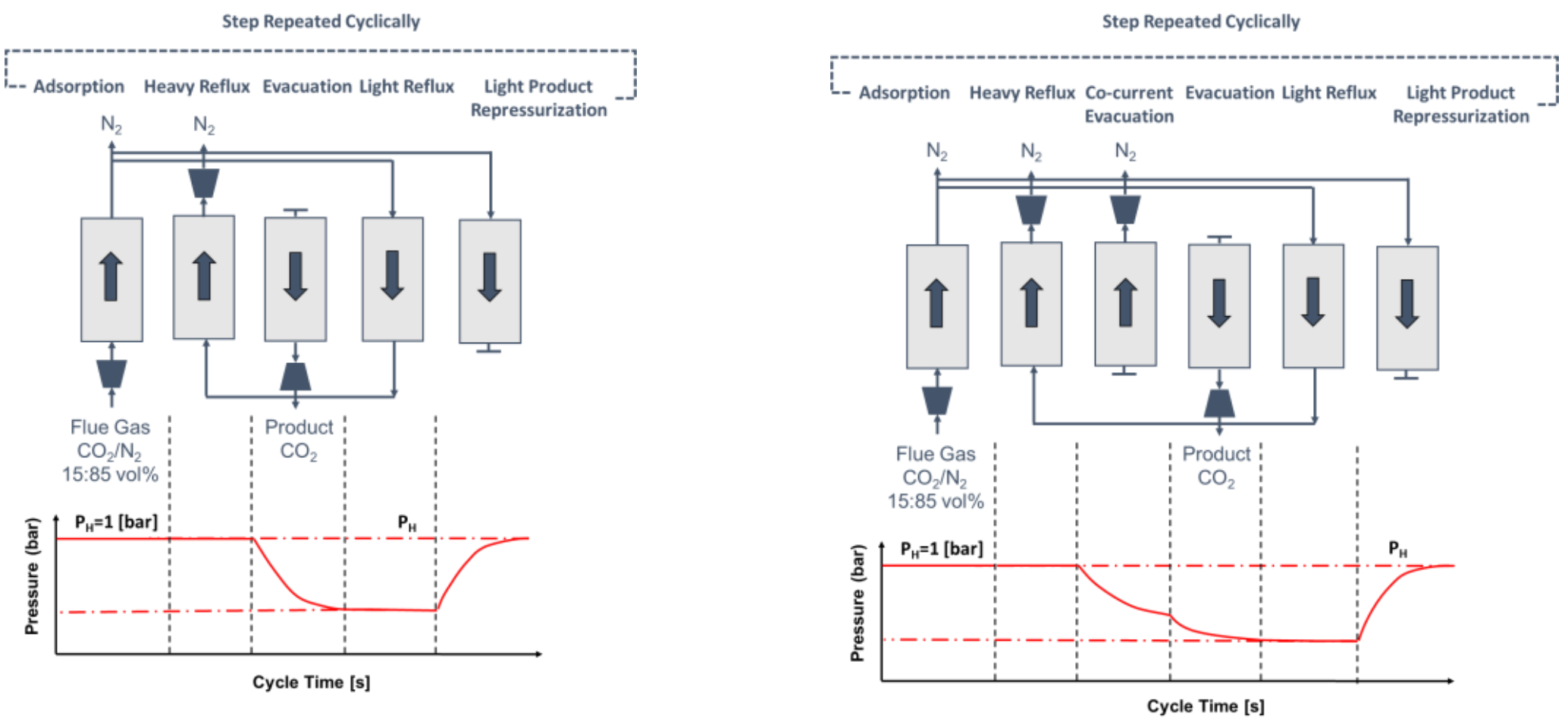

**Figure 20.** Examples of more complex process configurations: 5-bed 5-step cycle with heavy reflux from the counter current depressurization on the left; 6-bed, 6-step cycle with feed, recovery step using the outlet stream of heavy reflux, heavy reflux, heavy reflux with light reflux product, counter current depressurization, light reflux and light product pressurization, on the right [348].

Alternatively, we can consider more complex multi-bed multi-step configurations. Below we review several studies that explore more complex process configurations in the context of post-combustion carbon capture. In particular, Reynolds *et al.* [348, 349] studied the capture of $CO_2$ from a flue gas mixture containing 15% $CO_2$, 10% $H_2O$ and rest $N_2$ using potassium-hydrotalcite as the adsorbent. They had studied 9 different cycles with heavy and light reflux steps in 4-bed, 5-bed and 6-bed configurations. Two examples of such advanced PSA/VSA processes are shown in **Figure *20***. A parametric study was then carried out and the best performing cycle was the 5-bed 5-step cycle with heavy reflux from the counter current depressurization (shown in **Figure *20***). The purity and recovery values were 98.7% with a productivity of 0.11 mol/m$^3$·s. The next best cycle was cycle shown in **Figure *20*** on the right, which, although it showed a much lower recovery of 71%, had a high throughput of 1.11 mol/m$^3$·s. It should be noted that this was a parametric study which did not show the optimum performance of these cycles, in other words, detailed process optimization was not carried out to identify conditions corresponding to maximum productivity while meeting purity and recovery targets.

Zhang and Webley [338] constructed a pyramidal hierarchy of cycles. The pyramid consisted of cycles ranging from a simple 2-step cycle to complex cycles which included heavy reflux, light reflux, pressure equalization, *etc*. They carried out a parametric study on the effect of feed step duration, pressure equalization, rinse, evacuation and purge steps on the purity and recovery values. Although, their model was a simple one, nevertheless it provided some useful insights on the performance. In particular, they had studied the effect of feed time, light reflux, pressure equalization and heavy reflux steps on the purity, recovery and specific energy consumption. One of the main conclusions was that the addition of heavy reflux improved the purity from 85.7 % to 95.2% in their experiments.

A two-stage pressure swing adsorption process was studied with activated carbon by Shen *et al.* [350]. The first stage employed a Skarstrom cycle comprising of the following steps: pressurization with feed, adsorption, counter-current evacuation and light product purge. In the second stage a 5-step cycle comprising of pressurization with feed, adsorption, depressurizing pressure equalization, counter-current evacuation and pressurizing pressure equalization steps, was used. The first stage used a feed under ambient pressure while the second stage required a compression up to 3.5 bar. With a vacuum pressure of 0.1 bar, the two stage process produced a 95% pure $CO_2$ product with 74.4% $CO_2$ recovery. The specific energy and productivity values were 0.72 MJ/kg and 0.23 mol/$m^3$ ads/s. Deepening the vacuum to 0.05 and 0.03 bar improved the purity to 96.3 and 96.6% respectively, while the recovery increased to 80.7 and 82.9%, respectively. This also improved the productivity to 0.25 and 0.26 mol/$m^3$ ads/s. The increase in energy consumption was significant and the values were 0.83 and 0.9 MJ/kg $CO_2$ for pressure values of 0.05 bar and 0.03 bar.

Through a combination of experiments and simulations Wang *et al*. [351], studied $CO_2$ capture from a dry flue gas containing 15-17% $CO_2$. They used a two-stage process with the 1st stage containing 3 columns packed with Zeolite 13X and the second stage containing 2 columns packed with activated carbon. In the first stage, the cycle chosen was an 8-step cycle comprising of pressurization with feed, adsorption, co-current evacuation, heavy reflux, depressurizing pressure equalization, counter-current evacuation, light reflux and pressure equalization. The $CO_2$ product was collected from the counter-current evacuation and the reflux step. For the second stage, a six-step cycle comprising of pressurization with feed, adsorption, rinse, depressurizing pressure equalization, light reflux and pressurizing pressure equalization was used. The vacuum pressures of the first and second stage were 0.08 bar and 0.2 bar respectively. In the 1st stage, $CO_2$ purity of 70% was achieved. This stream containing 70% $CO_2$ was then compressed and sent to the second stage and here the product contained over 95% $CO_2$. The overall recovery was over 90%. From the experiments and the simulations, the values of the energy consumption were found to be 2.44 and 0.76 MJ/kg $CO_2$ respectively. The large differences in the reported energy consumption values could have been a consequence of the low pressure used in the first stage which would have resulted in a lower vacuum pump efficiency as shown by Krishnamurthy *et al.*[98].

Haghpanah *et al.* [299] also performed detailed process optimization on 7-different cycles to identify the optimal configuration of post-combustion $CO_2$ capture using zeolite 13X as adsorbent, ranging from 4 to 6 steps. The genetic algorithm-based optimization was carried out in two steps: a) to maximise purity and recovery; b) minimize energy consumption and maximize productivity for cycles satisfying 90% purity and recovery constraints. The decision variables were the step durations, evacuation pressures and the feed flow rate. The adsorption step pressure was kept constant at 1 bar and the feed was a 15% $CO_2$ and 85% $N_2$ mixture at 298K. The optimization results from the 1st step showed that 4-cycles namely 4-step cycle with LPP, 5 step cycle with light reflux (LR) and LPP and the two 6-step cycles satisfied the 90% purity-recovery targets. The next step was to minimize energy and maximize productivity and, in this case, the 4-step cycle with LPP was the best performing cycle in terms of the energy consumption. The minimum energy consumption was 0.47 MJ/kg $CO_2$ for a productivity of 0.57 mol/$m^3\cdot$s and with an evacuation pressure of 0.03 bar. The 6-step cycle with LR and HR was the best in terms of the purity and recovery and in this cycle over 98% purity and recovery were achieved. However, this cycle had much higher energy consumption. The 4-step cycle with LPP was also able to achieve the 95% purity and 90% recovery targets and in this case the energy and the productivity values were 0.554 MJ/kg $CO_2$ and 0.44 mol/$m^3\cdot$s respectively. The 4-step cycle with LPP was shown to meet the 95% purity and 90% recovery targets through a pilot plant study by Krishnamurthy *et al.* [98].

In a later study, Haghpanah *et al.* [225] studied $CO_2$ capture using a carbon molecular sieve by the optimization of 1-stage and 2-stage VSA processes. The two-stage process basically is the 4-step cycle with LPP carried out twice with the product from the counter-current evacuation step serving as the feed for the adsorption step in the second stage. In the one stage process, 5-step cycles with heavy reflux and with feed and light product pressurization were studied. It was seen that the 2-stage VSA process was the best in terms of energy and productivity. However, the productivity was about 50% smaller than that of Zeolite 13X mentioned above as carbon molecular sieve is a kinetically selective adsorbent.

It should be noted that in both the studies by Haghpanah *et al.* [225, 299] the evacuation pressures were from 0.01 to 0.05 bar and this, as we have already discussed, may not be industrially achievable. In a recent study by Khurana and Farooq [304], it was shown that with a 6-step cycle it was possible to achieve evacuation pressures of 0.1 bar and above. The 6-step cycle is essentially the 5-bed 5-step cycle with heavy reflux from light reflux product of Reynolds *et al.*, and with the addition of a co-current evacuation step [349]. The authors have compared the performance of this cycle with that of the 4-step cycle with LPP and used two adsorbents namely Zeolite 13X and UTSA-16. Through detailed process optimization, it was seen that the 6-step cycle was able to achieve similar productivity values as the four-step cycle but at a much higher evacuation pressure of 0.1 bar.

Another cycle that has shown promise for producing both $CO_2$ and $N_2$ in high purities is the dual reflux pressure swing adsorption cycle. The concept of the Dual reflux PSA (DRPSA) process first appeared in the 90's in the work by Diagne *et al*. [352, 353], who had obtained 95% $CO_2$ purity and recovery from a stream containing 20% $CO_2$. The DRPSA contains two columns one operating at high pressure and the other at low pressure at a given instance. Feed can be introduced from the bottom or from an intermediate position and both at low and high pressures. The enriched gas from low pressure feed is then compressed and sent to the column at high pressure to perform the heavy reflux. The light product from this heavy reflux step can be used to recover the heavy component simultaneously during the feed step. After this step, the column roles reverse, and the same sequence of steps are carried out. Over the years variations of the dual reflux PSA cycle have been studied by a few authors [328-330, 354]. One among them is that of Li *et al.* [329], who had studied the $CO_2$ capture from a binary mixture containing 15% $CO_2$ and 85% $N_2$ at 2 bar and 20°C feed using silica gel adsorbent. They were able to achieve over 99% purity and recovery for $CO_2$ and $N_2$. In a follow up work Shen *et al.* [330], used the same cycle and the adsorbent and achieved over 95% purity and recovery with energy consumption of 2.5 MJ/Kg.

In **Table 9** below we complement the summary of Abanades *et al.* [286], with the summary of the recent studies of various process configurations for post-combustion capture. We note that although the table contains two TSA cycles, in reality the productivity of these cycles will be low compared to the PSA/VSA processes: this would lead to columns simply too large (or to a large number of them) to be competitive in the post-combustion capture from the coal-fired plants. They may however find application in the carbon capture from natural gas fired power plants, where the concentration of $CO_2$ in the flue gas is much lower.

**Table 9.** Summary of selected process configurations studies for post-combustion capture in chronological order from 1993 to 2020.

| Process | Adsorbent | $y_{CO2}$ (%) | $P_{high}/P_{low}$ (kPa) | Purity (%) | Recovery (%) | Minimum specific energy (MJ/kg) | Source | Reference |
|---|---|---|---|---|---|---|---|---|
| 4-step VSA | AC, CMS | 17 | 120/10 | 99.99 | 68.4 | N/A | Sim | Kikkinides *et al.*, [355] |
| 7-step PSA | 13X | 16, 26 | 110/6.7 | 99 | 70 | N/A | Sim | Chue *et al.*, [356] |
| Dual reflux PSA | 13X | 15 | 101.3/8.1 | 95 | 95 | N/A | Exp | Diagne *et al.*, [353] |
| 2 bed PTSA/PSA | 13X | 15 | NA/5-15 | 99 | 90 | 2.02 mix | Exp | Ishibashi *et al.*, [357] |
| VSA | 13X | 10 | 115/6.7 | 50–70 | 30–90 | 0.9–1.1 | Sim | Park *et al.*,[358] |
| 2 bed 2 stage PVSA | 13X | 10.5 | 1st NA/6.7, 2nd NA/13 | 99 | 80 | 2.3-2.8 | Exp | Cho *et al.*, {Cho, 2004 #4} |
| 4-step PVSA | 13X | 15 | 652/10–70 | 88.9 | 96.9 | 1.5 | Sim | Ko *et al.*,[298] |
| TSA | 5A | 10 | 423K | >94 | 75–85 | 6.12–6.46 th | Lab | Merel *et al.*,[359] |
| VSA | 13X | 12 | 100/3 | 95 | >70 | 0.54 | Sim | Xiao *et al.*,[337] |
| 6-step PVSA (3 beds) | 13X | 12 | 130/5 | 82 | 60–80 | 0.34-0.69 | Exp | Zhang *et al.*, [360] |
| 9-step PVSA (3 beds) | 13X | 12.1 | 130/5 | 90–95 | 60–70 | 0.51-0.86 | Exp | Zhang *et al.*,[360] |
| 3-step PVSA | Zeolite 13X | 12% $CO_2$, 3.4% $H_2O$ | 118/4 | 72.4 | 60 | N/A | Exp | Li *et al.*, [361] |
| PVSA | 13X | 12.6 | 120/5–6 | 90–95 | 60–70 | 0.52–0.86 | Lab | Zhang and Webley, [338] |
| 5-bed 5-step step PVSA | Hydrotalcite | 15% $CO_2$, 10% $H_2O$ | 139.7/11.6 | 96.7 | 71.1 | N/A | Sim | Reynolds *et al.*,[349] |
| 6-step PVSA | Zeolite 13X, F200 alumina | 13% $CO_2$, 5% $H_2O$ | 140/3 | 89.6 | 74.9 | 0.72 | Exp | Zhang *et al.*, [362] |
| 9- step PVSA | Zeolite 13X | 13% $CO_2$, | 140/3 | 98.9 | 78.7 | 0.57 | Exp | Zhang *et al.*, [362] |
| | | 5% $H_2O$ | | | | | | |
| 9- step PVSA | Zeolite 13X | 13% $CO_2$, 7% $H_2O$ | 140/3 | 98.9 | 82.7 | 0.65 | Exp | Zhang *et al.*, [362] |
| 9- step PVSA | Zeolite 13X | 15% $CO_2$, 5% $H_2O$ | 140/3 | 86.1 | 60 | 1.07 | Sim | Zhang *et al.*, [362] |
| 9- step PVSA | Zeolite 13X | 15% $CO_2$, 7% $H_2O$ | 140/3 | 90 | 62 | 0.89 | Sim | Zhang *et al.*, [362] |
| 2-bed 4-step | 13X | 15 | 276/21.4 | 90.74 | 85.94 | 2.3 | Sim | Agarwal *et al.*,[363] |
| 4-step VPSA | AC | 15 | 202.6/10 | 93.7 | 78.2 | N/A | Exp | Shen *et al.*, [364] |
| 2-stage PVSA | 5A | 15 | 150/10 | 96.1 | 91.1 | 0.65 | Sim | Liu *et al.*, [321], |
| TSA | 5A | 10 | 433 K | 95 | 81 | 3.23 th | Sim | Clausse *et al.*, [365] |
| VSA | 13X | 13 | 100/2 | 93.8 | 91.5 | 0.43 | Sim | Delgado *et al.*, [366] |
| Two stage VPSA | AC | 15 | 350/10 | 95.3 | 73.6 | 0.73 | Sim | Shen *et al.*,[350] |
| Two stage VPSA | AC | 15 | 350/5 | 96.4 | 80.4 | 0.83 | Sim | Shen *et al.*,[350] |
| VTSA | 13X | 15 | 101/363K , 3 kPa | 98.5 | 94.4 | N/A | Exp | Wang *et al.*, [367] |
| 2-stage PVSA | 13X APG | 15 | 150/10 | 96.5 | 93.4 | 0.53 | sim | Wang *et al.*, [368] |
| 2-stage PVSA | 13X APG | 15 | 150/6 | 96.6 | 97.9 | 0.59 | Sim | Wang *et al.*, [368] |
| VSA | 5A | 15 | 101.3/5.5 | 71–81 | 79–91 | 2.64–3.12 | exp | Liu *et al.*, [369] |
| 2-stage VSA | 1st 13X APG 2nd AC beads | 16 | 1st 123/7.5, 123/20 | 95.2 | 91.3 | 0.76 | Sim | Wang *et al.*, [351] |
| 4-step VSA | 13X | 15 | 101/2 | 90 | 90 | 0.53 | Sim | Haghpanah *et al.*, [96] |
| VSA | 13X | 15 | 101/3 | 90–97 | 90 | 0.55 | Sim | Haghpanah *et al.*, {Haghpanah, 2013 #10} |
| VSA | 13X, AC, MOF-74, Chemisorbent | 15 | 120/5-10 | 45-95 | 35-95 | 0.95-1.2 | sim | Maring and Webley[97] |
| 1-stage and 2-stage VSA | CMS | 15 | 101/3 | 90 | 90 | 0.99 | Sim | Haghpanah *et al.*, [225] |
| 4-step PVSA | 13X | 15 | 150/2.2 | 95.9 | 86.4 | 1.7 | Exp | Krishnamurthy *et al.*,[98] |
| 4-step PVSA with LPP | 13X | 15 | 150/2.2 | 94.8 | 89.7 | 1.71 | Exp | Krishnamurthy *et al.*, [98] |
| 2 bed 4-step VSA with LPP | 13X, Silica gel | 15% $CO_2$, 3% $H_2O$ | 101/3 | 95 | 90 | 0.63 | Sim | Krishnamurthy *et al.*, [283] |
| 4-step VSA with LPP | 13X, rho-ZMOF | 15 | 101/3 | 95 | 90 | 0.56-0.7 | Sim | Nalaparaju *et al.*, [95] |
| 2 bed, 6-step VSA | 13X, AC, MOF-74 | 15 | 101/2 | 95 | 90 | 0.76-0.83 | Sim | Nikolaidis *et al.,* [370] |
| 4-bed, 9 step cycle | 13X | 15 | 105/3,5 | 70.5–92.4 | 62.9–91.3 | 0.22–0.3 | Exp | Ntiamoah *et al.*, [327], |
| Skarstorm cycle | 13X, HKUST, 5A, MOF-74 | 15% $CO_2$, 5.5% $H_2O$ | 1st 101/10 2nd 126/10 | 90 | 90 | 0.99-1.3 | Sim | Leperi *et al.,* [371] |
| 4-step VSA with LPP | 74 real and hypothetic | 15 | 101/2 | 95 | 90 | 0.43–0.53 | Sim | Khurana and Farooq, [67] |
| | al materials | | | | | | | |
| 4-Step VSA with LPP | UTSA-16, 13X | 15 | 101/0.02–0.1 | 95 | 90 | 0.43-0.86 | Sim | Khurana and Farooq, [304] |
| 4-Step VSA with LPP | UTSA-16, 13X | 15 | 101/2-10 | 95 | 90 | 0.56-1.85 | Sim | Khurana and Farooq, [304] |
| 4-step VSA | 13X, UTSA-16, AC, MOF-74 | 15 | 101/2–3 | 95 | 90 | 0.41-0.63 | Sim | Rajagopalan *et al.*, [16] |
| 4-step VSA | UTSA-16, 13X, Hypotheti cal material | 15 | 101/2 | 95 | 90 | 0.38-0.59 | Sim | Khurana and Farooq, [372] |
| 6-step VSA | UTSA-16, 13X, Hypotheti cal material | 15 | 101/10 | 95 | 90 | 0.41-0.66 | Sim | Khurana and Farooq, [372] |
| 4-step VSA | 13X | 15 | 100/1–2 | 95 | 90 | 0.57-0.85 | Sim | Farmahini *et al.,* [68] |
| 4-step TSA | 13X+ Alumina | 12% $CO_2$ ,1.5%, 3.1% and 4.5% $H_2O$ | 440 K | 95 | 90 | 4.86 Th | Sim | Hefti and Mazzotti, [373] |
| 4-step VSA | 13X, hypothetic al materials | 15 | 100/2 | 95 | 90 | 0.4-1.38 | Sim | Rajagopalan and Rajendran, [374] |
| 4-step VSA | 13X, Diamine | 15 | 100/3-10 | 95 | 90 | 0.51-0.63 | Sim | Pai *et al.*, [290] |
| | appended MOFs | | | | | | | |
| 4-step VSA with LPP | 13X, UTSA-16 and hypothetical materials | 15 | 101/3 | 95 | 90 | 0.8-0.9 | Sim | Burns *et al.*, [103] |
| 4-step VSA with LPP | 13X, UTA-16, AC, hypothetical material | 15 | 101/1-10 | 95 | 50–90 | 0.36-0.86 | Sim | Maruyama *et al.*, [375] |
| 4-step VSA with LPP | 13X, Silicalite, HKUST, Ni MOF-74 | 15 | 101/1–2 | 95 | 90 | 0.5-0.9 | Sim | Farmahini *et al.*, [101] |
| MBTSA | 13X, Ni MOF-74 | 5 | 101, 480 K, 405 K | 95.8, 98.9 | 98.2, 92.6 | 1.42, 1.89 th | Sim | Mondino *et al.*, [322]. |
| Modified Skarstrom, 5-step, and FVSA | 13X, 15 MOFs | 15 | 100–1000/10–50 | 90 | 90 | 0.55–2.5 | Sim | Yancy-Caballero *et al.*, [105] |
| 4-step VSA with LPP | 13X, UTSA-16, IISERP MOF2 | 20 | 100/1 | 95 | 90 | 0.55 – 0.72 | Sim | Subraveti, *et al.*, [71] |
| 4-step VSA with LPP | 36 materials | 0.05-0.7 | 100/1 | 95 | 90 | 0.42 | Sim | Pai *et al.,* [106] |
| 6-step VSA | Supported amine Sorbent | 15% $CO_2$, 5% $H_2O$ | 101/10 | 95 | 90 | 1 | Sim | Krishnamurthy *et al.,* [376] |

**Note 1 -** all energy values are electric *i.e.* the energy consumed by the vacuum pumps and the compressors.
**Note 2 -** th: "thermal" is the heat supplied to desorb the $CO_2$ in TSA/PSA process.
**Note 3 -** mix: electric+thermal is the electric energy consumed by vacuum pumps in the 2nd stage PSA process and heat needed to recover the $CO_2$ from the 1st stage PTSA process of Ishibashi *et al.* [357].
**Note 4 -** N/A: not available.

# 8. Existing Challenges and Open Questions in Performance-based Screening of Porous Materials

In this section, we outline what we believe are key challenges in the implementation of the multiscale workflows for performance-based material screening. Our awareness of these challenges evolved over time as, we being a collaborative group of molecular simulators and process simulators ourselves, navigated this emerging field of research.

## 8.1. Accuracy and Transferability of the Molecular Force fields

Accurate adsorption equilibrium models are the basis for equilibrium driven separations, such as those considered here for post combustion capture. From the perspective of a process modeller, the most immediate advantage of having access to accurate molecular simulation tools is gained by having the ability to predict multicomponent adsorption equilibria. This requires force fields for the mixture constituents that are validated against pure component adsorption data of good quality. As discussed in Section 6.2.5 about force fields, dispersion and electrostatic interactions are two important classes of interatomic interactions that are relevant to adsorption and diffusion phenomena. It was also noted that the ability of molecular simulations to correctly predict adsorption behaviour is limited by the ability of force fields to correctly model these interactions. In this section, we highlight the existing issues around accuracy of molecular force fields for predicting adsorption isotherms, and emphasize the need for consistent implementation of them in multiscale workflows for materials screening. **Dispersion interactions:** An important concern regarding the use of molecular force fields is availability and transferability of model parameters for calculation of dispersion interactions (*e.g.* Lennard-Jones parameters). Currently and for practical reasons, high-throughput screening of large materials databases heavily rely on the use of generic force fields [56-58, 65, 72, 78, 129, 377]. As mentioned before, the most commonly used generic force fields include DREIDING [243], UFF [240] and OPLS-AA [378]. Despite their wide spread, generic force fields fail to accurately reproduce experimental adsorption data in many cases [379-381] particularly for gas adsorption in MOFs with coordinatively unsaturated metal sites [260, 265, 382]. Even for the systems where generic force fields are deemed suitable, prediction of experimental adsorption isotherms is rather qualitative in which simulated isotherms only capture the general shape of their experimental counterpart [383-386]. Therefore, the use of generic force fields for screening of large and diverse databases of porous materials should be approached with caution and these issues have been raised in several excellent studies, see for example Refs. [383, 387, 388]. In fact, many groups have already started to develop specialized force fields for challenging systems such as those involving adsorption of water [381, 389] or MOFs containing open metal sites [261-265].

At this point it is also useful to reflect on what is considered to be a "good agreement" between a simulated adsorption isotherm and experimental data, as this terminology can mean different things for a molecular modeller and a process modeller. If correctly reproduced overall shape of the isotherm may seem a good achievement in molecular simulations, particularly for challenging cases, such as water or other polar species, in process modelling accurate Henry's constant, non-linearity/shape of adsorption isotherm, saturation capacity and other quantitative features of the adsorption data are important as they will impact separation performance of adsorbent materials (*i.e.* their energy-productivity or purity-recovery Pareto fronts).

While molecular simulations often focus on the behaviour of a single component of interest ($CO_2$ in carbon capture), in process modelling it is recognized that the separation performance depends on the behaviour of the mixture and accurate equilibrium data for all components is important. An illustrative example here is provided by Khurana and Farooq [67] where the influence of isotherm characteristics on minimum energy penalty and maximum productivity of a 4-step VSA-LPP cycle is demonstrated (**Figure *21***).

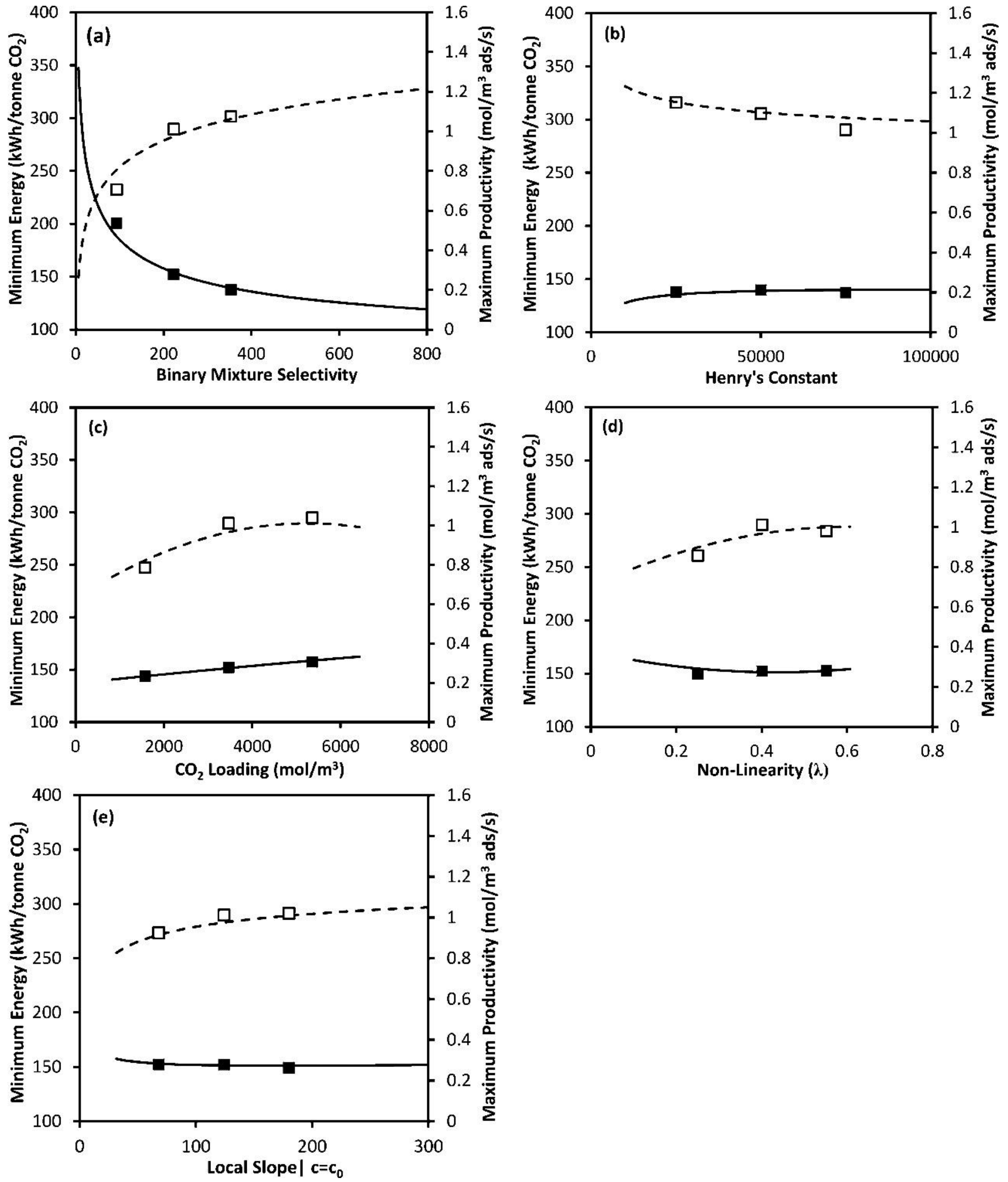

**Figure 21**. Variation of minimum energy penalty and maximum productivity in a 4-step VSA-LPP process with respect to different isotherm characteristics. The symbols (closed symbols for minimum energy and open symbols for maximum productivity) are associated with the results of detailed process modelling and optimization, and lines (continuous lines for minimum energy and dashed lines for maximum productivity) represent the predictions from a meta-models which is discussed in the original publication. Reprinted with permission from Khurana and Farooq, Industrial & Engineering Chemistry Research, 2016. 55(8): p. 2447-2460. Copyright (2016) American Chemical Society [67].

As shown in this figure, both minimum energy penalty and productivity of the process are highly sensitive to variation of binary mixture selectivity. For the case studied here, the local slope, nonlinearity, and Henry's constant of $CO_2$ isotherms seem to have rather small effects on the minimum energy and maximum productivity of the process. Nevertheless, one should consider these results with caution because in the above example only one isotherm characteristic has been allowed to vary for each case, while other isotherm characteristics are held constant. In reality, it is not rare to see GCMC simulated isotherms do not adequately reproducing multiple characteristics of experimental adsorption isotherm; hence the combined effect on the performance of the material at the process level will be larger than what is shown in the above figure. Finally, the effect of isotherm characteristics of nitrogen is not considered in the analysis provided in **Figure *21*** (i.e. this figure is only related to characteristics of $CO_2$ adsorption isotherm).

Several studies have shown that adsorption of nitrogen plays a significant role in separation performance of the PSA/VSA processes for post-combustion carbon capture [16, 67, 68, 72, 97, 103, 104, 374]. In fact, an important concern regarding the quality of available force fields is associated with the role of nitrogen in process simulation. To date, significant efforts have been made to develop more reliable force fields for adsorption of $CO_2$, however the impact of other components in the flue gas mixture has been somehow overlooked. A quick review in the literature shows that specialized QM-derived force fields for nitrogen adsorption are scarce [259, 390, 391], although the accuracy of generic force fields for prediction of nitrogen adsorption is not satisfactory for many materials. Some examples of this include adsorption of nitrogen in STT, CHA all silica zeolites [392], FAU and MFI type zeolites with different Si/Al ratio [68, 392], Mg-MOF74 [259] and Ni-MOF74 [101], ZIF-68 [393], Zn-MOF and Cu-BTC [394].

From a process simulation perspective, it is the nitrogen adsorption behaviour that most significantly determines whether or not the process can produce $CO_2$ with 95% purity and 90% recovery [16, 72, 103, 374]. A recent study by Rajagopalan and Rajendran [374] has clearly demonstrated that purity-recovery Pareto fronts obtained using a 4-step VSA-LPP cycle for separation of $CO_2$ and $N_2$ are very sensitive to variation of nitrogen adsorption. This observation is illustrated in **Figure *22***.

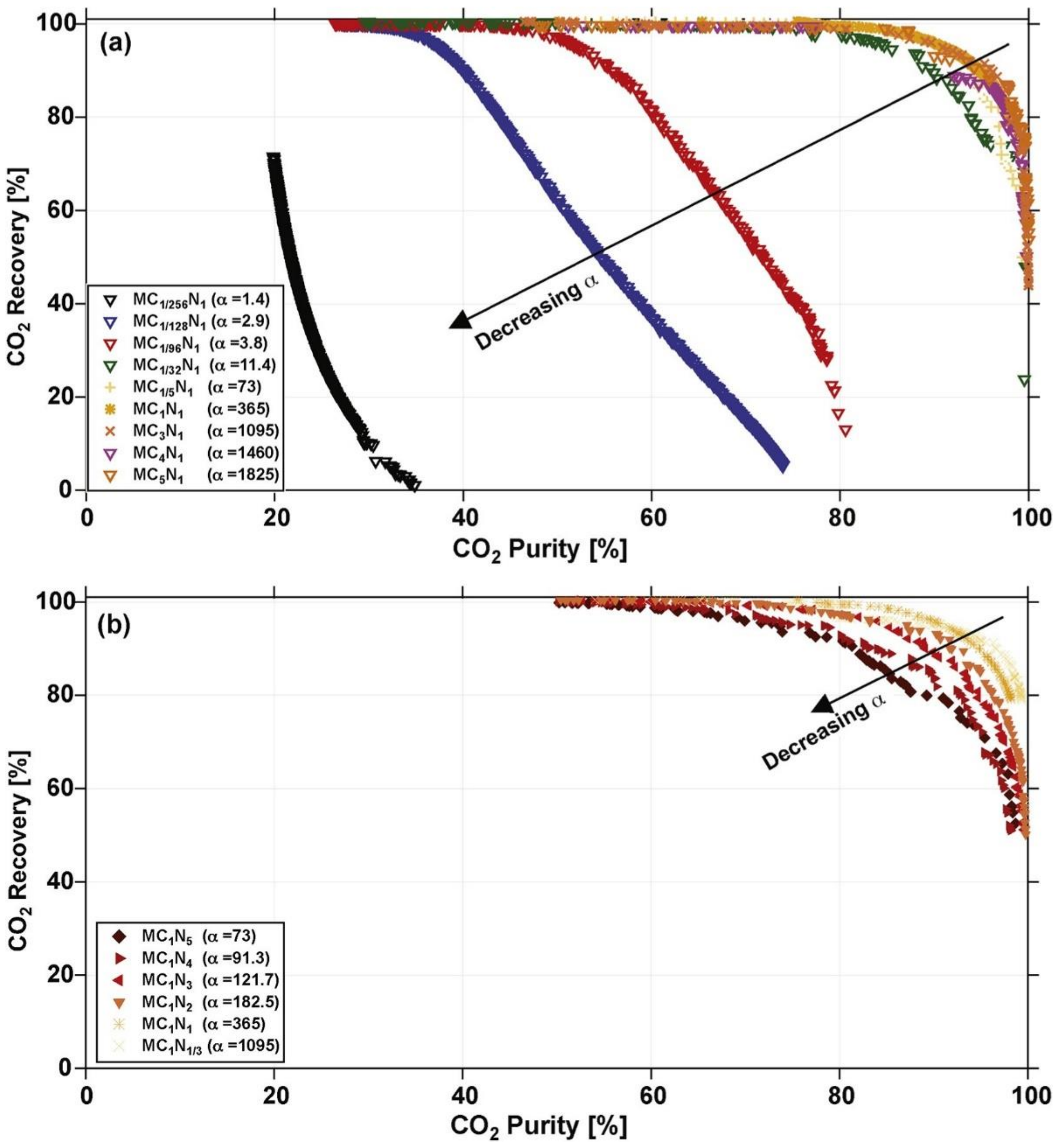

**Figure 22**. Purity and recovery Pareto fronts for different hypothetical adsorbents ($MC_cN_n$) with fixed N2 isotherm (a) and fixed $CO_2$ isotherm (b). $MC_1N_1$ represents Zeolite 13X as a reference material, $\propto$ is binary selectivity of $CO_2/N_2$ for each material, and ratio of $^{c}/_{n}$ is equal to selectivity of $MC_cN_n$ normalized by selectivity of Zeolite 13X ($\propto_{13X}= 365$). Reprinted with permission from Rajagopalan and Rajendran, International Journal of Greenhouse Gas Control, 2018. 78: p. 437-447. Copyright (2018) Elsevier [374].

As shown here, reducing selectivity by more than 160 times (from 1825 to 11.4) through changing $CO_2$ adsorption isotherms, while $N_2$ isotherm is held constant (**Figure *22*** (a)) has less impact on purity and recovery of the process compared to the case where selectivity is reduced only by 15 times (from 1095 to 73) through changing $N_2$ adsorption isotherm, while $CO_2$ isotherm is fixed (**Figure *22*** (b)) [374].

In separate study from Leperi *et al.* [72], it is shown that heat of adsorption of nitrogen plays a crucial role on the maximum $CO_2$ purity that can be achieved in a fractionated vacuum pressure swing adsorption (FVPSA) cycle. It is shown that if $N_2$ heat of adsorption is greater than 16 kJ/mol, the process cannot produce $CO_2$ with 90% purity [72]. This is illustrated in **Figure *23***.

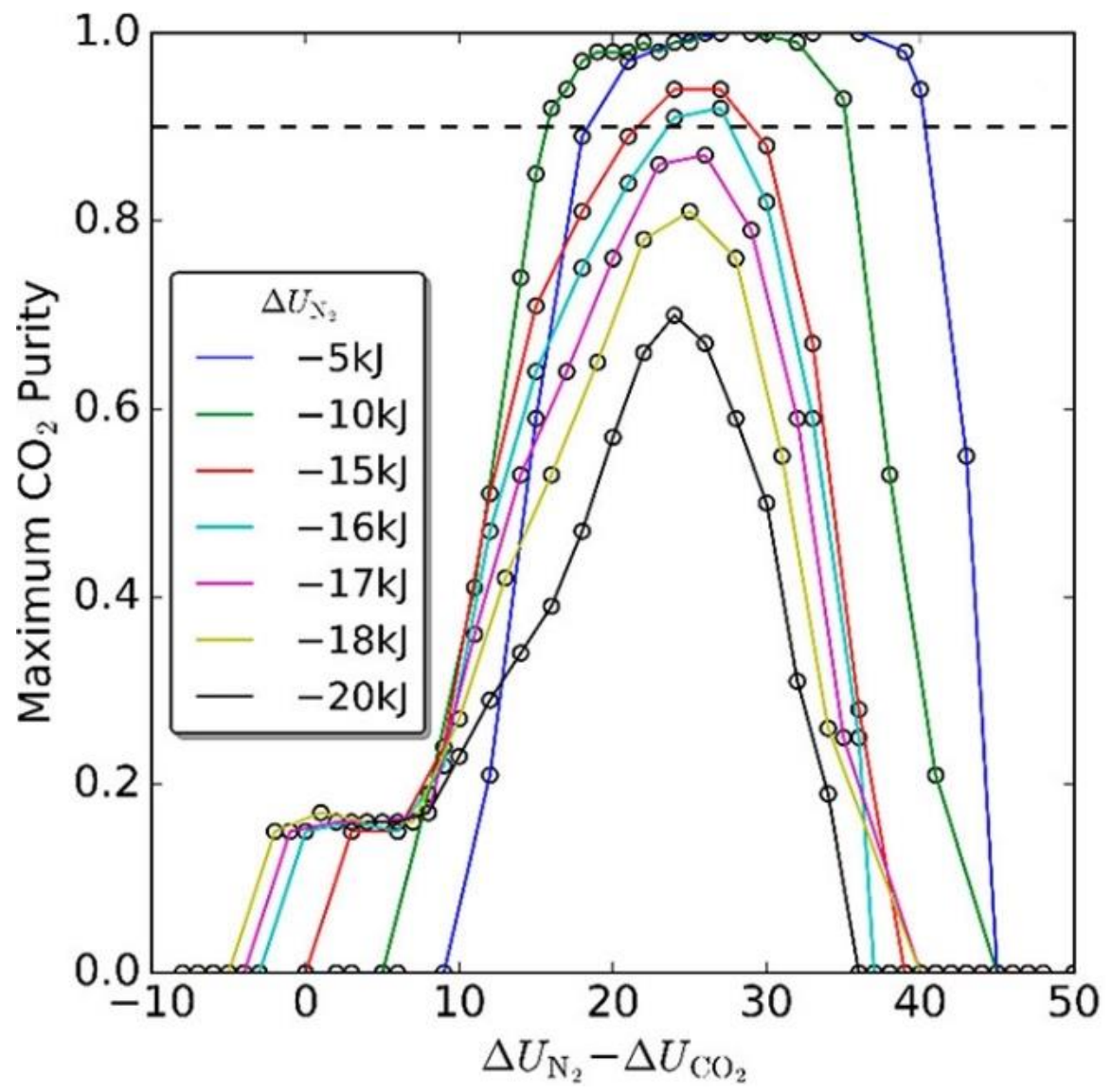

**Figure 23.** Optimal heats of adsorption for the FVPSA cycle using a generic adsorbent with density of 1.1 g/cm$^3$. Each point represents the highest $CO_2$ purity that can be achieved while recovering 90% of $CO_2$. Each line depicts a different $N_2$ internal energy of adsorption. The dashed horizontal line is $CO_2$ purity of 90%. Reprinted with permission from Leperi *et al.*, ACS Sustainable Chemistry & Engineering, 2019. 7(13): p. 11529-11539. Copyright (2019), American Chemical Society [72].

Even for those materials that meet 95%-90% criterion target for purity and recovery of $CO_2$, inaccurate prediction of $N_2$ adsorption data is shown to affect energy and productivity of the process [68]. These observations in turn highlight the importance of having access to accurate molecular force fields for more accurate prediction of nitrogen adsorption data using molecular simulations, the topic whose importance has been overlooked so far. Force field development for nitrogen is, however, a challenging task for two related reasons. Initially, one would consider QM methods to develop a detailed picture on the potential energy surface for nitrogen in various porous materials including MOFs, in a similar fashion that has been done for several $CO_2$-MOF systems. What is important to remember is that nitrogen is a weakly adsorbing component (heat of adsorption 10-20 KJ/mol, but likely to be closer to 10-12 kJ/mol). Relative error in QM estimates of energy of binding is likely to have a much stronger impact on nitrogen adsorption than on stronger adsorbing carbon dioxide. For a similar reason, the uncertainty in the experimental adsorption isotherms of nitrogen (which are used for validation of QM-based force fields) is also greater, considering the amount adsorbed tends to be much smaller for $N_2$ compared to $CO_2$ under the conditions of interest.

**Electrostatic interactions:** As mentioned in Section 6.2.5, there are several computational schemes to assign partial charges to the atoms of the porous materials.

The overwhelming majority of material databases constructed so far do not include partial atomic charges and there is no universal agreement in the scientific community on what scheme should be adopted to assign these charges. Effect of framework atomic charges on gas adsorption have been already demonstrated in several studies [247, 395]. It is shown that application of different charge calculation schemes in molecular simulation can lead to a substantial variation in the adsorption data [247, 395]. For example, Sladekova *et al.* [247] have compared adsorption isotherms of carbon dioxide and water in several MOFs using different set of partial atomic charges which are obtained from different charge calculation methods such as DDEC [245], ChelpG [246], LoProp, EQeq [92], and REPEAT [93]. As shown in **Figure *24***, the use of various charge calculation schemes can lead to significantly different adsorption behaviour of $CO_2$ and water in these materials. This in turn will have a profound impact on separation performance of these materials at the process level. As we previously mentioned, the flue gas contains water and if we wished to include water adsorption in the process model, obtaining accurate equilibrium data from molecular simulations would be challenging.

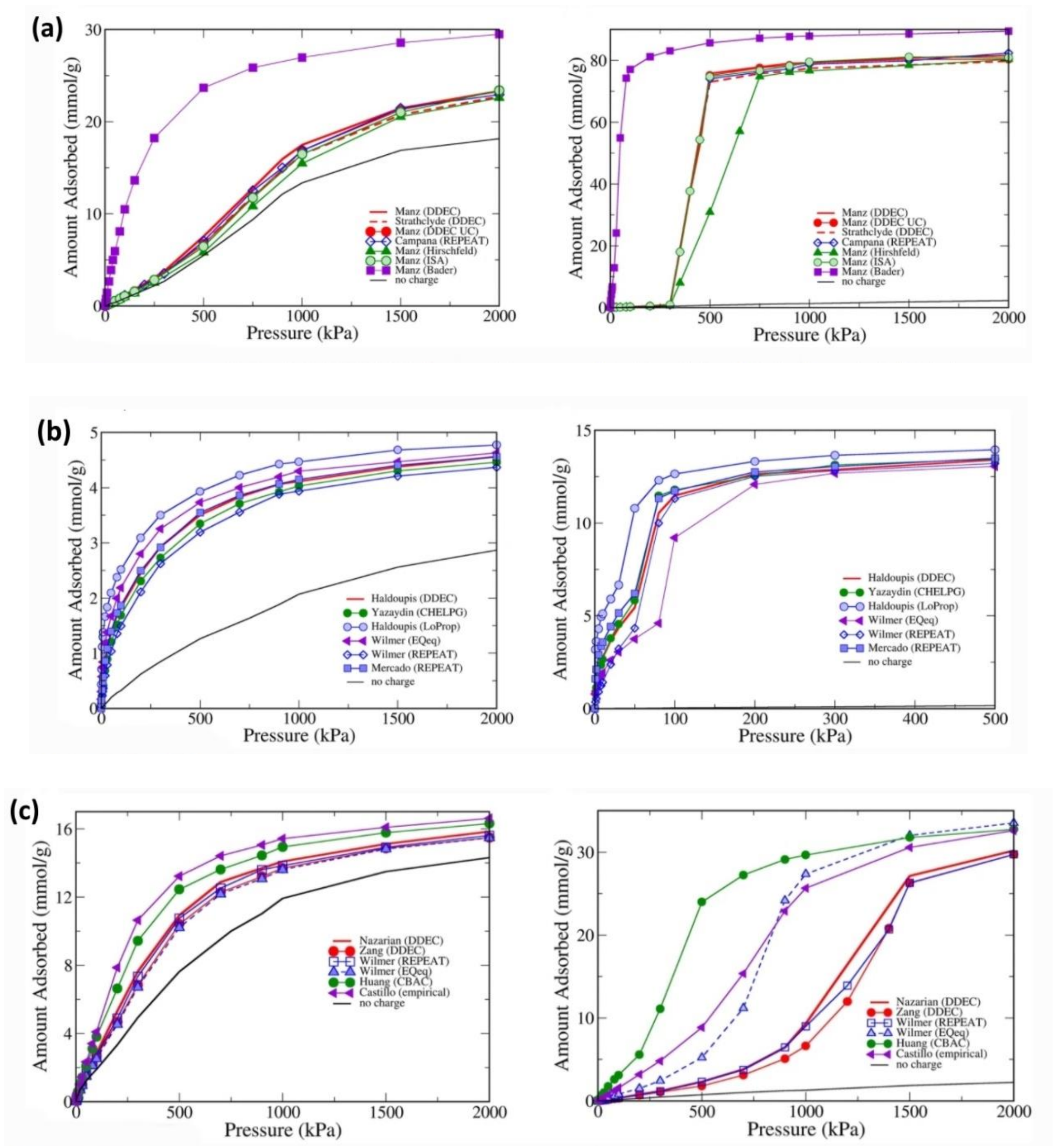

**Figure 24.** Adsorption isotherms of $CO_2$ and water in IRMOF-1 (a), Co-MOF-74 (b) and Cu-BTC (c) as obtained from different charge calculation schemes. Reprinted with permission from Sladekova *et al.*, Adsorption, 2020. 26(5): p. 663-685. Copyright (2020), Springer Nature [247].

Li *et al.* [58] provide a another good example for the use of two different charge calculation techniques for screening of MOFs for $CO_2$ capture in the presence of water. They use the extended charge equilibration (EQeq) method [92] as well as the repeating electrostatic potential extracted atomic (REPEAT) method [93] to compute atomic partial charges of MOFs for adsorption of $CO_2/H_2O$ and $CO_2/H_2O/N_2$ mixtures in a large group of MOFs, which are selected from the CoRE-MOF database. It is demonstrated that water adsorption behaviour can be greatly influenced by the choice of methods used for calculation of electrostatic interactions [58]. This is evident in **Figure *25*** which compares $CO_2/H_2O$ selectivities of various MOFs based on two different set of partial charges obtained from EQeq and REPEAT methods.

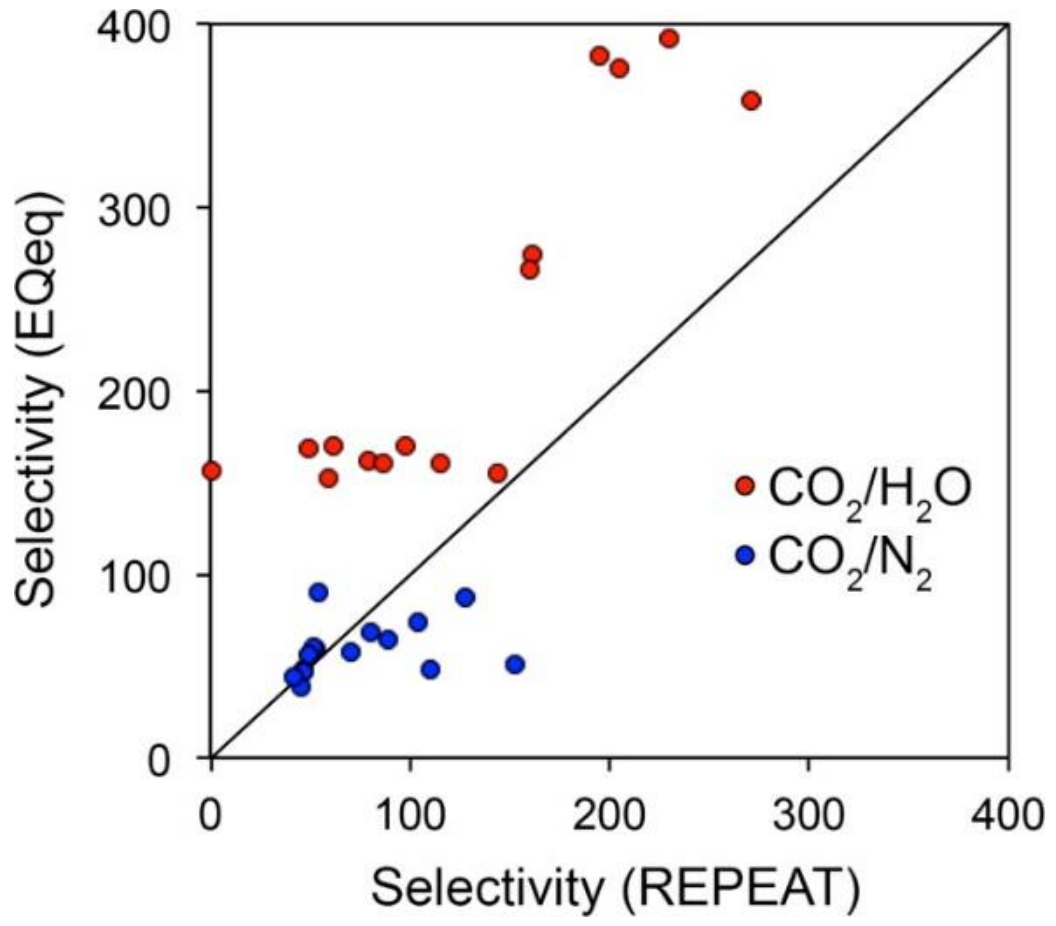

**Figure 25.** Comparison between the selectivities based on the ratio of Henry's law constants obtained for the structures with EQeq partial atomic charges (y-axis) and that obtained for the structures with REPEAT partial atomic charges (x-axis). Red circles are related to $CO_2/H_2O$ selectivity, and blue circles are associated with $CO_2/N_2$ selectivity. Reprinted with permission from Li *et al.*, Langmuir, 2016. 32(40): p. 10368-10376. Copyright (2016) American Chemical Society [58].

In a more recent study, Altintas and Keskin [396] discussed the role of partial charge assignment methods in high-throughput screening of MOFs for $CO_2/CH_4$ separation. They employ a quantum-based density-derived electrostatic and chemical charge method (DDEC) [245] and an approximate charge equilibration method (Qeq) [134] to predict adsorption of $CO_2/CH_4$ mixtures in 1500 MOFs. The authors demonstrate that gas uptake, working capacity, selectivity, adsorption performance score (APS), and regenerability of MOFs vary considerably depending on the charge assignment methods used in molecular simulations, as shown in **Figure *26*** and **Figure *27*** [396]. They also report that the ranking of the best-performing MOFs are also different depending on the method used for charge calculations [396].

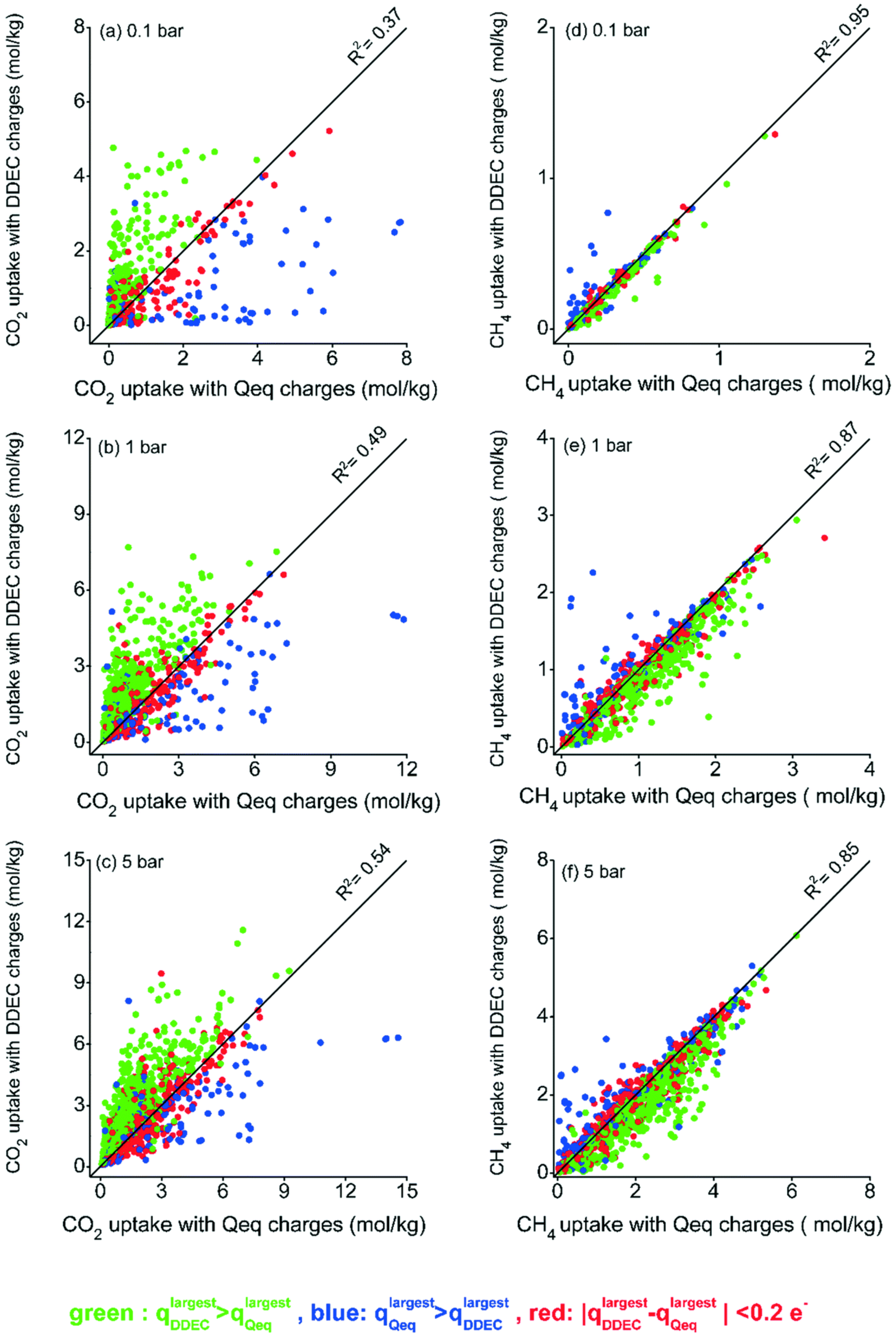

**Figure 26.** $CO_2$ (a - c) and $CH_4$ (d–f) uptakes obtained from simulations using Qeq and DDEC charges at 0.1, 1.0, and 5.0 bar for the 10%-90% $CO_2$/$CH_4$ mixture. Reprinted with permission from Altintas

and Keskin, Molecular Systems Design & Engineering, 2020. 5(2): p. 532-543. Copyright (2020) Royal Society of Chemistry [396].

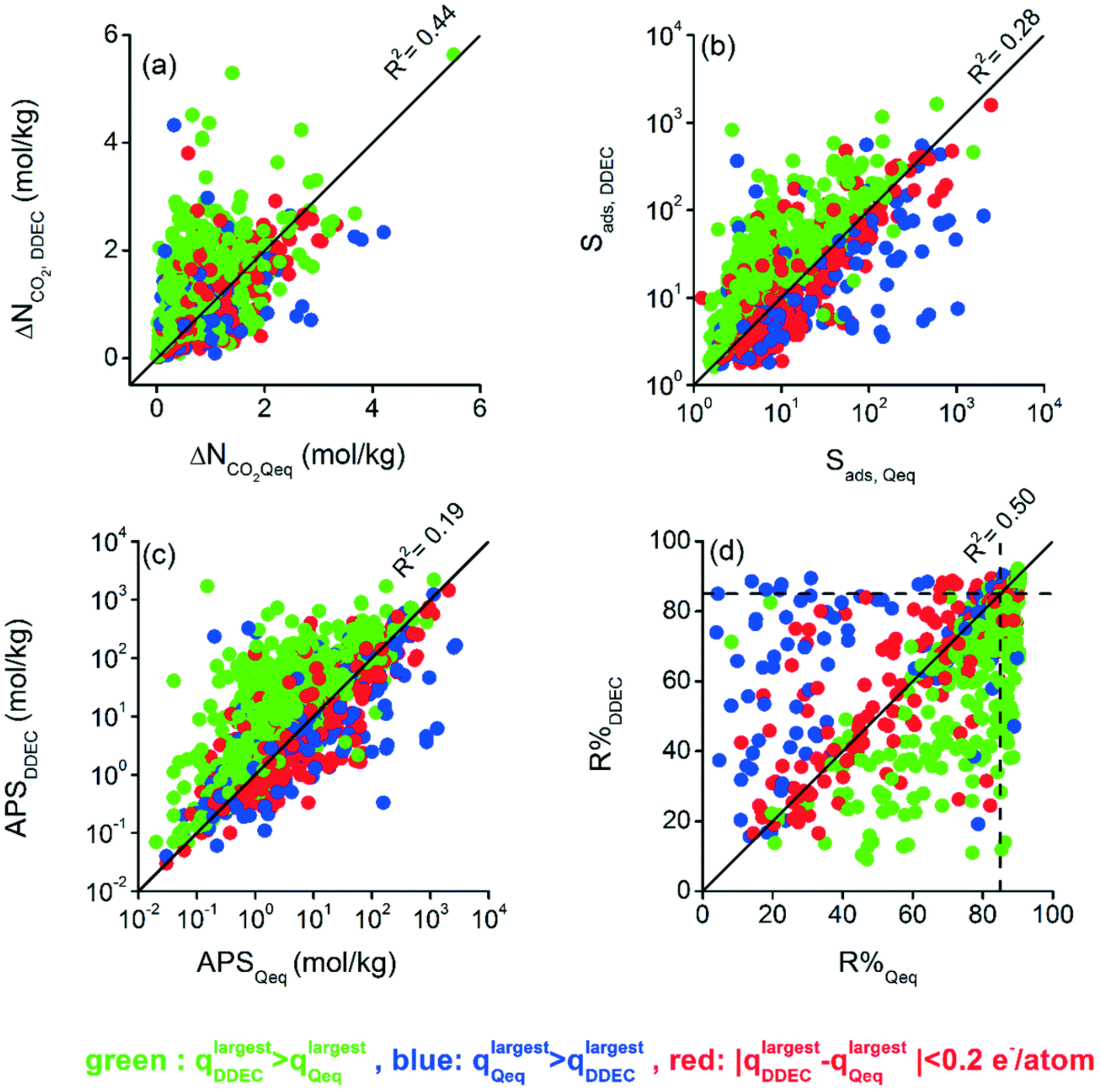

**Figure 27.** Calculated adsorbent performance evaluation metrics using Qeq and DDEC charges under VSA conditions at 298 K for 10%-90% CO2/CH4 mixture: (a) $CO_2$ working capacity, (b) selectivity, (c) adsorbent performance score and (d) regenerability. Reprinted with permission from Altintas and Keskin, Molecular Systems Design & Engineering, 2020. 5(2): p. 532-543. Copyright (2020) Royal Society of Chemistry [396].

## 8.2. Data Availability and Consistent Implementation of Multiscale Screening Workflows

Imagine, two articles have been published by two different academic groups ranking two sets of porous materials for a particular separation using the multiscale workflows discussed in this review. Are these rankings compatible and consistent with each other? In other words, can the results of the two studies be combined in one ranking, and therefore the best material be identified out of the combined group of materials in both articles? As we will see in Section 9.1, in general, the answer is "no"! This is because different groups employ different hierarchies of models and assumptions, use different sources and values of parameters, and apply different conditions of the process. Here in this section, we would like explore the issue of consistency in materials screening simulations in more detail.

As has been discussed throughout the article, not all of the data required to set up multiscale process simulations is available from molecular simulations. This is in fact a limiting factor for multiscale materials screenings to be performed fully *in silico*. Some of the data required for process modelling such as, for example, diffusion models for the bulk gas phase, can be constructed using appropriate, well-established theories and models. For other data however, some assumptions have to be made. In order to be able to compare different materials rankings consistently, similar assumptions for implementation of the models and estimation of input parameters should be used. Recent studies have probed the influence of some of these assumptions on the actual process performance predictions as outlined below:

*Heat capacity of the adsorbent:* Traditionally, it has been assumed that heat effects play a secondary role in the adsorption column and will not significantly influence the performance of the process or ranking of the material. Adsorbent heat capacity is also a property scarcely measured or available for the new classes of porous materials, such as MOFs. As a pragmatic approach, some studies have assumed the heat capacity of all materials to be equal to the heat capacity of a reference material, such as Zeolite 13X [67, 95]. Recent preliminary sensitivity study of the influence of this parameter, painted a somewhat different picture [101]. As shown in **Figure *28***, performance of HKUST with the value of the heat capacity equal to that of Zeolite 13X was considerably different from the performance of the same material, using an experimental value of this property [101].

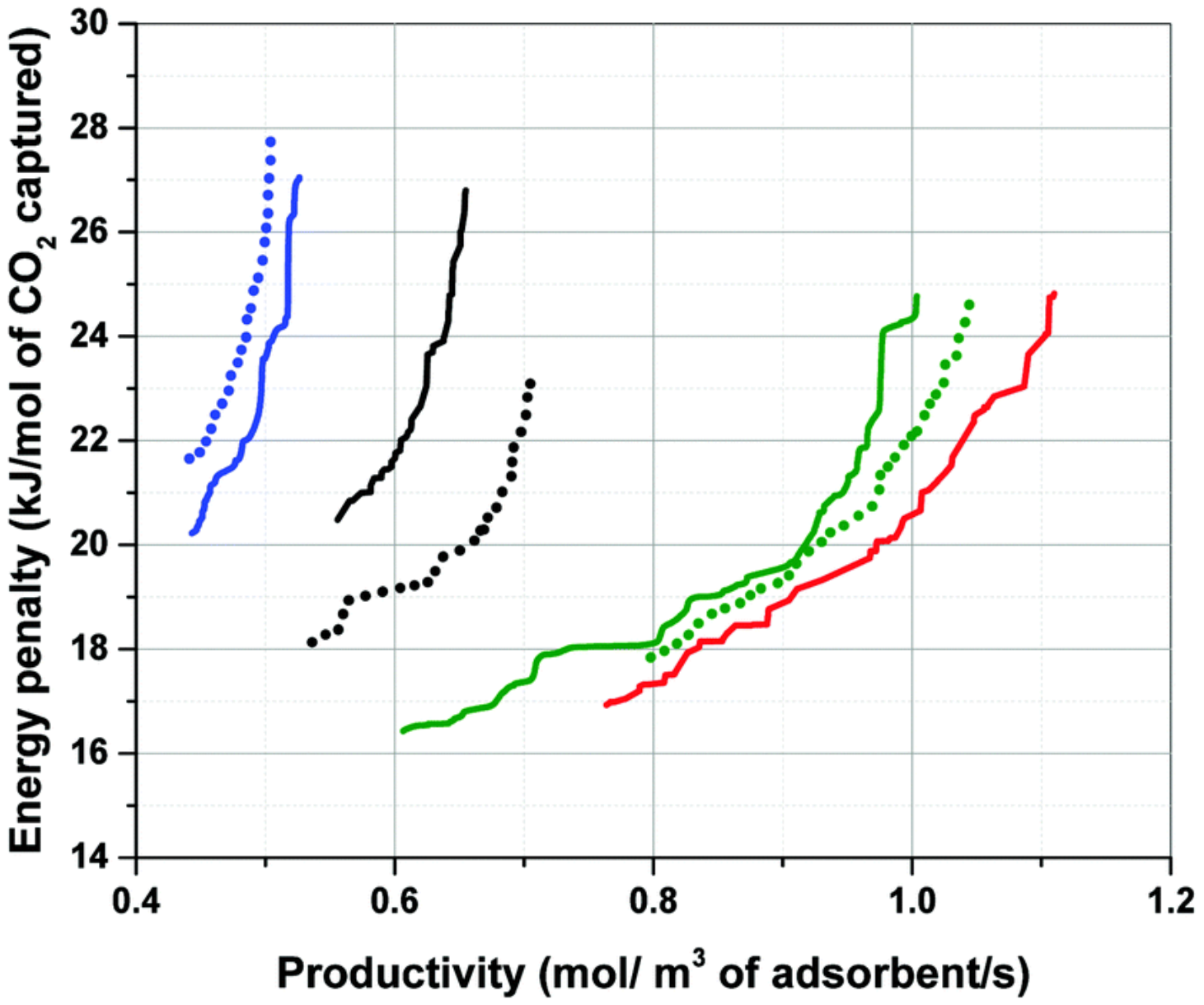

**Figure 28.** Effect of specific heat capacity ($C_p$) on position of the Pareto fronts for Cu-BTC (black), MOF74-Ni (green), Silicalite-1 (blue) and Zeolite 13X (red) obtained from optimization of a 4-step VSA-LPP cycle. Dashed lines illustrate a case where experimental $C_p$ of each material is used for process simulations. Solid lines represent another case where $C_p$ for all materials is assumed to be equal 920 J $kg^{-1}.K^{-1}$. Reprinted with permission from Farmahini *et al.*, Energy & Environmental Science, 2020. 13(3): p. 1018-1037. Copyright (2020) Royal Society of Chemistry [101].

Recently, Danaci *et al.* [104] have also analysed sensitivity of three process-level performance indicators namely purity, recovery, and capture cost to variation of specific heat capacity for MOF74-Mg and UTSA-16 using a 0-dimensional equilibrium-based PVSA model. As shown in **Figure *29***, all of the above indicators show considerable sensitivity to variation of specific heat capacity for MOF74-Mg. For UTSA-16 however, the sensitivity to variation of heat capacity is negligible.

## (a) Purity

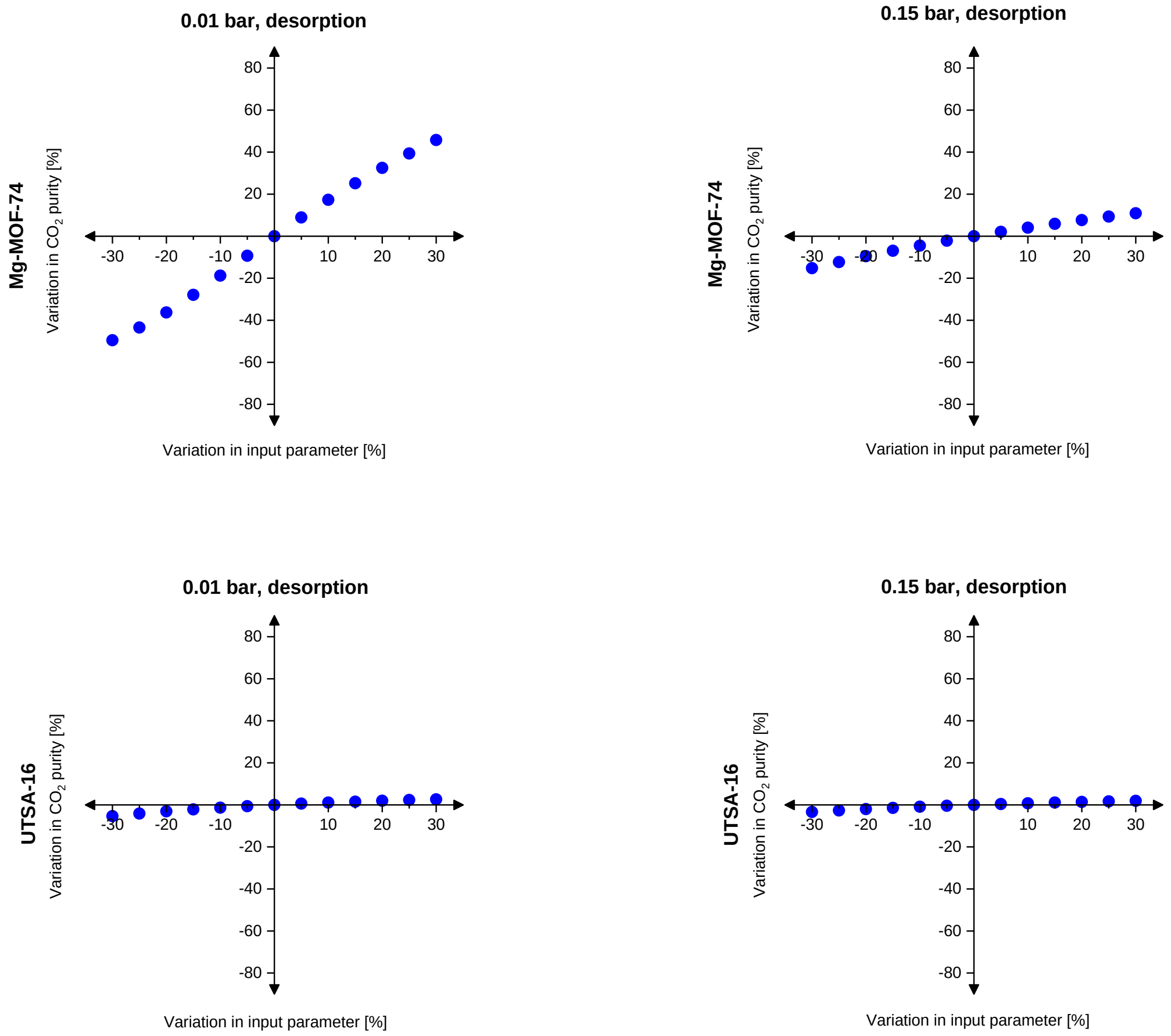

## (b) Recovery

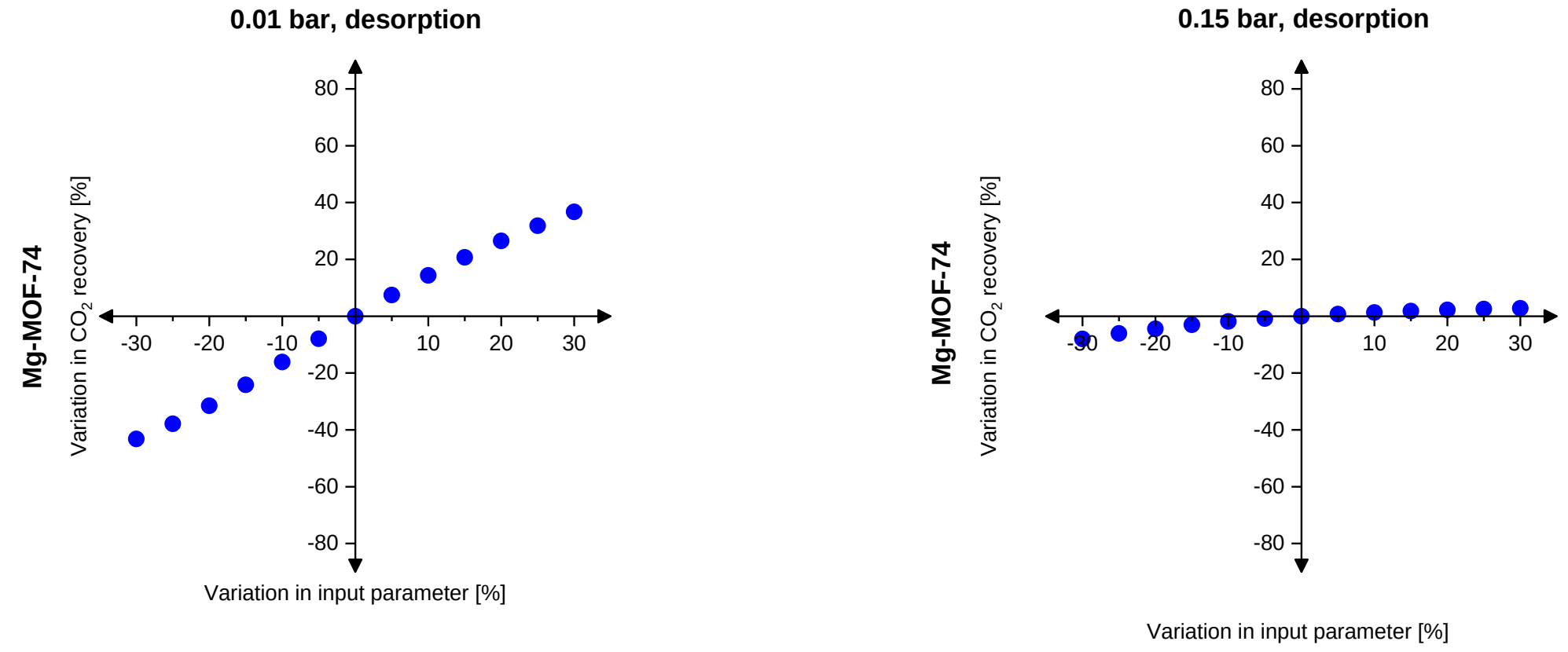

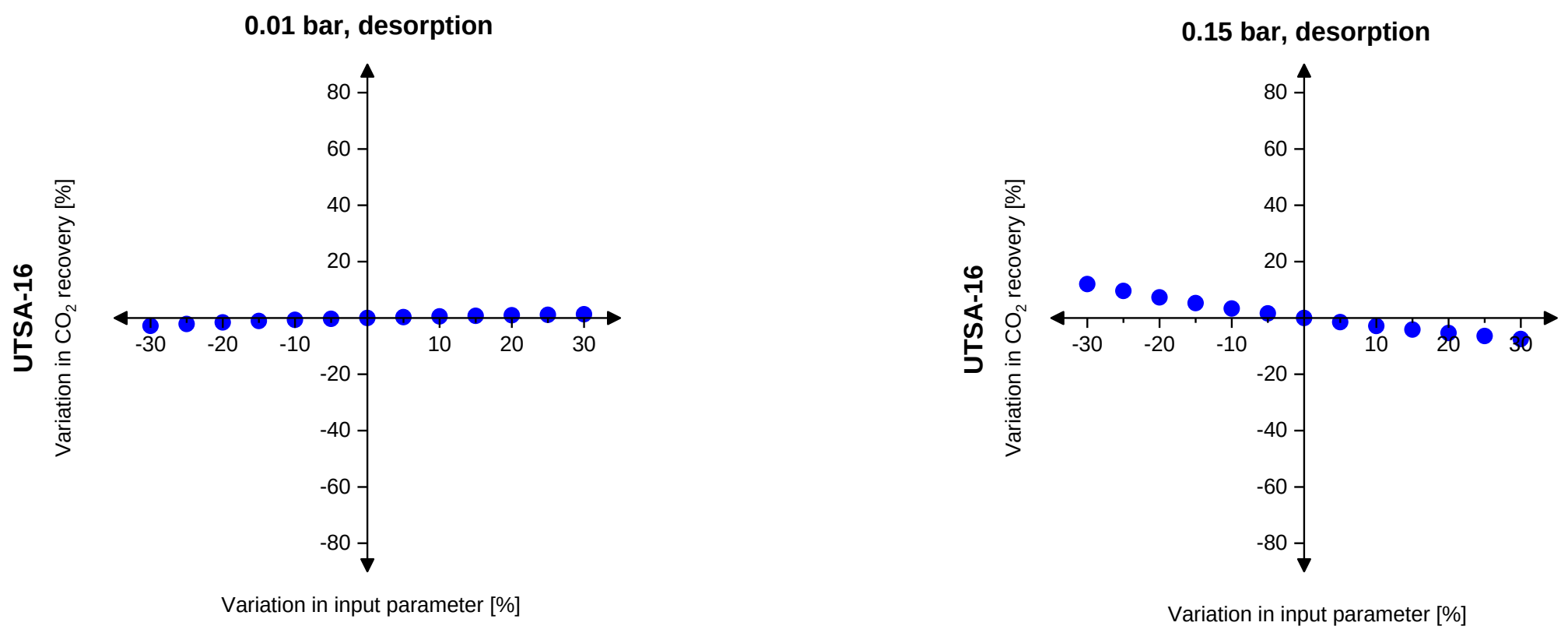

## (c) Capture cost

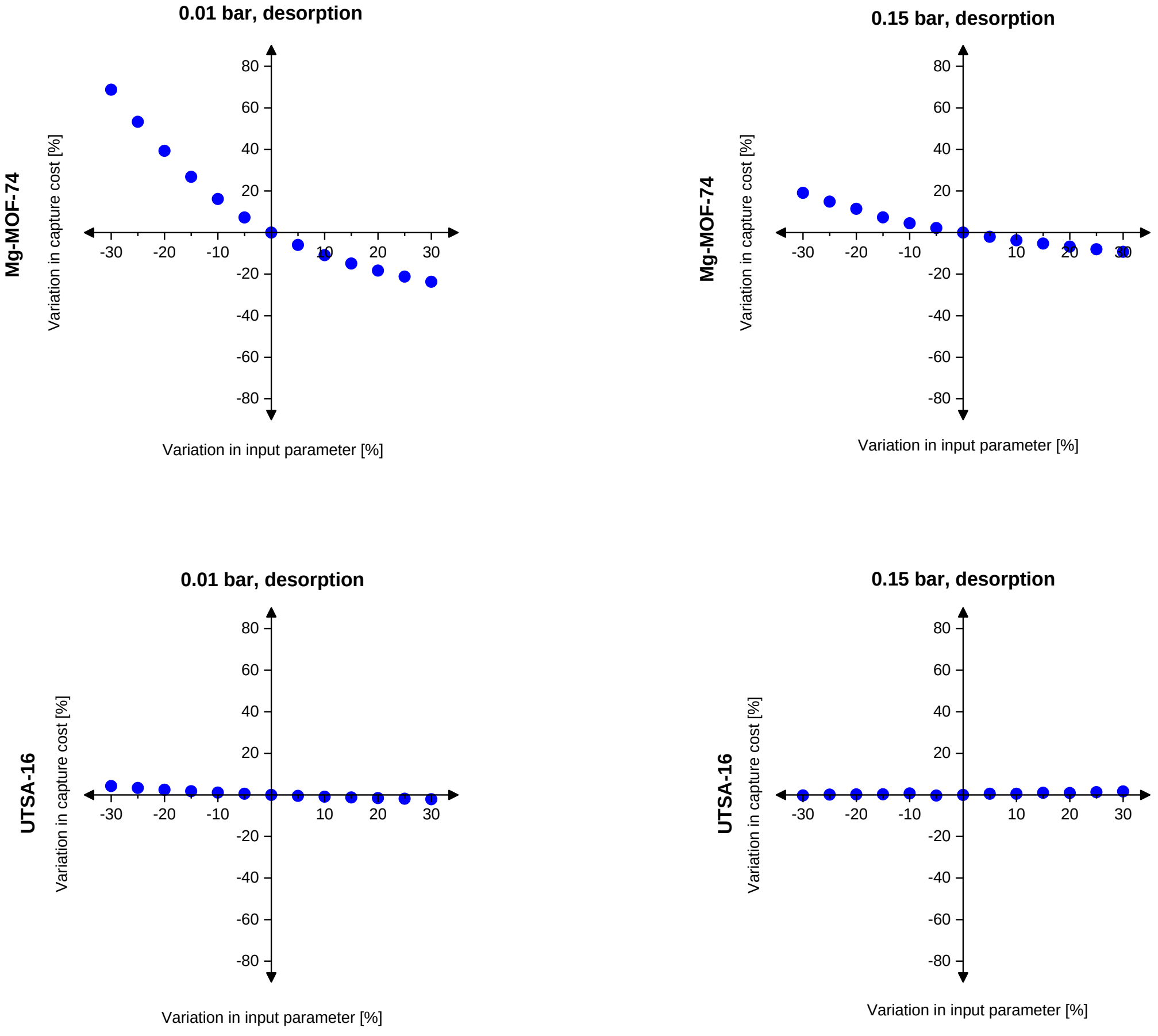

**Figure 29.** Sensitivity analysis purity (a), recovery (b), and capture cost (c) to variation of heat capacity for Mg-MOF-74 and UTSA-16 at 0.01 and 0.15 bar desorption pressure in a PVSA process. Reproduced with permission from Danaci *et al.*, Molecular Systems Design & Engineering, 2020. 5(1): p. 212-231. Copyright (2020) Royal Society of Chemistry [104].

Therefore, it seems for a diverse group of porous materials (MOFs, zeolites), it would be prudent to procure more heat capacity data, explore the heat effects in more detail and use more accurate values of $C_p$ in process modeling. How can this be accomplished in purely *in silico* workflows? With some compromise we can use empirical group contribution methods where heat capacity of an adsorbent are estimated by summing the molar fraction contributions of its functional groups (*e.g.* metal nodes and organic ligands in MOFs) [104, 397, 398]. Kloutse *et al.* [398] have recently calculated specific heat capacities of MOF-5, Cu-BTC, Fe-BTC, MOF-177 and MIL-53 (Al) at a single temperature (323 K) using this method. The results are compared with experimental heat capacity data showing relative difference errors between 2.58% and 14.77% [398]. Nevertheless, in the absence of experimental data for many porous adsorbents, the results from this method should considered with caution especially for flexible materials such as MOFs. As correctly pointed out by Danaci *et al.* [104], the group contribution method for calculation of heat capacity does not take into account the contribution of vibrational modes of crystalline materials. This will lead to underestimation of the heat capacity. In opposite, atomic vibrations of ligands may be reduced upon coordination which is due to the loss of some degrees of freedom. This is also not considered in the group contribution method resulting in overestimation of the heat capacity [104]. To fully evaluate reliability of the group contribution method, a separate computational study is needed to investigate the relative contributions of these factors for a range of different materials including MOFs.

Another technique for estimation of heat capacity is accurate analysis of the vibrational modes of porous lattice/crystal or the so called phonon analysis methods [273, 274, 399, 400]. This, however, will add yet another level of complexity to the already long chain of simulations that must be performed to compute the prerequisite properties for multiscale simulation of PSA/VSA systems. Analysis of phonon properties of porous solids could require time-consuming quantum mechanical calculations, considering that currently available classical force fields may not necessarily be able to capture the full range of harmonic/anharmonic properties pertinent to thermal properties of new classes of materials [400, 401]. Therefore, the remaining challenge here is to develop affordable computational techniques by which thermal properties of porous materials can be calculated without compromising the accuracy of the predictions.

*Pellet size and pellet porosity:* In the traditional process modelling literature, the values of pellet size and pellet porosity are typically obtained from experiments for a specific sample of a material under consideration. The values of these characteristics cannot be provided by molecular simulations or from some approximate theories. Again, the pragmatic approach, adopted in the previous studies, has been to use the values available for a reference material, such as Zeolite 13X. However, in the context of the ranking of porous materials, a question can be asked whether optimal performance of a material can be achieved only at material-specific values of pellet size and pellet porosity (within the range of feasible experimental values)? In this case, shall the ranking be performed under the constraint of specific values of pellet size and porosity, or shall these properties also become some optimization parameters? Farmahini *et al.* [101] have recently explored these questions in a recent study, and observed that depending on the protocol (pellet size and porosity are constrained to the reference values of Zeolite 13X versus being free optimization parameters) the order of performance of materials does change [68, 101]. This effect is illustrated in **Figure *30*** for flour materials including Cu-BTC, MOF74-Ni, Silicalite-1 and Zeolite 13X.

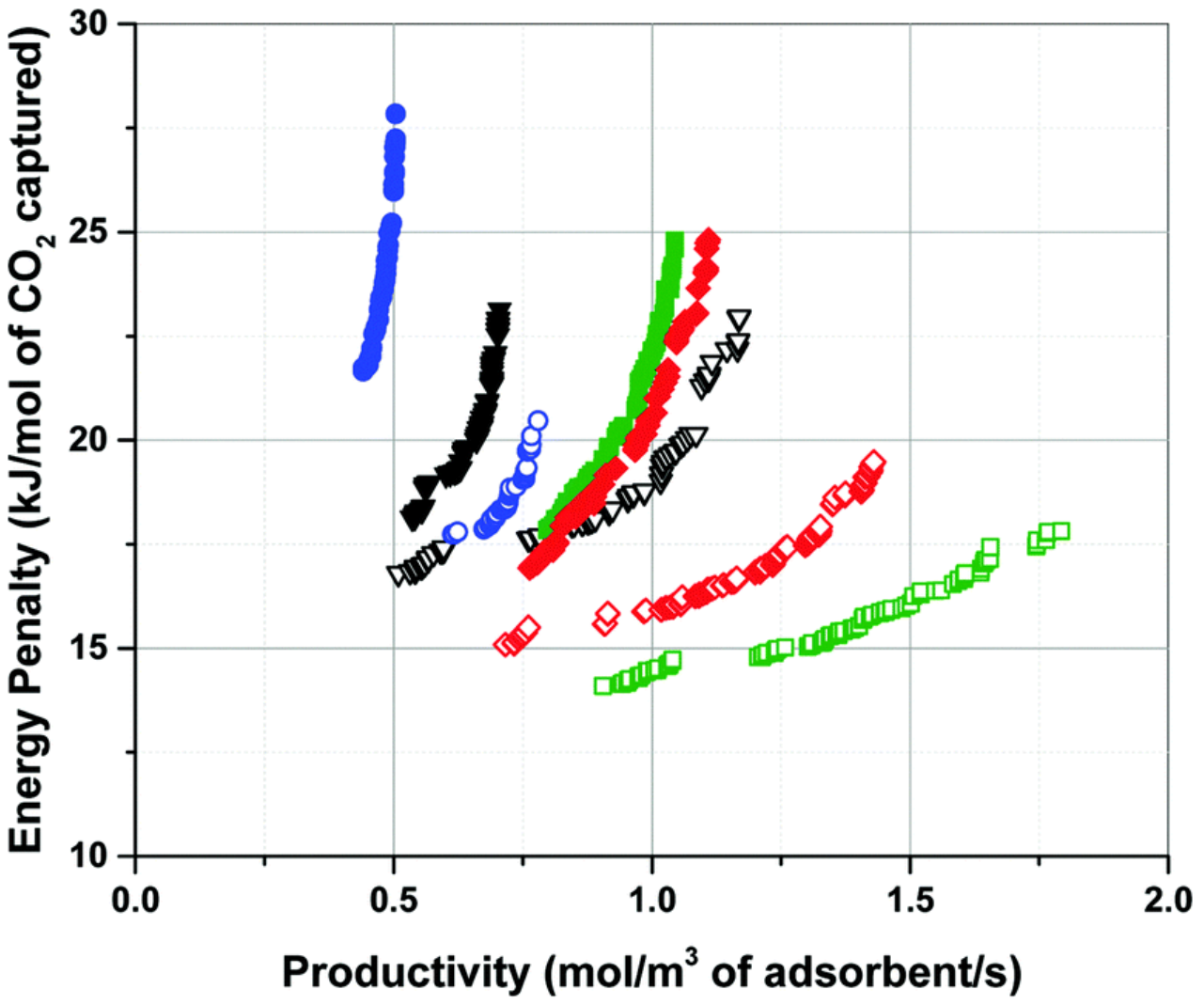

**Figure 30.** Pareto fronts of Cu-BTC (black triangle), MOF74-Ni (green square), Silicalite-1 (blue circle) and Zeolite 13X (red diamond) obtained using fixed values of pellet size and pellet porosity (solid symbol) and optimized values of pellet size and pellet porosity (open symbol) in process optimization of a 4-step VSA-LPP cycle. Reprinted with permission from Farmahini *et al.*, Energy & Environmental Science, 2020. 13(3): p. 1018-1037. Copyright (2020) Royal Society of Chemistry [101].

Therefore, consistent comparison of the materials' performance at the process level must take into account opportunities for the geometry optimization of the column packing.

*Numerical models for adsorption isotherms:* Adsorption models for process simulations are trained on the available experimental or simulation data and they should be able to give consistent and accurate representation of multicomponent adsorption equilibria. This is however not always the case and two different models trained on the same data may give different predictions of the binary (or multicomponent) adsorption equilibria (depending on the training protocols adopted) and hence, process level predictions [68, 402]. For example, Farmahini *et al.* [68] have recently demonstrated that the use of different recipes for fitting adsorption data to the DSL model can affect position of the energy-productivity Pareto fronts obtained from the process optimization. The authors have therefore proposed and validated a rigorous numerical protocol for consistent fitting of the widely used DSL model, in which the choice of temperature range, fitting constraints, and calculation of Henry's constant are standardized [68]. Similarly, several other studies have attempted to establish consistent routines for fitting equilibrium adsorption data [402-404], nevertheless none of the proposed procedures have been universally adopted by other groups, as a result of which consistency of various materials rankings that have been produced so far remains uncertain. As has been also discussed by Farmahini *et al.* [68], the ultimate test of the analytical models used in the process level simulations is their ability to predict binary and multicomponent adsorption data. This can be easily done using molecular simulations, as simulations of multicomponent systems come with relatively small additional effort compared to the simulation of single component systems. Hence, we propose this validation step to be adopted as a routine practice in the simulation community in order to probe the accuracy of the analytical models used for describing adsorption isotherms before performing any process simulation.

*Sensitivity analysis and propagation of errors:* From the studies reviewed so far, it is clear that overall process performance and ranking of porous materials depend upon calculation of a large group of parameters and model assumptions at both molecular and process level of descriptions (see **Table 6**). Despite some studies on the sensitivity of process performance to various input parameters and data [68, 101, 104, 105, 374, 405], it is yet to be established what level of accuracy is required for the full spectrum of parameters and models to guarantee consistent and comparable ranking of porous materials between different studies. One crucial element of such a study would be the investigation of errors propagation from molecular level all the way through to process modelling and optimization. For example, we can understand how the errors arising from the use of inaccurate molecular force fields in GCMC simulations for prediction of adsorption isotherms are combined with the errors resulting from the use of numerical models for fitting adsorption data, and what impact they will have together on the overall performance of the process. This analysis should be expanded to contain all sources of errors and uncertainties in the multiscale workflow so that the impact of the combined error on variation of purity-recovery and energy-productivity Pareto fronts can be identified. In this case, separation performance of each material will be represented by a range of Pareto fronts, rather than a single Pareto front. The results from this analysis is likely to change our perspective on what is currently perceived as the top performing candidates for post-combustion carbon capture in VSA/PSA processes.

*Consistency between various simulation packages:* One important aspect in developing consistent multiscale workflows for materials ranking is having access to widely-used and validated open-source software and case studies. As mentioned in Sections 6.1.2, 6.2.2 and 6.2.4, there are a number open-source molecular simulation packages for which several benchmark and case studies have been produced [169, 183, 226]. However, this has not been done for process simulation packages mainly because the majority of these software are not available as open-source. In fact, among all the software introduced in Section 6.3.6, only one code has been released as open source [105]. As shown by the molecular simulation community, having access to open-source software and clear case studies facilitates expedited development of new generations of software and tools for material screening studies. We believe this can be also true for the process simulation community and hope that the current review has been successful in demonstrating the importance of any efforts which can address the current gap.

## 8.3. Validation of Multiscale Screening Workflows

Despite recent advances in development of more sophisticated multiscale screening workflows, validation of the materials rankings produced by these frameworks is still an outstanding issue. As illustrated throughout this review, multiscale screening workflows have a modular structure in which various computational modules are put together to perform different types of simulations. The simplest workflow contains three modules in order to perform (1) GCMC simulation, (2) process modelling, and (3) process optimization. This can be further extended, if one decides to include quantum mechanical or MD simulations in the workflow. Normally, results of each module can be validated separately. For example, adsorption isotherms generated using GCMC simulation are routinely compared against equilibrium adsorption data obtained from experiments to ensure the accuracy of the predictions. At the process level, validation tests are conventionally carried out by reproducing column breakthrough curves, temperature, pressure, and concentration profiles from experiments [98, 99, 311, 406, 407] example of which is illustrated in **Figure 31** for a basic 4-step VSA cycle.

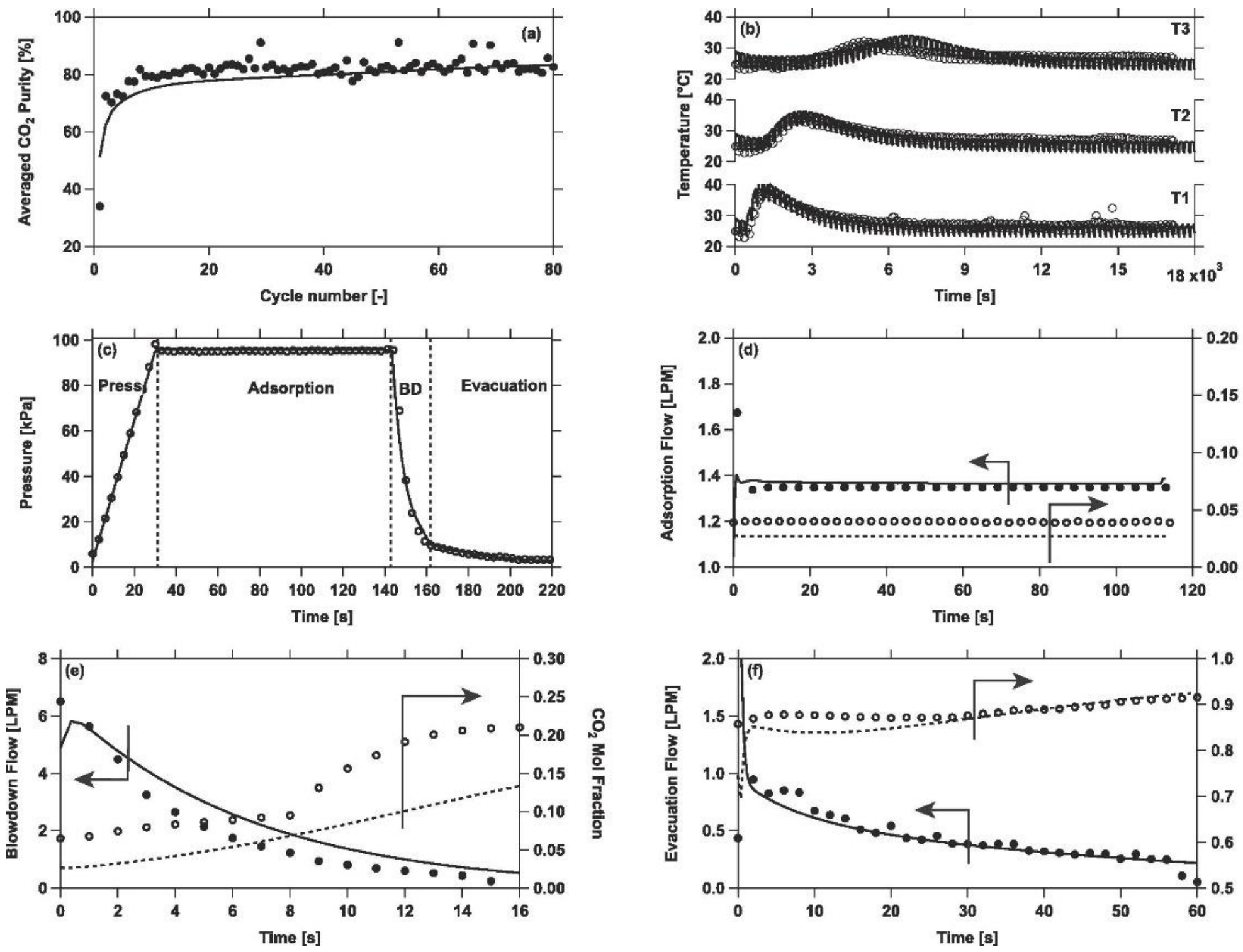

**Figure 31.** Example of validation of simulated transient histories against experimental data for adsorption of $CO_2$ and $N_2$ in Zeolite 13X using a basic 4-step VSA cycle. Evolution of $CO_2$ purity (a), temperature histories at three locations in the column (b), pressure history for one cycle at CSS (c), $CO_2$ composition and flow rate at the outlet of the adsorption step (d), $CO_2$ composition and flow rate at the outlet of the blowdown step (e), $CO_2$ composition and flow rate at the outlet of the evacuation step (f). Symbols represent experimental data and lines indicate numerical simulations. Reprinted with permission from Estupiñan Perez *et al.*, Separation and Purification Technology, 2019. 224: p. 553-563. Copyright (2019) Elsevier [99].

Efforts for validation of genetic algorithms which are used for multi-objective optimization of PSA/VSA processes are more recent and less wide-spread. In one recent example, the ability of multi-objective optimization techniques to guide the design of PSA/VSA processes have been shown by Perez *et al.* [99]. In this study, purity and recovery of $CO_2$ predicted through numerical optimization of a basic 4-step VSA cycle and a 4-step VSA cycle with LPP were replicated by experiment for post-combustion carbon capture using zeolite 13X as adsorbent [99] which is shown in **Figure *32***.

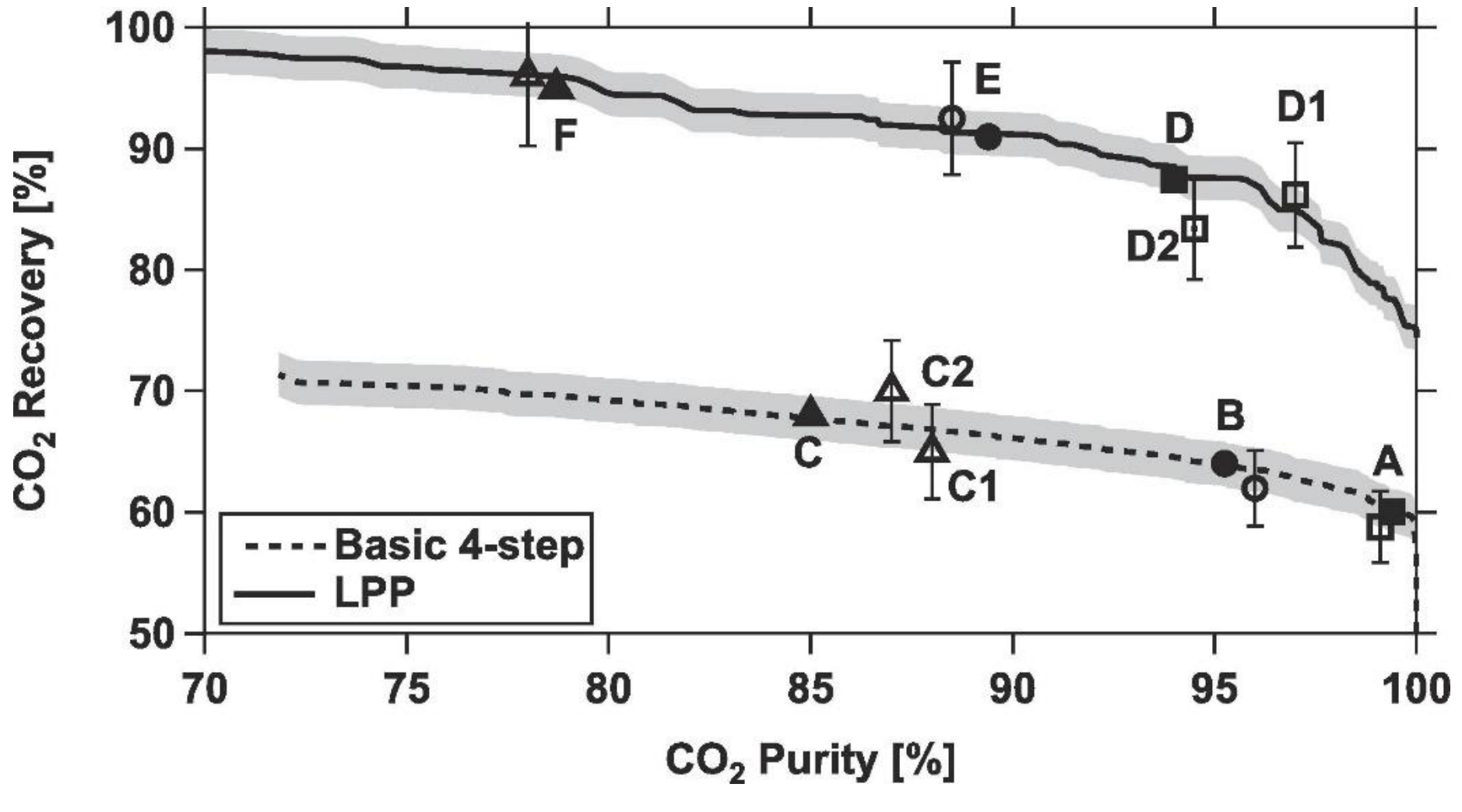

**Figure 32.** Pareto fronts corresponding to a basic 4-step VSA cycle (dashed line) compared to a 4-step VSA-LPP cycle (solid line). Closed symbols correspond to the optimized set of operating conditions that were implemented in experiment. Open symbols represent the corresponding purity and recovery values obtained from the experiment. The shaded region around the Pareto front denotes the 1.8% points confidence region arising due to a 10% uncertainty of selected model inputs. Reprinted with permission from Estupiñan Perez *et al.*, Separation and Purification Technology, 2019. 224: p. 553-563. Copyright (2019) Elsevier [99].

Unfortunately, it was not possible in this study to carry out any measurement to verify total energy consumption of the process against experimental data [99]. A pilot plant study conducted by Krishnamurthy *et al.* [98] for $CO_2$ capture using the same 4-step processes with Zeolite 13X report significant discrepancies between theoretical and experimental values of energy consumptions, while their analyses show overall quantitative agreement for purity and recovery, and somewhat modest agreement for productivity data (**Figure *33***) [98].

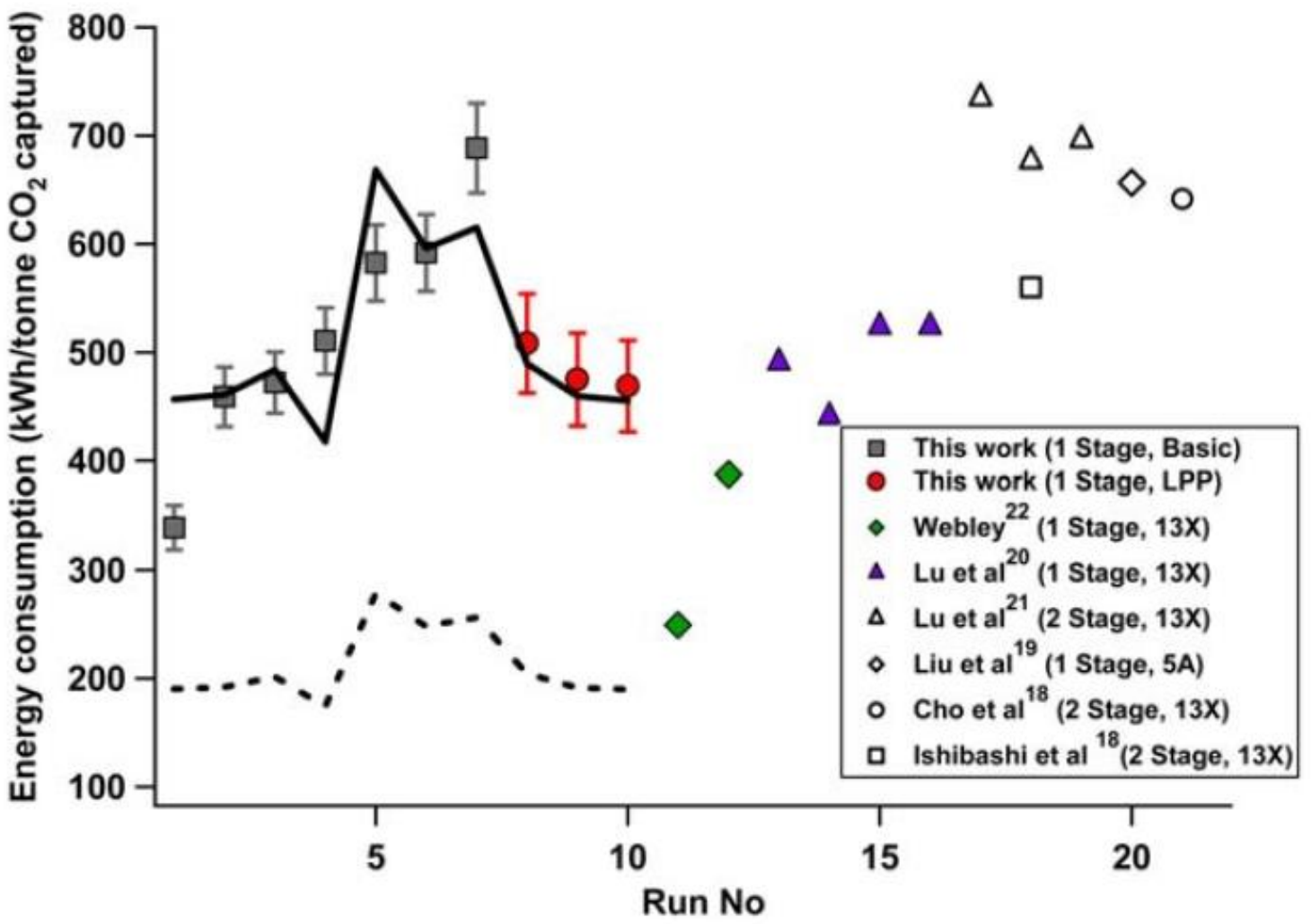

**Figure 33.** Energy consumption of pilot plant experiments conducted by Krishnamurthy et al. [98] compared with other data extracted from literature. The dotted line represents an efficiency of 72% for the VSA process, while the solid line denotes an efficiency of 30%. Note that all the experiments shown in this figure resulted in different purity-recovery values, thus care should be taken in comparing their corresponding energy values directly (The references in the inset are available from the original publication). Reprinted with permission from Krishnamurthy *et al.*, AIChE Journal, 2014. 60(5): p. 1830-1842. Copyright (2014) John Wiley and Sons [98].

Other pilot- or lab-scale studies also report reasonable agreement between experimental and theoretical values of purity and recovery at the given feed concentrations for separations of $CO_2$ using VSA/PSA processes [368, 369], while the discrepancy reported for the total energy consumption is still considerable [369]. This becomes especially crucial, if we remember that energy-productivity Pareto fronts obtained from multi-objective optimization of the process play a central role in performance-based ranking of porous materials. Here, it should be noted that estimation of total energy consumption of the process from simulation is particularly problematic due to difficulty of including exact characteristics of the valves, heat losses across the system, and the performance at a variable flow rate of the vacuum pumps and compressors.

With the surge in development of machine-learning approaches for modelling and optimization of separation processes, one would also need to consider additional tests for validation of these novel techniques. Some recent studies have reported promising cases where machine-learning based surrogate models developed for the optimization of PSA/VSA processes have been validated against Pareto fronts and cyclic steady state (CSS) column profiles that are mainly obtained from detailed process simulations [106, 312, 316] but some also from lab-scale experiments [312].

Despite all these efforts for validation of different computational modules of multiscale screening workflows, there is no single material ranking study in which the order of top performing materials has been confirmed experimentally. In fact, unless this final level of validation is achieved, it is unlikely that the top-performing materials proposed by various computational screening studies are going to found their way into any industrial application.

## 8.4. Other Challenges

### 8.4.1. Improving Efficiency of Process Optimization for Comprehensive Screening of Materials Space

Multiscale simulation of PSA/VSA processes for screening of large databases of porous materials requires extensive computational resources. In the screening workflow, process optimization is usually considered as a bottleneck where significant computational efforts are incurred [317]. Attempts have been made to improve computational efficiency of process optimization, through reducing dimensionality of the variable space in process optimization, by development of novel machine-learning methods (Section 6.3.5), or by hierarchical approaches, where the simplified process models are used first in the pre-screening studies, while the full process optimization focuses on a smaller subset of cases. Together, these methods pave the way not only for faster screening of large databases of porous materials but also for identifying the most efficient process configurations for a particular separation process. This can lead to better understanding of the material-process-performance relationships. Nevertheless, the remaining challenge is yet to tackle the magnitude of the material-process phase space. Currently, these methods have only been tested for screening of small sets of porous materials (<2,000) [103, 106] which is infinitesimal compared to the huge number of materials that has been discovered so far, as mentioned in Section 6.1.1. Also, experimental evidence for validation of numerical techniques that are used for expedited optimization of PSA/VSA processes are still scarce [312] and it is for the future studies to address this important limitation.

### 8.4.2. Multiscale Workflows for Unconventional Adsorbents

In addition to what has been discussed in this section, development of more advanced multiscale workflows for PSA/VSA/TSA processes can be envisioned where behaviour of more complex materials is simulated. An important example of such cases is the prediction of separation performance of many novel porous materials [266, 267, 408], and in particular materials with gaiting effects and phase-change adsorbents exhibiting step-shaped adsorption isotherms [289, 409-411]. Atomistic structures of these materials undergo considerable structural changes in response to external stimuli such as heat, pressure, humidity and adsorption of guest molecules [181, 409]. Simulation of adsorption process in this class of materials must capture the interplay between presence of adsorbate molecules and structural deformation of the framework using computational methods that go beyond conventional GCMC (*e.g.* the osmotic Monte Carlo method or hybrid MC/MD methods) [168, 412]. In addition to simulation of structural flexibility of these materials that must be handled at the molecular level, it is also crucial to develop more sophisticated analytical adsorption models that can capture stepwise shape of the isotherms in these materials as required for process simulation [289, 413, 414]. These two issues alone pose a significant challenge to the development of the future generation of multiscale simulation workflows for screening of flexible materials.

# 9. Current Perspective and the Future Outlook

## 9.1. Current Perspective

In this article, we reviewed the recent progress in the application of performance-based multiscale workflows for material screening in post-combustion carbon capture. To make it useful for our wide range of audience consisting of material scientists, computational modellers, and chemical engineers we introduced the basic principles involved in each element of the workflow and provided references to the available computational tools. We outlined what data are required at each level and showed how they can be calculated computationally without resorting to experiment. We also highlighted the issue of availability and completeness of the data, as well as the consistency of implementations for multiscale workflows. The article also summarized all the recent studies in the field, and as such can serve as a starting point for further developments. Before we can close this review with our concluding remarks, it is important to highlight the current perspective of the field and explain what actually the multiscale materials screening approach has achieved. Naturally, two questions emerge here:

*What processes and materials have we identified so far as the most promising candidates for post-combustion carbon capture?*

and:

*What is the impact of these findings on the industrial practices?*

To answer these questions, we have compiled the top performing material candidates which have been identified so far by the most comprehensive screening studies using detailed process modelling and optimization for post-combustion carbon capture from binary $CO_2/N_2$ flue gas mixtures. This is summarized in **Table *10*** and **Table *11*** where the top 10 materials from each study have been listed in order of performance. The tables contain results of 7 screening studies performed using 5 different PSA/VSA process configurations. **Table *10*** summarizes top-performing materials based on minimum energy consumption of the process, while **Table *11*** lists the candidates based on their maximum productivity.

**Table 10**. Ranking of top 10 materials based on minimum energy penalty

| Index | 4-step VSA with LPP (Khurana and Farooq[a]) [67] | 4-step Skarstrom PSA (Park et al.[b]) [73] | 4-step VSA with LPP (Balashankar and Rajendran[a]) [100] | 4-step VSA with LPP (Burns et al.[a]) [103] | Modified Skarstrom (Yancy-Caballero et al.[c]) [105] | FVSA (Yancy-Caballero et al.[c]) [105] | 5-step PSA (Yancy-Caballero et al.[c]) [105] |
|---|---|---|---|---|---|---|---|
| 1 | h8155527 | TASXIW | h8116500 | IISERP-MOF-2 | UTSA-16 | UTSA-16 | UTSA-16 |
| 2 | NAB | BIBXUH | h8297545 | IGAHED02 | Zeolite 13X | Cu-TDPAT | Ti-MIL-91 |
| 3 | UTSA-16 | TERFUT | h8210285 | XAVQIU01 | SIFSIX-3-Ni | Zeolite 13X | Cu-TDPAT |
| 4 | NaA | FAKLOU | h8180594 | YEZFIU | Ti-MIL-91 | Ti-MIL-91 | Zeolite 13X |
| 5 | h8124767 | MODNIC | h8116694 | NaA | Cu-TDPAT | SIFSIX-3-Ni | SIFSIX-3-Ni |
| 6 | ZIF-36-FRL | ZESFUY | IZA-WEI | ZIF-36-FRL | Ni-MOF-74 | Zn-MOF-74 | Zn-MOF-74 |
| 7 | h8291835 | RAXCOK | h8329775 | UTSA-16 | SIFSIX-2-Cu-i | | Mg-MOF-74 |
| 8 | ZIF-82 | SENWIT | IZA-BIK | HUTTIA | Zn-MOF-74 | | |
| 9 | ZIF-78 | CUHPUR | ZIF-Im-h8127937 | QIFLUO | Mg-MOF-74 | | |
| 10 | ZIF-68 | SENWOZ | IZA-MON | GAYFOD | | | |

[a] 4-step VSA with LPP using a packed-bed adsorbent system. Feed composition: 15% CO2/85% $N_2$ at 298 K. Optimization constraints: 95% $CO_2$ purity and 90% $CO_2$ recovery.

[b] 4-step PSA using a hollow fiber adsorbent model. Feed composition: 14% CO2/86% $N_2$ at 243 K. No optimization constraint imposed on purity and recovery.

[c] PSA cycles using packed-bed adsorbent system. Feed composition: 15% CO2/85% $N_2$ at 313 K. Optimization constraints: 90% $CO_2$ purity and 90% $CO_2$ recovery.

**Table 11.** Ranking of top 10 materials based on maximum productivity

| **Index** | **4-step VSA with LPP (Khurana and Farooq[a]) [67]** | **4-step Skarstrom PSA (Park et al.[b]) [73]** | **4-step VSA with LPP (Balashankar and Rajendran[a]) [100]** | **4-step VSA with LPP (Burns et al.[a]) [103]** | **Modified Skarstrom (Yancy-Caballero et al.[c]) [105]** | **FVSA (Yancy-Caballero et al.[c]) [105]** | **5-step PSA (Yancy-Caballero et al.[c]) [105]** |
|---|---|---|---|---|---|---|---|
| 1 | UTSA-16 | SENWOZ | h8315144 | GAYFOD | UTSA-16 | Cu-TDPAT | UTSA-16 |
| 2 | NaA | SENWIT | h8328529 | WUNSII | Zeolite 13X | UTSA-16 | Zeolite 13X |
| 3 | h8155527, h8124767 | WONZOP | IZA-MON | IISERP-MOF-2 | Cu-TDPAT | Zn-MOF-74 | Cu-TDPAT |
| 4 | ZIF-36-FRL | UTEWUM | IZA-RRO | UTSA-16 | Ni-MOF-74 | Zeolite 13X | Zn-MOF-74 |
| 5 | NAB | OJICUG | IZA-JBW | YEZFIU | Mg-MOF-74 | Ti-MIL-91 | Ti-MIL-91 |
| 6 | CaX | BIBXUH | h8206103 | IGAHED02 | SIFSIX-3-Ni | SIFSIX-3-Ni | SIFSIX-3-Ni |
| 7 | ZIF-78 | SENWAL | IZA-WEI | XAVQIU01 | SIFSIX-2-Cu-i | | Mg-MOF-74 |
| 8 | h8272272 | FEFDAX | h8313037 | NaA | Ti-MIL-91 | | |
| 9 | Zn-MOF-74 | RAXCOK | ZIF-Im-h8055553 | HUTTIA | Zn-MOF-74 | | |
| 10 | MgX | CUHPUR | ZIF-Im-h8164555 | ZEGSUB | | | |

[a] 4-step VSA with LPP using a packed-bed adsorbent system. Feed composition: 15% $CO_2$/85% $N_2$ at 298 K. Optimization constraints: 95% $CO_2$ purity and 90% $CO_2$ recovery.

[b] 4-step PSA using a hollow fiber adsorbent model. Feed composition: 14% $CO_2$/86% $N_2$ at 243 K. No optimization constraint imposed on purity and recovery.

[c] PSA cycles using packed-bed adsorbent system. Feed composition: 15% $CO_2$/85% $N_2$ at 313 K. Optimization constraints: 90% $CO_2$ purity and 90% $CO_2$ recovery.

Evidently, the answer to the question raised at the beginning of this section does not seem to be straightforward. This is because different screening studies have employed various process and cycle configurations for assessment of materials performance at the process level. We also need to be aware that these studies draw their candidates form different databases of materials, so the fact that a particular material does not appear in a toped ranked group, may simply indicate that it was not included in the original screening set. For the studies listed in **Table *10*** and **Table *11***, this includes 4-step VSA cycle with LPP, 4-step Skarstrom-based PSA cycle, 5-step PSA cycle, and FVSA cycle. The use of different process configurations inevitably makes direct comparison of materials performance problematic. For example, according to Yancy-Caballero *et al.* [105], Mg-MOF-74 appears to be among the top performing materials in the modified Skarstrom and 5-step PSA cycles but not in the FVSA cycle (as seen in **Table *10*** and **Table *11***). Nevertheless, the issue is beyond the use of different process configurations, because the hierarchy of materials rankings are not consistent even within those studies that have used a similar cycle (*e.g.* 4-step VSA with LPP). A prominent example here, is the position of UTSA-16 in **Table *10*** in which UTSA-16 outperforms NaA according to the ranking by Khurana and Farooq [67], however its performance is found to be poorer compared to the same material according to the study conducted by Burns *et al.* [103]. The same is true if we compare position of UTSA-16 with ZIF-36-FRL in the two studies mentioned above in **Table *10***. This could be due the use of different model assumptions at the molecular or process levels (*e.g.* different force fields used for molecular simulations, different numerical protocols employed for fitting adsorption isotherms, *etc*). These observations clearly demonstrate our point about importance of consistent implementation of multiscale screening workflows which is highlighted throughout this review and particularly discussed in Sections 8.1 and 8.2.

On the other hand, in the tables above and in the discussion, we deliberately did not put the actual values of the energy penalty or productivity as we believe it would be an inconsistent comparison of studies done on vary different basis. However, without the actual numbers, we also need to be careful in our criticism of the consistency of ranking – the data presented in **Tables *10*** and ***11*** do not tell us *how close* materials are in terms of the numerical performance. Therefore, a more comprehensive discussion of the meaning of the ranking is not possible without the accompanying analysis of the propagation of uncertainties.

So what is the impact of the studies reviewed above? From an engineering point of view, they have identified several candidates that are very promising for post-combustion carbon capture. From the above tables, many materials have energy consumption and productivity values that are superior to those of Zeolite 13X which is the current industrial benchmark. Examples of these materials include NAB, UTSA-16, NaA, and ZIF-36-FRL [67]. IISERP-MOF-2 is another example whose energy consumption is less than that of amine-based absorption technology, while its productivity surpasses Zeolite 13X [103]. This new MOF is known to have excellent stability against moisture and acid gas environments [415]. The above list of top-performing candidates also include other MOFs with high kinetic stability in presence of water such as UTSA-16 and SIFSIX-2-Cu-i [416]. As a LTA zeolite, NaA is another promising candidate which is currently synthesized in industrial scales [103, 417]; hence its application for carbon capture from dried flue gas can be more economic compared to other candidates that are not currently mass produced [103].

In addition to identifying promising materials and processes, application of performance-based screening strategies have led to important learning outcomes, a prominent example of which is the role of nitrogen adsorption for material performance. Now, we know that we do not need to limit our search for ideal adsorbents to materials with high CO2 capacity, but rather we should look for the candidates which have low nitrogen uptake.

Considering what was discussed in this section, it is reasonable to state that performance-based screening of porous materials have significantly improved our ability to realistically identify a range of promising materials that can be considered for lab-scale and pilot-plant examinations. This will be especially true if the research community focuses on addressing the challenges that were identified and discussed in Section 8 of this review to improve consistency and accuracy of the multiscale screening workflows.

## 9.2. Future Outlook

After reflecting on the state-of-the-art in the field, here we provide our concluding remarks and proposals for the future direction of the field:

**Beyond post-combustion carbon capture:** In this article we focused on the post-combustion carbon capture as it is a very challenging, societally relevant and most investigated process. However, we believe the multiscale screening approaches reviewed here will become a new way to design and appraise material options in other separation applications. Decarbonization of the chemical industry by 2050 cannot be achieved with carbon capture from power plants alone and will require a wider range of technologies. These technologies will deal with different process conditions (primarily different levels of carbon dioxide concentration), and will be operating on a relatively small scale of the processes, compared to post-combustion capture (meaning, smaller amounts of the material will be required). For these processes, it is likely that faster cycles will be used to reduce the footprint of the units, especially in retrofit applications. This will in turn lead to larger effects of mass transfer and heat transfer limitations – precisely the challenges that need to be explored within the multiscale framework.

Air separation is a very useful case to consider for benchmarking multiscale modelling approaches. Production of oxygen is an equilibrium driven separation where the light component is produced. Therefore, simpler process configurations will work well in this case and advances are more likely to be in the definition of the ideal structural properties of the formed materials. Production of nitrogen is a kinetic separation that requires materials with small pore openings. Again, although this process is well-established, the data accumulated over the years can provide a benchmark to understand whether accurate a-priori predictions based on force fields that are efficient in equilibrium calculations can be used also in predicting diffusivities.

Finally, we envision that other separation processes, such as membrane separations, where the performance of the process is defined by the material used, will also benefit from multiscale screening workflows in producing more realistic, performance-based rankings of the available materials.

**The role of ML methods will grow:** It is already evident that the scope of multiscale screening methods will be expanding along with the range of available materials. This, combined with a large number of parameters, leads to multidimensional material-process configuration-performance space, which is very challenging for conventional optimisation and design tools. Machine learning methods have already been used successfully to accelerate the optimisation. The growing availability of data across all scales opens exciting opportunities to use ML not only to extend the search space even further but also to use ML for other aspects of the multiscale workflow such as the design of force fields and the prediction of the best material structure [418]. This direction is both very promising and still widely unchartered. Hence, there is a strong incentive to more fully explore potential of the ML models to accelerate process-level screening workflows as well as material properties.

**Quality data, reproducibility of results and consistency of comparison:** we believe these aspects will be a singular, most important barrier for the multiscale approaches to make an actual impact through identifying both better *and* realistic options for carbon capture. The molecular simulation community has already produced a substantial number of screening studies for carbon capture. Similarly process level community has been examining various options for both processes and materials (but not in a large scale screening mode) for this task. The multiscale methods emerging from the combination of these two realms have been reviewed here. However, these studies use different assumptions, models and conditions which makes systematic comparison of their results difficult. One possible proposal for the simulation community would be an open call for the systematic comparison of the currently existing process modelling codes (including commercial ones) and model assumptions using a reference case study. This will be a significant step towards building confidence in ranking of the materials.

**Techno-economic analysis and scale-up of the process:** Development of multiscale screening studies should eventually go beyond the process-level. This is because, similar to any technology, the ultimate driver for commercialization of adsorption-based carbon capture is also the cost. Therefore, the screening studies at the process level must be linked with techno-economic analyses where the ultimate design objective is to reduce the overall cost of $CO_2$ capture and concentration (CCC) at industrial scales. Although, there has been some attempts at this direction [69-71, 104], there is still a dire need for integrated adsorbent-process optimizations that are properly linked with techno-economic assessments of the CCC technology. Such studies, must realistically assess capital and operating costs of the process including the cost of adsorbent, operational lifetime of key components of the cyclic process, realistic efficiencies of vacuum pumps, process scheduling, and finally the costs associated with scale-up of the technology and its footprint requirements [69, 71, 104].

From few techno-economic studies conducted so far, it seems, at the end of the day, the cost of the $CO_2$ capture using VSA/PSA technology is generally higher than that of the current technological benchmark which is the amine-based solvents absorption separation, despite promising values of energy penalty presented here [71, 104]. Recent studies have consistently noted that that the cost of adsorbent has a major impact on the ultimate cost of the $CO_2$ capture process [71, 104]. It is not difficult to anticipate this for materials (such as MOFs) whose synthesis is expensive and limited for large-scale productions. In the case of VSA process, another major challenge is the limitation of maximum feed velocities that can be employed in beaded/pelletized adsorbents [71]. Apparently, this results in the requirement of a large number of adsorption columns and multiple parallel trains which in turn poses other technological challenges associated with practicality of deploying large and complex capture plants [69, 71]. From this perspective, future efforts in the domain of adsorption-based carbon capture technology must focus on addressing the following technological barriers:

- Development of monolith adsorbents [419, 420] or parallel-passage contactors [421] that can increase productivity of the process through reducing pressure drop and enhancing kinetic [69, 71].

- Development of better and cheaper adsorbents that can be economically mass produced.

- Development of adsorbents with improved thermal stability and resistance to moisture that can operate under rapid cyclic conditions of PSA/VSA with reasonable operational lifetime.

In this context, the following should be particularly undertaken by the simulation community for future materials screening studies using the multiscale workflows:

- Materials screening using structured adsorbents [419] such as monolith (as opposed to pelletized adsorbents) in PSA/VSA systems. These adsorbents do not suffer from fluidization at high feed velocities, and can be used for cycle intensification and increasing productivity. A number of recent studies have already developed new models for implementation of structured adsorbents in PSA/VSA simulations [303, 346].

- Focusing the computational efforts on screening of materials that are known to meet the essential criteria stated in this section (*e.g.* low price, water resistant, and thermally stable materials).

**The ultimate challenge in post-combustion carbon capture still remains:** It is important to recognize, that despite 15 or so years of computational materials screening studies for carbon capture, there is no pilot-scale plant that is designed to operate using a MOF or ZIF as an adsorbent. While the aim of developing an *in-silico* route to finding optimal materials is a sound aspiration, there is the need to include in the selection process also the ability to synthesize the new materials and assess their stability against thermal cycling and exposure to contaminants and moisture. Predicting the stability of the materials is a challenging area. There are also other technical issues associated with scale-up that were mentioned earlier in this section. Hence, it is clear that there are still significant challenges towards industrial implementation of carbon capture technologies based on novel materials. Notwithstanding, we believe development of more advanced mutliscale methods (such as those reviewed in this article) is an important step for accelerating our progress towards this objective.

**Acknowledgements**

This work has been supported by the UK Engineering and Physical Sciences Research Council (EPSRC), grant EP/N007859/1. The project has made use of the computational resources provided by the Edinburgh Compute and Data Facility (ECDF) (http://www.ecdf.ed.ac.uk). We would like to thank Dr. Miguel Jorge for designing the schematics in Figure 12 and Figure 13, and Dr. David Danaci for sharing the original data for **Figure 29**. We would also like to thank Prof. Arvind Rajendran and Dr Mauro Luberti for insightful discussions about different sections of this review. We would like to thank Mr Olivier Baudouin for additional comments on ProSim DAC software. SINTEF's contribution for this manuscript was supported by the SINTEFs internal publication project number 102005015-63.